\documentclass[sigconf]{acmart}

\AtBeginDocument{%
  }

\copyrightyear{2026}
\acmYear{2026}
\setcopyright{cc}
\setcctype{by}
\acmConference[CHI '26]{Proceedings of the 2026 CHI Conference on Human Factors in Computing Systems}{April 13--17, 2026}{Barcelona, Spain}
\acmBooktitle{Proceedings of the 2026 CHI Conference on Human Factors in Computing Systems (CHI '26), April 13--17, 2026, Barcelona, Spain}
\acmDOI{10.1145/3772318.3790781}
\acmISBN{979-8-4007-2278-3/2026/04}

\acmSubmissionID{5481}

\usepackage{float}
\usepackage{graphicx}

\begin{document}

\title{Teen Vigilance: 
Navigating Risky Social Interactions on Discord
}

\author{Elena Koung}
\orcid{0009-0006-2946-840X}
\affiliation{%
  \institution{The Pennsylvania State University}
  \city{University Park}
  \state{Pennsylvania}
  \country{USA}
}
\email{elenakoung@psu.edu}

\author{Yunhan Liu}
\affiliation{%
  \institution{The Pennsylvania State University}
  \city{University Park}
  \state{Pennsylvania}
  \country{USA}
}
\email{liuyunhan@psu.edu}

\author{Zinan Zhang}
\affiliation{%
  \institution{The Pennsylvania State University}
  \city{University Park}
  \state{Pennsylvania}
  \country{USA}
}
\email{zzinan@psu.edu}

\author{Xinning Gui}
\affiliation{%
  \institution{The Pennsylvania State University}
  \city{University Park}
  \state{Pennsylvania}
  \country{USA}
  }
  \email{xinninggui@psu.edu}

\author{Yubo Kou}
\affiliation{%
  \institution{The Pennsylvania State University}
  \city{University Park}
  \state{Pennsylvania}
  \country{USA}
  }
\email{yubokou@psu.edu}

\renewcommand{\shortauthors}{Koung et al.}

\begin{abstract}
  Teenagers are avid users of Discord, a fast-growing platform for synchronous communication where they often interact with strangers. Because Discord combines private DMs, semi-private voice channels, and public servers in one place, it creates a hybrid environment that can produce complex—and underexplored—safety risks for teenagers. Drawing on 16 interviews with teenage Discord users, this study examines their strategies for navigating risky social interactions in the platform. Our findings reveal that when teenagers encounter risks during social interactions, they exercise vigilance by evaluating suspicious interactions before forming friendships, using safety tools, and engaging in controlled risk-taking to safeguard their privacy and security. At the community level, they mitigate risks through selective participation in servers, a practice supported by vigilant governance structures. We discuss how vigilance enables teenagers to act during risky encounters to protect themselves, advancing understanding of teenagers’ agency in risk navigation and informing teen-centered designs for safer online environments.
\end{abstract}


\keywords{Social Media/Online Communities, Teens, Empirical study that tells us about people, Qualitative Methods}

\maketitle

\section{Introduction}
Social platforms have become integral to teenagers’ social lives because they offer unique environments for meaningful online social experiences, providing opportunities for play, community, and identity exploration ~\cite{kennedyShiftOfflineOnline2016, perez-torresSocialMediaDigital2024}. Recently, the most commonly used social platforms for teenagers are YouTube, TikTok, and Instagram ~\cite{faverioTeensSocialMedia2025} Each of these platforms are known for their user-generated content in algorithmic feeds ~\cite{alluhidanTeenTalkGood2024}, broadcasting to large public audiences ~\cite{boydSocialNetworkSites2007, schlosserSelfdisclosureSelfpresentationSocial2020, zhangBurstCollaborativeCuration2025} and curated posts ~\cite{balleysSearchingOneselfYouTube2020, jangTeensEngageMore2016, shutskoUserGeneratedShortVideo2020, thomasInstagramToolStudy2020}. In contrast, Discord is designed for gamers to socialize through conversations and interactions with other users ~\cite{DiscordOurMission}. Discord stands out from other gaming-related social platforms: 35\% of Gen Z (ages 13--25) use Discord—more than Reddit (30\%) or Twitch (24\%)~\cite{backlinkoteamDiscordUserFunding2023}. Its prominence is reinforced by its popularity among teen gamers, where 44\% of teens gamers use Discord~\cite{jeffreygottfriedTeensVideoGames}.

The widespread, fast-paced interactions within Discord servers make platform moderation and governance central to shaping user’s safety. Discord’s moderation infrastructure is significantly different from other commonly used social platforms, where platform safety relies on algorithmic moderation to remove harmful content ~\cite{hePlatformGovernanceAlgorithmBased2025, maHowAdvertiserfriendlyMy2021}. In each Discord server, moderators have extensive control over content moderation and governance of members ~\cite{leeMappingCommunityAppeals2025, robinsonGovernanceDiscordPlatform2023, yoonItsGreatBecause2025}. This decentralized moderation infrastructure is similar to Reddit \cite{wuHowYouQuantify2023}, where volunteer community moderators take on the responsibility of removing problematic and harmful content from forums ~\cite{chandrasekharanCrossmodCrossCommunityLearningbased2019, jhaverDoesTransparencyModeration2019}. However, Discord servers’ real-time voice and text interactions makes harmful exchanges harder to monitor and moderate ~\cite{jiangModerationChallengesVoicebased2019}, which could expose users to harassment or inappropriate content. 

Given Discord’s large, diverse young user base \cite{kumarDiscordUsersMarket2025} and its moderation challenges \cite{jiangModerationChallengesVoicebased2019}, understanding how teenagers navigate risky online social interactions is critical. Recently, one teenager went missing to meet people she had interacted with on Discord \cite{larsonMissingFremontTeen} and Discord servers may host illegal activities and the distribution of Child Sexual Abuse Material \cite{leeMappingCommunityAppeals2025}. In an extreme case, Discord was linked to the violent “Unite the Right” rally in Charlottesville, prompting scholars to associate its usage with hate groups \cite{heslepMappingDiscordsDarkside2024, brownHatewareOutsourcingResponsibility2019}. In response to increased scrutiny from journalists and academics between 2017 and 2019, Discord strengthened its policies and shifted towards more transparent safety practices \cite{discordsafetyDiscordTransparencyReports}.

While teenagers’ online safety has been addressed on multiple social media platforms, Discord remains as an underexplored site to understand how teenagers navigate risky online social interactions. Prior research has examined safety concerns on Instagram around self-presentation \cite{hernandez-serranoAnalysisDigitalSelfPresentation2022, yauItsJustLot2019, vanouytselAdolescentsSexySelfPresentation2020} and exposure to harmful content \cite{aliGettingMetaMultimodal2023}. Notably, Alluhidan et al. highlighted numerous risks and negative experiences on popular social media platforms, such as social drama on TikTok, body shaming and cyberbullying on Instagram, and mental health struggles \cite{alluhidanTeenTalkGood2024}. Despite increasing media attention on safety risks teenagers face on Discord—such as grooming, sexual exploitation, and exposure to explicit content \cite{mcfaddenDiscordAppExposes2025, gogginDiscordServersUsed2023, kellyDarkSideDiscord}—little research has examined how teenagers themselves mitigate these risks in online social experiences. We propose this research question to address the emerging concerns around teenagers’ safety on Discord: 

\textbf{How do teenagers navigate risky social interactions on Discord?}

We addressed this RQ by conducting semi-structured interviews with teenagers (aged 13-17) in the United States who use Discord and have experienced risky social interactions. We used Reflexive Thematic Analysis (RTA) \cite{braunDoingReflexiveThematic2022} to examine the ways teenagers have experienced risky interactions online and what can be learned from their existing practices. Finally, we established how teenagers’ vigilance enables controlled exploration of unfamiliar online spaces. 

Teenagers navigated these interactions by balancing privacy and security concerns with social motivations, seeking friendship, collaboration, and entertainment while encountering risks such as account scams or uncomfortable interactions. At the community level, teenagers evaluated community cultures, used platform features to set boundaries, and sometimes contributed to moderation processes to reduce risks. Together, these practices illuminate how teenagers’ vigilance supported their safe participation across dyadic and community contexts. 

Despite Discord being a widely used platform among teenagers, there is still very limited empirical research on how teens experience and manage safety within it. This study directly contributes to discussions about teenagers’ online safety by presenting how their \textbf{vigilance during risky social interactions is a proactive strategy that differs from previously studied resilient methods}. Then, we discuss how teenagers' \textbf{vigilant governance of Discord maintains safe communities by enabling timely reaction to risk}. Lastly, this study identifies design implications for social platforms to improve safety settings for teenagers.
\section{Background}
Discord was established in 2015 to provide an online space for gamers to talk with friends while gaming \cite{DiscordOurMission}. The platform requires users to be 13 years old for account creation and currently supports over 259 million monthly active users worldwide, with the largest market residing in the United States \cite{shewaleDiscordUsersStatistics2025}. While over 90\% of Discord users play games \cite{DiscordOurMission}, Discord has expanded beyond its game-focused purpose and has become a broader social platform where users continue conversations beyond gameplay, exchange information, and meet new people. 

Discord users can socialize with new people by joining \textbf{servers} and interacting with their members. Servers are a starting point for interpersonal discovery, where each server functions as a community space structured through a set of interactive channels and tools. Servers enable users to bond over community interests by engaging in both real-time and asynchronous interactions, largely related to gaming, memes, and unconventional interests for community connection \cite{joshiContextualizingInternetMemes2024, DiscordOurMission}. Given that Discord supports multiple communication modalities, interactions may unfold through text-based channels or voice-based channels.

Conversations on Discord are restricted within a server where only other server members can engage with and participate in the server discussion. While other forums like Reddit may not require membership to view public discussions, Discord interactions require membership in the server, which is established through invitation. For instance, in Figure 1, ‘username1’ and ‘teenager’ are friends and have been chatting through Discord Direct Messages. One day, ‘username1,’ who is a member of ‘ServerName,’ sends a server invite link to ‘teenager’ so they may join ‘ServerName’ to socialize with more friends of ‘username1.’ Any Discord user can create their own server, resulting in a diverse landscape of community interests and population, ranging from small groups of friends to organizations with thousands of members \cite{jiangModerationChallengesVoicebased2019}. 

\begin{figure*}[h]
    \centering
    \includegraphics[width=\textwidth]{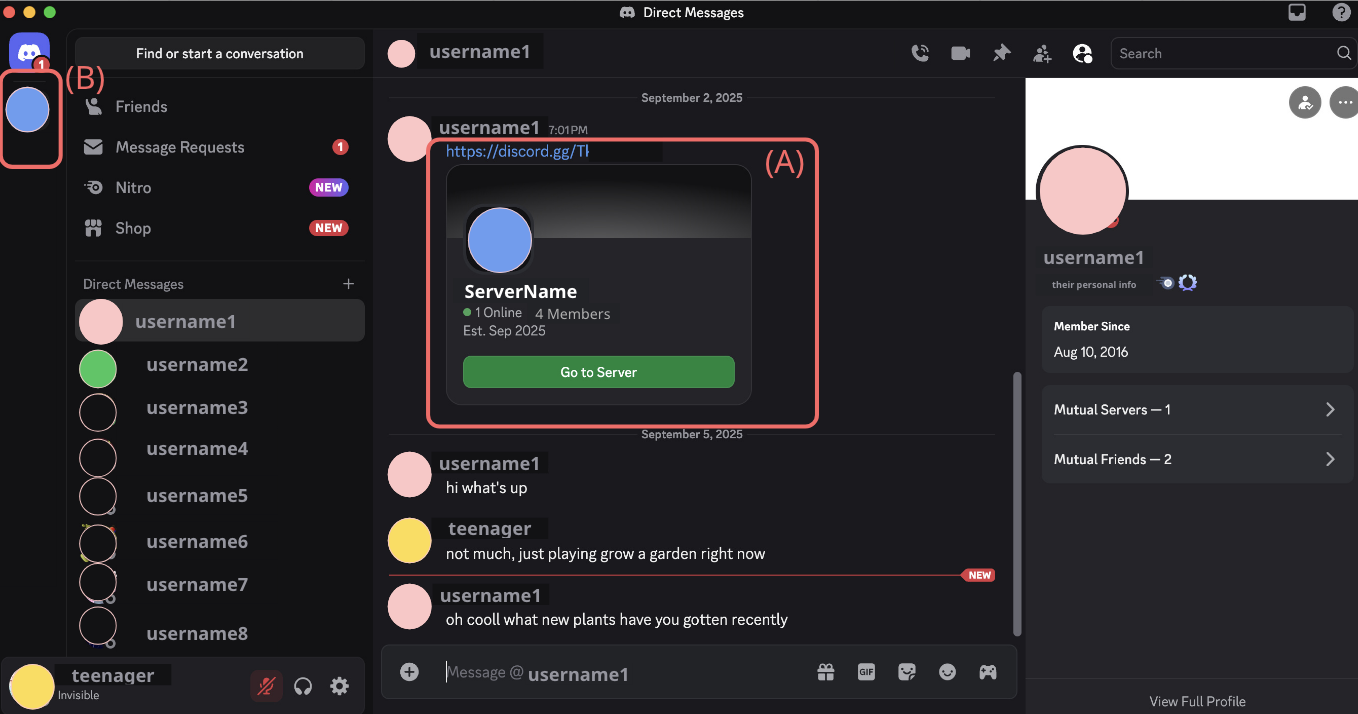}
    \caption{Discord’s Direct Message Interface allows users to talk to others privately. “username1” sends an invite to the server: “ServerName”(A). Then, “teenager” accepts the invite by clicking on the link so that they can further interact in a shared server space. Users can navigate to joined servers by clicking on the server icon (B).}
    \label{fig:1}
\end{figure*}

Within a server, \textbf{text channels} serve as forum-like spaces where members engage in asynchronous conversation, post images and memes, and participate in ongoing discussions. These channels allow users to revisit message history, enabling interactions that unfold over time and support both casual chatter and more persistent topic-based threads. Social interactions on Discord relies on server membership and members’ deep engagement with others in the server, which are often sustained through multiple conversations. Figure 2 illustrates a conversation between three members of ‘ServerName’, where they create a welcoming environment and invite each other to play games together. This sets Discord’s social interactions apart from other platforms, where common social interactions with strangers on such platforms often take place as public engagement in the form of views, likes, and comments. 

\begin{figure*}[h]
    \centering
    \includegraphics[width=\textwidth]{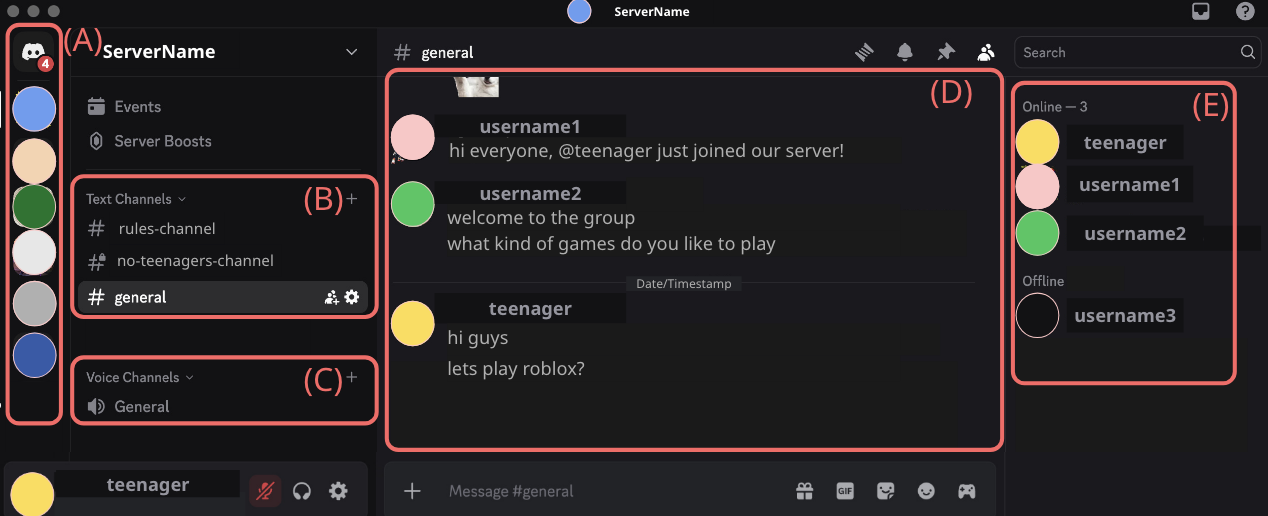}
    \caption{Discord server of “ServerName.” Server list can be seen in the left sidebar (A). Text Channels (B), one for rules, one locked channel, and one general chat channel. It includes one Voice Channel (C). Members can chat together in general chat (D). Member list can be seen in the right sidebar (E).}
    \label{fig:2}
\end{figure*}

Alongside text channels, servers typically include \textbf{voice channels}, which facilitate synchronous, real-time communication. In these spaces, members can talk freely, share their screens, and use features such as soundboards to play audio clips during conversations. The immediacy of voice channels often leads to dynamic interactions in group calls that resemble in-person hangouts. For example, voice channels enable first-time interactions with unfamiliar users through real-time voice chat. Moreover, there is a special feature in voice chat where users can play sound clips from a soundboard (Figure 3) to react to conversations and other live activities \cite{DiscordSoundboardGuide2025}. Other voice channel features, such as low-latency voice chat and screen sharing, were introduced to enhance multiplayer gaming experiences \cite{discordDiscordOpensIts}.

\begin{figure*}[h]
    \centering
    \includegraphics[width=\textwidth]{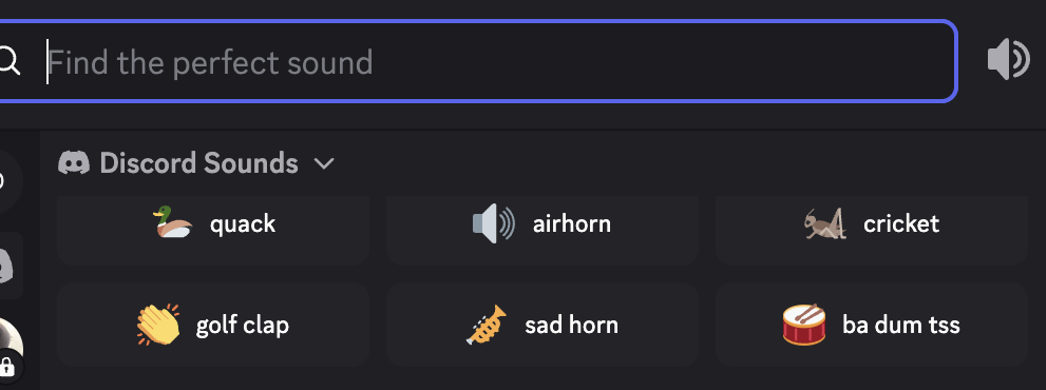}
    \caption{Soundboard menu featuring various sound clips. After joining a voice channel, users can navigate to the soundboard and click on sound clips that plays for everyone in the call to hear.}
    \label{fig:3}
\end{figure*}

Each server also displays a \textbf{member list}, allowing users to see who else is present in the community. Membership is generally visible to all server participants, and clicking on a username reveals a user profile with basic information such as status, roles, and activity. Importantly, servers have designated member moderators who hold elevated permissions to manage behavior, enforce community rules, and oversee the health of interactions within the space. Bots may also be server members that can respond to user commands and play music, host minigames, and share memes to create an interactive and entertaining experience \cite{seeringDyadicInteractionsConsidering2019}.

Discord’s broader interface offers a \textbf{server list} and allows users to navigate across multiple communities. This is illustrated in Figure 2 as the left vertical bar of icons. These servers often reflect users’ interests, such as gaming, creative hobbies, or fandoms, and provide access to distinct social groups. In addition to server-based interactions, Discord supports private conversations through direct messages and small group chats, enabling more individualized or intimate communication outside of the public server environment.

Discord’s conversation-centered design downplays traditional forms of user-generated content, such as images or short- and long-form videos. Instead, \textbf{user-generated content on Discord} primarily consists of users contributing through text and voice interactions, sharing memes, stories, and experiences in ongoing dialogue. User-generated content on Discord is disseminated and discovered by users through exploring servers and their multi-modal channels. Unlike TikTok or Instagram, which use algorithmic content recommendations to distribute user-generated content, Discord does not algorithmically surface channels or content to users. Instead, Discord presents featured servers on a Discover page (Figure 4) that showcases multiple interest communities, ranging from gaming and entertainment to tech and education. 
\begin{figure*}[h]
    \centering
    \includegraphics[width=\textwidth]{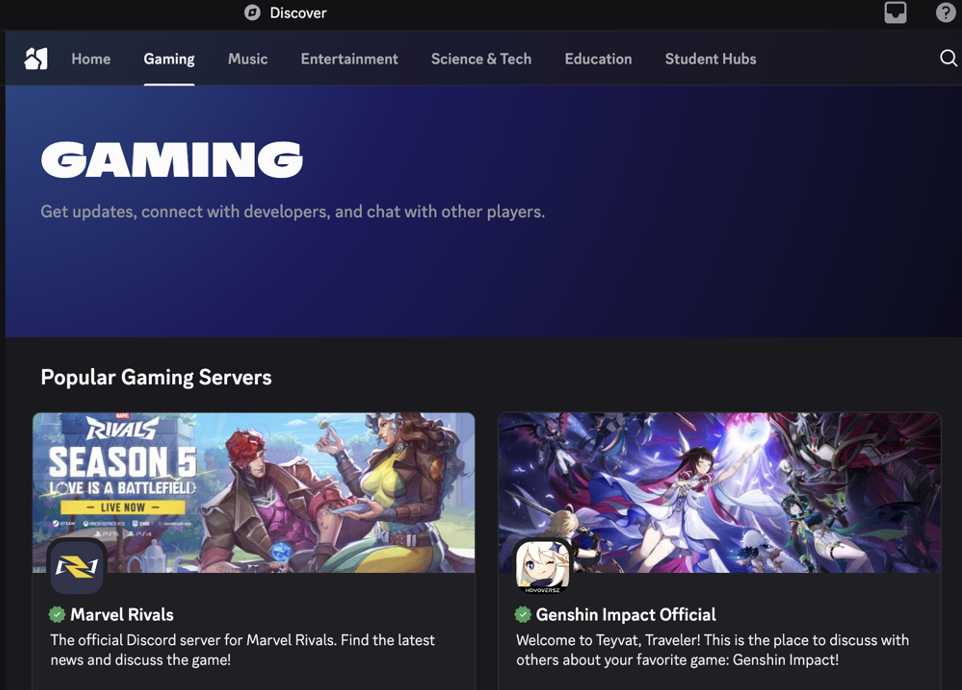}
    \caption{Discover menu featuring various official servers moderated by the developer team and community managers. The top menu bar allows users to navigate and explore categories of interest.}
    \label{fig:4}
\end{figure*}

Discord’s features taken together allow users to explore millions of communities to discover and socialize through their communication preference (voice, text, video). Discord’s infrastructure creates a layered communication ecosystem that blends forum-style exchanges, real-time voice interactions, and private messaging, shaping the social dynamics and governance structures that characterize Discord communities. All screenshots and descriptions of Discord’s major features described reflect its latest version, 304.0.

\section{Related Work}
\subsection{Online Social Platforms Pose Safety Risks for Teenagers}
Teenagers encounter various safety risks in online spaces, particularly when interacting with others on social media and online gaming environments. Social interactions foster the development of online communities, as meaningful conversations strengthen belonging and sustain relationships \cite{choiSocializationTacticsWikipedia2010, farzanSocializingVolunteersOnline2012, krautInformationDevelopingRelationship2008, prinsterCommunityArchetypesEmpirical2024}. However, online social interactions pose various risks and safety issues for teenagers \cite{aliUnderstandingDigitalLives2022, bozzolaUseSocialMedia2022, fireOnlineSocialNetworks2014}.

Online risks for teenagers in social platforms have been understood through Livingstone’s classification of online risks into the 4Cs: contact, content, conduct, and contract \cite{livingstone4CsClassifyingOnline2021}. Researchers have addressed these risks in their specific forms relevant to teenagers’ online social interactions, such as contact risks that emerge when minors interact with unknown users in unmoderated or semi-private spaces \cite{cernikovaYouthInteractionOnline2018, obokataAssociationsOnlineCommunication2023, pinterAdolescentOnlineSafety2017, wisniewskiDearDiaryTeens2016}, content risks arise through users sharing and exposing others to harmful material, such as self-harm and inappropriate content \cite{hashishInvolvingChildrenContent2014, stanickeChoseBadYouths2023}, conduct risk are shaped by the social dynamics of online communities where teens face cyberbullying \cite{ashktorabDesigningCyberbullyingMitigation2016, singhTheyBasicallyDestroyed2017} or seek sexual experiences \cite{alsoubaiHumanCenteredApproachImproving2023, raziLetsTalkSext2020}, and contract risks stem from opaque data practices or commercial features that minors may not fully understand when exploited by scams or gambling games \cite{kouHarmfulDesignMetaverse2023, liEthicalIssuesVideo2025, zhangDangerousPlaygroundsChild2025}. These risks are not isolated; teenagers often experience them simultaneously, such as when negative peer interactions (conduct risk) are intertwined with harmful content exposure \cite{livingstone4CsClassifyingOnline2021}.

There is extensive research about teenagers’ safety risks in social media contexts. For instance, Twitter, YouTube, and Reddit were some platforms involved in spreading content about encouraging suicide for teenagers \cite{sumnerTemporalGeographicPatterns2019}. Teenagers may struggle with depression after problematic social media usage \cite{okeeffeImpactSocialMedia2011} and exhibit problematic online behaviors when they seek online social interactions to compensate for lower levels of offline social support \cite{benvenutiTeensOnlineHow2024}. This position is exploited by online groomers who target teenagers’ desire for support and meaningful connections on social media \cite{renesesHeFlatteredMe2024}. Thus, social media interactions between teenagers and strangers pose many risks \cite{alsoubaiFriendsBenefitsSextortion2022, cernikovaYouthInteractionOnline2018}, where they encounter harmful strangers with sexual or criminal intentions \cite{obokataAssociationsOnlineCommunication2023}.

Risks are unavoidable in social media interactions, affecting teenagers’ personal information disclosure \cite{agostoDontBeDumb2017, parkTeensPrivacyAlgorithms2025, rocheleauPrivacySafetySocial2022}. A study on teenagers’ disclosure on MSN and YouTube showed how they understand that privacy risks were often out of their control \cite{bryceRoleDisclosurePersonal2014}. On Instagram, where its focus on image sharing, teenagers are often exposed to sexual abuse and explicit content \cite{alsoubaiFriendsBenefitsSextortion2022} and privacy risks from family, friends, and strangers \cite{zhaoUnderstandingTeenagePerceptions2022}. Privacy violations may be severe for teenage girls, who may become traumatized from leaked photos or harassed by men on Instagram and newer platforms like Snapchat that emphasize ephemeral media \cite{charterisSnapchatYouthSubjectivities2018, obokataAssociationsOnlineCommunication2023}. However, for both Instagram and Snapchat, teenagers control privacy risks by adjusting their information-sharing behaviors based on their perceptions of the risk \cite{adorjanNewPrivacyParadox2019}. For instance, on various social media, including Instagram, teenagers often decide what to share based on their relationship with the requester, avoiding disclosure to strangers while comfortably sharing personal details with best friends and romantic partners \cite{costelloVulnerableSelfdisclosureCodevelops2024, rocheleauPrivacySafetySocial2022}.

Online gaming exposes players to interpersonal risks \cite{kwonEffectsEscapeSelf2011, bhagatRoleIndividualsNeed2020}. The gaming environment allows players to collaborate and meet new people \cite{domahidiDwellGamersInvestigating2014}, even continuing social interaction off the game \cite{laiOnlineStrangersOffline2020}. Notably, online gaming contexts are where teenagers’ communication with strangers is most prevalent \cite{greenfieldTeensInternetInterpersonal2006}, motivated by their desire to seek information and gaming entertainment with others \cite{peterCharacteristicsMotivesAdolescents2006, rocheleauPrivacySafetySocial2022}. In multiplayer online games, players often participate in toxic cultures, where insults to identity and hate speech are ingrained in community culture and a critical safety risk \cite{koungGenderedToxicityCompetitive2025, beresDontYouKnow2021, thomasSoKHateHarassment2021}. Gaming also exposes teenagers to content risks in violent games \cite{fergusonVideoGameViolence2014} and privacy risks \cite{sasInformingChildrenPrivacy2023}, where players may be unaware and encounter economic exploitations, including “online fraudulent transactions” \cite{vanderhofChildsRightProtection2020}. For instance, Zhang et al. identified how some teenagers encountered risks as scams within user-generated games \cite{zhangDangerousPlaygroundsChild2025}.

Discord, uniquely situated between social media and gaming contexts, provides a game-adjacent space where teenagers can simultaneously maintain conversations with peers they meet in-game. In fact, risks on Discord may be perpetrated by teenagers’ peers, engaging in hateful and inappropriate “memetic rhetoric” made up of obscure jokes that were not well comprehended by adults \cite{johnsonEmbracingDiscordRhetorical2022, sparbyDigitalSocialMedia2017}. Teenagers using Discord are often exposed to threats because servers may bring teenagers and adults together into conversation \cite{kellyDarkSideDiscord}, including undesired contact \cite{parkTeensPrivacyAlgorithms2025}. Kou et al. revealed risky situations where adults may leverage their status in harmful ways, including grooming minors who were seeking game information \cite{kouSystemMadeInherently2025}. This makes it important to understand how teenagers navigate risky interactions with others on Discord through their participation in servers and development of interpersonal relationships. As teenagers’ online communities are considered “a huge part of their social life” \cite{leeMappingCommunityAppeals2025}, understanding their risk navigation may be instrumental to teenagers’ development. Thus, Discord presents a timely context for examining the gap in understanding teenagers’ strategies to deal with risks associated with online socialization.

\subsection{Online Safety Strategies for Teenagers}
Safety strategies vary depending on the context of a risky encounter and the stakeholder involved in teenagers’ online safety. The mitigation of online risks for teenagers largely depends on external measures, including strategies such as parental mediation \cite{dedkovaParentalMediationOnline2023}, adult intervention \cite{padilla-walkerProtectiveRoleParental2018}, or the design of new safety technologies \cite{aghaStrikeRootCodesigning2023, parkTeensPrivacyAlgorithms2025}. Existing literature on online safety practices reported adults’ perspectives \cite{redmilesJustWantFeel2019} and teachers’ perspectives on pre-teens’ risk navigation \cite{maqsoodTheyThinkIts2021}, while studies on teenagers have focused on broad attitudes toward online privacy, emphasizing protection through non-disclosure \cite{agostoDontBeDumb2017} and coping through resilience \cite{wisniewskiMovingFearRestriction2025}. Scholars have examined online safety through understanding multiple stakeholders and their strategies for risk mitigation, such as parents and parental mediation \cite{dedkovaParentalMediationOnline2023, padilla-walkerProtectiveRoleParental2018}, social platforms and platform governance \cite{dasPlatformGovernancePresent2022, hePlatformGovernanceAlgorithmBased2025}, and teenagers’ resilience and agency \cite{ashktorabDesigningCyberbullyingMitigation2016, wisniewskiMovingFearRestriction2025}.

Parents are often the most involved in their children’s online safety because they are responsible for teaching, monitoring, and managing their children’s usage of social media \cite{aghaConductingResponsibleResearch2020, yardiSocialTechnicalChallenges2011}. Online safety strategies may also be systems that allow parents to remain in control, protecting their child’s privacy and security using age-verified content restriction \cite{nairSafeguardingTomorrowFortifying2024}. Research on parental control apps reveals that parental mediation strategies need to dynamically consider children’s perspectives and needs to effectively safeguard them from harm \cite{wangProtectionPunishmentRelating2021}. Wisniewski et al. highlighted that existing strategies are still focused on parental mediation, but also proposed Teen Online Safety Strategies (TOSS) to demonstrate how parental control and teenagers’ own strategies may work together for their online safety \cite{wisniewskiParentalControlVs2017}. This dynamic interaction is necessary as restrictions through parental mediation do not enhance teenagers’ resiliency towards online risks \cite{vandoninckOnlineRisksCoping2013}.

Research on teenagers’ online risk mitigation strategies often emphasizes their capacity for resilience and learning after harm \cite{livingstoneMaximizingOpportunitiesMinimizing2017, wisniewskiPrivacyParadoxAdolescent2018}. Resilience refers to an individual’s trait or their process by which they recover from adversity and achieve better outcomes, such as an individual’s nature or, more holistically, positive functioning despite adverse harm \cite{bredaCriticalReviewResilience2018}. Particularly, using coping strategies to reduce feelings of harm after experiencing online risk can build teenagers’ resilience \cite{wisniewskiMovingFearRestriction2025, mchughMostTeensBounce2017}. Teenagers may cope by using humor \cite{mchughMostTeensBounce2017, brooksChildrenRiskFostering1994}, avoiding rumination, and seeking social support \cite{wisniewskiDearDiaryTeens2016}. In the context of teenagers’ online safety, resilience is often used to describe how to empower teenagers to use coping strategies after experiencing harm \cite{wisniewskiMovingFearRestriction2025}. However, not all teenagers are resilient by nature. Teenage victims of online risk could be classified into three groups: resilient, those who were able to spot risks, reject, and report inappropriate approaches, and interpret safety messages, typically due to having more secure backgrounds; at-risk, those who have “gregarious online personas” by being confident and outgoing in seeking out risks, such as being complicit to sexual interactions; and vulnerable, those who are considered to have lower self-esteem, struggle with loneliness, and often already victims of abuse \cite{websterEuropeanOnlineGrooming2012}. Thus, overemphasizing resilience in coping with online risks may inadvertently harm at-risk and vulnerable teenagers when their backgrounds are overlooked \cite{bredaCriticalReviewResilience2018, livingstoneWhenAdolescentsReceive2014, savoiaAdolescentsExposureOnline2021}. Additionally, resilience, when combined with restrictive tools and attitudes, does not sufficiently reduce teenagers’ risk exposure \cite{aghaStrikeRootCodesigning2023}.

Previous work acknowledges that teenagers use tools to protect their privacy and recognize their limitations in mitigating risks online \cite{freedUnderstandingDigitalSafetyExperiences2023, parkTeensPrivacyAlgorithms2025}. Scholars have emphasized teenagers’ agency in their own safety practices \cite{pinterAdolescentOnlineSafety2017} and focused on teen-centric design-based interventions \cite{aghaStrikeRootCodesigning2023, aghaCaseStudyUser2022}. This is because teenagers are most familiar with their own safety needs and act accordingly. For instance, they use risk-specific strategies \cite{freedUnderstandingDigitalSafetyExperiences2023, vandoninckOnlineRisksCoping2013} before, during, and after encountering risk \cite{aghaStrikeRootCodesigning2023}. Teenagers may practice agency by reducing risk exposure before it occurs when they avoid talking to strangers on Discord \cite{rocheleauPrivacySafetySocial2022} and downplay harm after risk exposure, which may signal protection through resilience but also raises concerns about desensitization and future problematic behaviors \cite{wisniewskiDearDiaryTeens2016}. They may also demonstrate agency choosing to protect themselves after experiencing risks through self-regulation, responding in positive or neutral ways \cite{sonialivingstoneRisksSafetyInternet2011, staksrudCHILDRENONLINERISK2009}. Teenagers’ agency is enhanced by the use of timely interventions during risk exposure, such as nudges, that help them make decisions about their own safety \cite{aghaTrickyVsTransparent2024}. Moreover, teenagers struggle to navigate the interpersonal risks of sharing personal information, as social and privacy harms, such as damage to their reputation or exposure of personal information, depend on trusting others to handle these aspects responsibly \cite{marwickNetworkedPrivacyHow2014}.

Teenagers use safety and privacy settings on social media to protect personal information \cite{farrukhYouthInternetSafety2014}, adjust disclosure \cite{gruzdPrivacyConcernsSelfDisclosure2018, taddeiPrivacyTrustControl2013}, and even manage content exposure \cite{parkTeensPrivacyAlgorithms2025}. Although many teenagers are aware of safety settings on social media, few older teenagers utilize these settings \cite{agostoDontBeDumb2017}. Notably, the usage of a variety of privacy and safety settings on social media can be stressful for teenagers, especially when the systems are unsuccessful \cite{weinsteinTheirScreensWhat2022}. When teenagers fail to manage their online safety, they often blame themselves and even seriously consider withdrawing from social platforms \cite{livingstoneCanPlatformLiteracy2025}.

Together, scholars have highlighted preventive and proactive strategies for teenagers’ online safety. Some of these strategies are focused on preventing risks from harming teenagers, such as practicing parental mediation, avoidance, and education \cite{elsaesserParentingDigitalAge2017, simona-nicoletavoicuPREVENTIVEONLINESAFETY2023, doringConsensualSextingAdolescents2014, wolakOnlinePredatorsTheir2010}. Preventive strategies may protect teenagers by fostering resilience, enabling them to recover from risky experiences and develop skills to navigate future risks \cite{agostoDontBeDumb2017, wisniewskiPrivacyParadoxAdolescent2018, marwickNetworkedPrivacyHow2014}. Recent research focuses on proactive strategies that extend beyond merely recognizing and reacting to risks, emphasizing action from all stakeholders involved in teenagers’ online safety \cite{wisniewskiPreventativeVsReactive2015, boydConnectedConcernedVariation2013, pirExplorationHowChildren2024, haimeExploringMentalHealth2025}, such as tools for teenagers to control unwanted messages \cite{luriaYoungUsersStrategies2023}.

These strategies reveal tensions in online safety practices and underscore the need for involvement from multiple stakeholders. For instance, teenagers may misunderstand privacy risks and avoid sharing information based on fear of punishment rather than genuine safety concerns \cite{kumarNoTellingPasscodes2017} while parental mediation can be inconsistent, shaped by varying levels of digital literacy \cite{wisniewskiAdolescentOnlineSafety2014}. While these approaches highlight important protections, scholars increasingly call for efforts that also empower teenagers’ own agency in navigating online spaces \cite{parkTeensPrivacyAlgorithms2025, wisniewskiPrivacyParadoxAdolescent2018}. Although prior research has examined online safety strategies for teenagers in various ways, strategies that balance the tension between teenagers’ desire to socialize and the need for safety in online interactions are still underexplored. Therefore, this study investigates teenagers’ practices when navigating risky socialization on Discord. 

\section{Methodology}
We obtained Institutional Review Board (IRB) approval from our institution before proceeding with data collection. The recruitment instrument included a Qualtrics survey where interested teenagers registered or parents registered interest on their child’s behalf, posted on social media and local public spaces. The Recruitment and Data Collection occurred between June and July 2025. We opted for semi-structured interviews because they allow discussions to adapt to teenage participants’ narratives \cite{bassettToughTeensMethodological2008, dearnleyReflectionUseSemistructured2005}. Interviews were conducted by the first author over Zoom following parental verbal consent and analyzed using RTA \cite{braunDoingReflexiveThematic2022}. 

\subsection{Recruitment and Data Collection}
Recruitment messages seeking teenagers (aged 13-17) who live in the United States and use Discord were posted on X, Instagram, Facebook, LinkedIn, and local public spaces. We attempted recruitment on various Discord servers (local Discord servers, University Discord servers, and teen-centric servers) and Reddit (r/AskTeens and r/Teenagers) but failed to recruit participants. Additionally during recruitment, we received an overwhelming response from “imposter participants” who faked their identity to register their interest \cite{roehlImposterParticipantsOvercoming2022}. We determined imposter participants in multiple ways: the respondent’s location was outside of the United States and birthdays that were outside of the age range, despite self-identifying as a teenager. Of the 151 responses received, only 24 were valid and were extended study invitations. Among these, 5 did not respond. The resulting participants were recruited from these platforms: X (n=3), Viva Engage (n=5), Nextdoor (n=1), snowball sampling (n=6), and locally from a flier (n=1). The participants were diverse in gender (7 male, 8 female, 1 non-binary) and background (10 Asian, 5 White, 1 Mexican/European). Detailed demographics are provided in Appendix C. 

The first author reached out to every participant to schedule a Zoom call to obtain informed verbal consent from the parent and determine eligibility from the teenager. Parents and teenagers appeared on a video call together and were read a script detailing study procedures and consent. Once consent was obtained, the first author asked the parents to leave the room so that she could proceed with eligibility questioning with the teenagers privately. This was done to allow teenagers to speak freely without their parents around. We conducted a rigorous eligibility verification process before the formal interview, during which the first author discussed requirements with teenagers, reviewed their personal accounts, and screened their experiences. This was done because IRB restrictions from our institution prevented the collection of teenagers’ eligibility information before parental consent and the number of imposter participants \cite{roehlImposterParticipantsOvercoming2022}. To be eligible for the study, the teenager needed to be 1) a current Discord user and 2) have experienced risky social interactions through Discord.

The first author verified the first criterion of being a current user from interested teenagers by requesting them to share their Discord username, clarifying this was solely for verification. All teenagers were comfortable sharing their usernames for the study. She checked account creation dates to confirm they had joined before recruitment. During this process, she also showed her profile, which helped build rapport by demonstrating her use of the platform. 

The first author verified the second criterion by allowing teenagers’ own perceptions of risk and harm to guide our definition of risky social interactions. We intentionally allowed participants to ultimately decide which interactions were risky to them and shared their experiences through the interview. Teenagers were asked questions such as, “Have you ever encountered any situations on Discord that felt uncomfortable or risky to you?” and “Have you interacted with anyone who made you feel weird on Discord?” In line with previous research, the term “risky” was used to refer to a spectrum of “uncomfortable events” or unsafe experiences \cite{alsoubaiProfilingOfflineOnline2024}. These experiences were not labeled as “harmful” and teenagers were encouraged to share what they perceived to be risks and harms without feeling judged for their behaviors. Teenagers’ definitions aligned with prior research on risky online situations, including but not limited to sexual harassment \cite{coppOnlineSexualHarassment2021, ybarraHowRiskyAre2008}, cyberbullying \cite{sampasa-kanyingaSocialNetworkingSites2015, valkenburgOnlineCommunicationAdolescents2011}, and encounters with unfamiliar people \cite{livingstoneTakingRisksWhen2007}.

After questioning to determine study criterion, 3 teenagers were determined to be ineligible for the study because they were unable to recall or describe any risky interactions on the platform that they experienced on Discord, even after follow-up probing. 16 teenagers who confirmed they had experiences with risky social interaction were deemed eligible and invited to interview. 

The interviewer was experienced with semi-structured interviewing, where the conversation organically unfolds and guided by participant’s experience \cite{dearnleyReflectionUseSemistructured2005}, so a pilot interview was not conducted. We designed an interview protocol with open-ended questions (Appendix B). The interview process began after eligibility was determined or rescheduled for a more convenient time on request. Participant assent was obtained before beginning interviews. All interviews were recorded and transcribed via Zoom. Participants were compensated \$25 Amazon gift cards upon completion. The average duration of the interviews was 35 minutes.

\subsection{Data Analysis}
Each transcript was manually and inductively analyzed by the research team, following the six steps of RTA \cite{braunUsingThematicAnalysis2006, braunDoingReflexiveThematic2022}. In RTA, meaning is developed through analyzing shared experiences and meaningfully capture patterns in participants’ stories \cite{braunDoingReflexiveThematic2022}. We adopted RTA to share why teenagers’ experiences with risky social interactions on Discord matter. In data collection and analysis, the research team prioritized data richness instead of determining data saturation, as its realist assumptions are not aligned with RTA \cite{braunThematicAnalysisPractical2021}.

To begin RTA, three coders independently reviewed and coded transcripts for quotes related to how teenagers navigated risky social experiences. Although coding in RTA does not produce a codebook \cite{braunThematicAnalysisPractical2021}, the initial coding process resulted in 490 unique codes, which were all considered for theme generation. Once codes were produced using semantic and latent coding aligned with RTA \cite{braunThematicAnalysisPractical2021}, each coder reviewed the others’ codes to create a mutual understanding of coding consistency and held a meeting to discuss interesting patterns while resolving disagreements. On Miro, the first author arranged similar codes into manageable groups for manual analysis, consistent with RTA \cite{braunThematicAnalysisPractical2021}. After this, the entire research team collaborated in multiple discussions to highlight interesting codes and discussed potential themes \cite{pattonQualitativeResearchEvaluation2014}. All authors contributed by iteratively refining identified themes to construct meaningful interpretations of the data \cite{braunDoingReflexiveThematic2022}. To demonstrate our approach to RTA, we provide the first author’s analysis and reflection: 
\begin{quote}
“I don’t usually talk in the 1st week. I just sit there. Let [the chat] roll so I know who the target audience is there. And then if I feel safe. I’ll talk more.” P3
\end{quote}

The first author noticed how the participant mentioned how they do not participate in a server right after joining and instead, observe the chat to understand the members. Therefore, the initial code became: ‘After joining a new server, observe and only speak if feel safe.’ This code resonated with the author’s own personal experiences of lurking in channels to learn what others’ interests were before participating. She found this meaningful connection in other codes and grouped related initial codes into a second-level code: ‘Observe server culture.’ She noticed that other second-level codes also described a pattern of how they actively gained awareness about community norms in servers, which led to the subtheme ‘Learning server norms.’ After analyzing several more subthemes, she synthesized their stories and meanings into the final theme: ‘Assessing Community Culture through Selective Participation in Servers. ’

The theme names were finalized after assessing internal homogeneity and external heterogeneity \cite{pattonQualitativeResearchEvaluation2014}. The team deliberately decided on the most representative quotes for each theme through continuous refinement \cite{braunDoingReflexiveThematic2022}. Once the research team agreed on the final themes and representative quotes, a thematic map was created by the first author. To protect teenagers’ privacy, we will refer to them with code names (P1, P2, …). 

Our study reveals two major themes answering the RQ: 1) Balancing Risk Exposure when Pursuing Interpersonal Connections on Discord and 2) Engaging in Discord Servers for Community-Level Risk Reduction. 

Following qualitative research practices that view coding as an iterative process of theme development \cite{mcdonaldReliabilityInterraterReliability2019}, we did not calculate inter-rater reliability. In RTA, themes are created through reflexive interpretation rather than generated through objective agreement.

\subsection{Researcher Positionality Statement}
The research team has varying levels of experience with using Discord. The first author is a long-time and active user of Discord, starting when she was a teenager, to communicate with friends while playing games. While interviewing teenagers could be challenging due to the first author’s position as a researcher, her insider position allowed her to relate understandings of Discord activities with the participants \cite{bassettToughTeensMethodological2008}. When conducting the interviews, the first author was likened to being an “older sister.” The second and third author use Discord less frequently but have participated in servers related to their personal interests. The other authors frequently use Discord for communication in professional settings.

\subsection{Ethical Considerations}
We prioritized protecting minors’ privacy by ensuring informed consent and minimizing potential risks associated with direct communication between researchers and underaged participants. The IRB at our institution did not permit researchers to contact teenagers directly, so recruitment was conducted by posting in public venues for teenagers to come across organically or to target parents who could register interest on behalf of their child. Recruitment forms were designed to allowed limited collection of personally identifiable information to protect teenagers’ privacy: their birthday, legal names of the participant and parent, and parent contact information. 

Parental involvement was minimal but essential, as minors required parental consent to participate. To establish trust and transparency, the first author conducted Zoom introductions with parents to explain the study, address questions, and obtain verbal consent for teenager’s participation. While it is possible that parental awareness of the study could influence how openly participants shared their experiences, this consideration reflects a broader methodological tradeoff. Decisions about parental involvement required the research team to balance ethical requirements for transparency and safety with the need to create conditions where teenagers can speak freely. To strike this balance, we took several steps to minimize any potential impact of parental awareness, including emphasizing confidentiality, conducting interviews in private settings chosen by participants, and reinforcing that parents would not have access to what they shared. These measures appeared to support candid participation: every teen described their experiences supported with many rich details, unprompted by the researcher.

Participant protection measures included informing them that their responses would remain confidential, and no identifying information (names, usernames, or personal details) would appear in publications. The researcher emphasized that participation was voluntary, and participants could skip any questions or end the interview at any time without consequence. The first author observed that participants were engaged and energetic throughout, showing no signs of discomfort or hesitation. She prioritized participants’ emotional well-being by asking thoughtful follow-up questions when discussing potentially uncomfortable topics. Audio recordings and transcripts were stored on password-protected devices, accessible only to the research team and deleted upon completion of analysis.

\section{Findings}
We found that participants navigated social interactions with others on Discord using various strategies that balanced concerns about privacy and security with social motivations. First, teenagers made sense of risky situations in their social pursuits by recognizing suspicious cues in account details and others’ language. Then, teenagers managed their interactions with risky people by restricting connections or engaging further for entertainment. Teenagers also used their participation in Discord servers to become aware of community-level risks that inform further engagement with others. Together, these practices illustrate how teenagers navigated both dyadic and community contexts on Discord to maintain their safety.

\subsection{Balancing Risk Exposure when Pursuing Interpersonal Connections on Discord}
Participants described how they seek interpersonal connections on Discord differently from other social media, where emphasis is on curating a personal profile for wide visibility. Teenagers on Discord focused on building situational, activity-driven connections such as voice chatting while gaming and trading in-game items. However, participation in these environments also introduces potential risks. Many participants experienced how risky Discord users may behave suspiciously towards them, such as attempting to gain access to teenagers’ account information and contacting participants with intentions that are misaligned with friendship. To navigate these risks, participants interpreted and assessed interactions carefully, balancing the need to protect themselves with the desire to engage in positive social connections. In some instances, they may choose to engage in risky scenarios to obtain personal benefits, such as entertainment.

\subsubsection{Recognizing Suspicious Users through Investigating Deceptive Methods}
Some participants monitored messages from strangers and checked for suspicious details when interacting with others on Discord to avoid being hacked or scammed through deceptive methods. Their awareness of deceptive links and how scammers in game may use them, helped them examine a suspicious interaction more closely. P10 recalled how he often dealt with scammers and suspicious interactions on Discord because he played Roblox and wanted to participate in game trades. 
\begin{quote}
    “The first thing… I would check when their account is made since a lot of scammers are new accounts. The second thing is check the link that they send... So you hover over it and see if it’s an actual link. With Roblox, scammers can make realistic links. But if you hover over, [the link] is different, just like videos online… they also try to teach you how to prevent it from happening as well.” P10
\end{quote}

Discord profiles publicly display the date of when the account was created. P10 developed the ability to interpret account age as a salient cue for identifying potential scams by drawing on both personal encounters with scammers and knowledge gained from online videos. The deceptive link looks genuine in the message, so P10 avoids scam risks by hovering over the link to reveal the fraudulent site, making sure to not click into it.  

While a few participants believed that suspicious users on Discord might be easy to recognize, P5 encountered an incredibly deceptive scam because it abused his trusted connections. P5 and his friends used Discord as their main platform to communicate with each other daily. One day, P5 received a message that included a link to a common website from his close friend on Discord, whose account had been hacked without his knowledge.
\begin{quote}
    “It was a link [with a Steam game video preview]. When I clicked on it… I logged in to see what he was trying to send me, I lost [my account]… I wasn’t even thinking of it, because we’ve been so close for so long that I was like, ‘Well, this is just a normal thing’… [we send each other] a lot of website videos that we want each other to watch.” P5
\end{quote}

This method spreads phishing links by directly messaging the hacked user’s friends to steal Steam accounts and personal information. A suspicious actor using P5’s friend’s Discord account created an illusion of a genuine connection and abused P5’s trust in his friend to send entertaining content. As a result of not being able to identify suspicious activity, he lost his Steam account. P5 described feeling “very sad, because I spent a lot of money on that account on games.” P12 also encountered an elaborate scam that steals Discord accounts. 
\begin{quote}
   “When I join, it tells me to verify, and they even make it look real by having a lot of people, but everyone in its account is hacked. They also have a guy waiting in the voice chat for you to make it look real before you’re allowed to have access to the server. And then the pop-up shows. ‘Verify your account before you can use the server’ with this bot that looks official. [I thought to myself] This is pretty weird… then I searched it up. I saw this forum said, ‘Don’t do the Discord verification scams, they’ll steal your account.’” P12
\end{quote}

P12’s experience demonstrated how server bots may be crafted deceptively to draw in unsuspecting users who joined the server to volunteer their Discord account information through “verifying” and granting the bot permission to their account. In addition, P12 explained how scammers deceive users by creating a genuine Discord server, simulating an interactive environment by displaying a high member count and appearing to be waiting for his presence in the server. In a typical Discord server, it would be expected to interact with server rules and be “allowed” permission to further access a server’s channels. While the deceptive bot acted as a gatekeeping mechanism that prevented P12 from further participating in the server, it also acted as a signal for P12 to realize that it was not expected from a typical server interaction. Due to his vigilance in recognizing suspicious details, P12 avoided having his Discord account hacked through deceptive methods. 
Although some participants experienced hacks and scams, they were also able to protect themselves by investigating how suspicious users were being deceptive. 

\subsubsection{Evaluating Intentions During Connection Development Using Linguistic Cues}
Many participants described how they evaluated others’ intentions using linguistic cues in conversations to guide their development of social connections on Discord. It was typical for many participants to desire chatting with strangers because they were seeking connections for gaming or making friends out of school. P2 is a competitive gamer and often connected with strangers on Discord to play games together. He described how he managed new connections with strangers by starting conversations that allowed him to understand their intentions and behaviors.
\begin{quote}
    “When I start connections or they start… we talk for a bit, and then if I want really want to know them, I’ll say ‘Hi, good morning!’ or ‘How’s your day?’ And then if I think they’re weird usually… it’s the things they’re doing. So they might be talking bad about another person. I’m like, ‘Oh, I don’t really want a part of that.’… I would just ignore them or just unfriend them. And I don’t really care if they want to talk to me or not because they’re just weird…” P2
\end{quote}

By establishing a friendly connection, he could assess linguistic cues during conversation that might signal bad intentions, such as negative gossip, and set boundaries when interactions conflicted with his expectations of safety and respect. Some participants also restricted further contact with strangers, like P2, through means such as ignoring, unfriending, and blocking \cite{parkTeensPrivacyAlgorithms2025}. Beyond these restrictive measures, a few participants, like P16, intentionally revealed her personal information to observe others' reactions. 
\begin{quote}
   “I usually say my age, just so that people know. So they’re not trying to get any ideas, because it deters people... but I’ll usually still try and be friends with them. I usually say my age. I usually say my gender. [Others react to my age] ‘Oh, you’re way too young!’ And I’m like, ‘Yeah, don’t talk to me in that way [romantically]’ or some people who are like older, they just don’t want to catch a case and don’t wanna risk it. So they’ll just be like, ‘Yeah, never mind.’ And then they unadd me.” P16  
   
\end{quote}
She strategically used her age identity as a signal to prompt others to reveal their intentions. P16 described how using this strategy allowed her to better understand others’ expectations through linguistic cues while giving them the opportunity to adjust their behavior, such as discouraging romantic advances. Since her own intention was friendship, she was not bothered by the absence of those who wanted more than friendship and those who set boundaries against being friends with teenagers. Overall, observing strangers’ behaviors and reactions provided many participants an understanding of how linguistic cues can reveal risky intentions.

\subsubsection{Applying Restrictive Functionalities to Selectively Manage Direct Interactions}
Many participants described how they used various levels of restrictions on direct messages that enabled selective interaction. Discord’s “Social Permissions” feature includes toggles to allow “DMs” and “Message Requests,” a feature that filters incoming messages from people not on a user’s friend list into a “Message Request” or “Spam” folder \cite{MessageRequests2025}. Figure 5 shows how messages are filtered into “Requests” and “Spam.” This allows users to view and decide whether to accept a pending DM request and continue conversing with an unfamiliar user. 
\begin{figure*}[h]
    \centering
    \includegraphics[width=\textwidth]{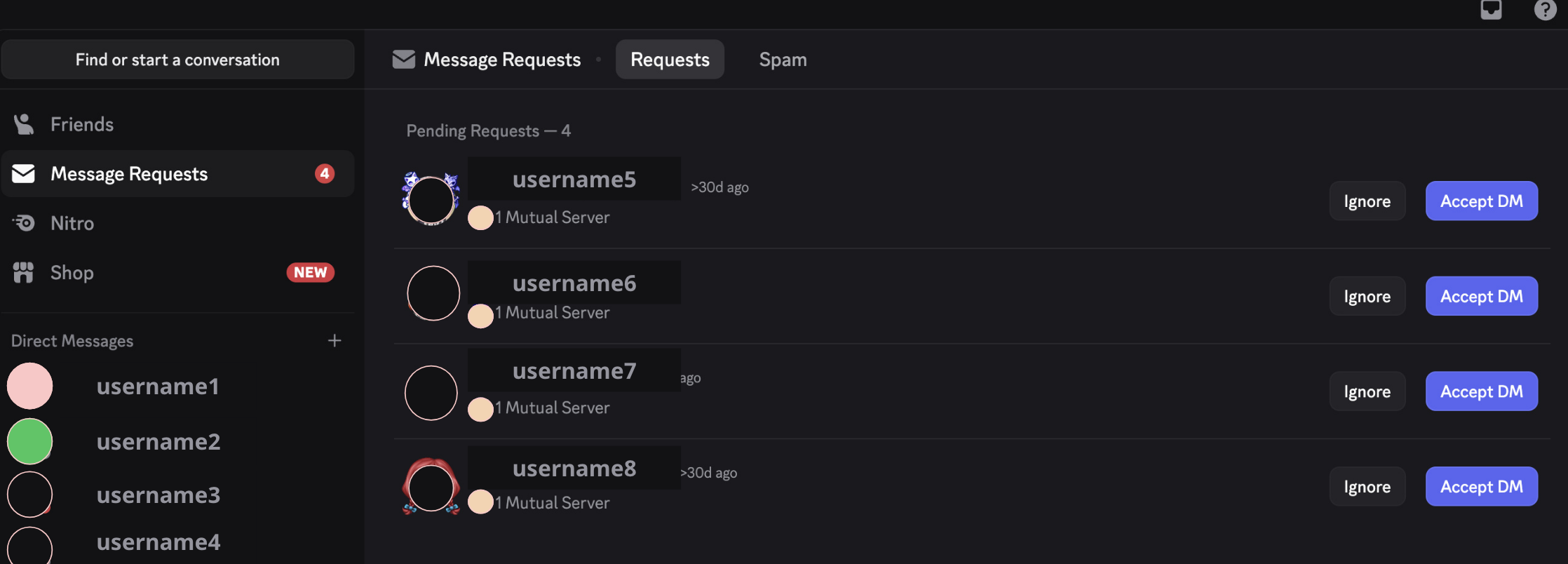}
    \caption{Screenshot demonstrating the “Message Request” feature. Shows how “Message Requests” are separate from DMs and further separated into spam. On this page, users have the choice to “Ignore” or “Accept DM.”}
    \label{fig:5}
\end{figure*}

By default, Discord allows DMs from other members in a shared server. Participants refer to these settings together as “DM filters.” P3 currently uses DM filters to restrict contact from others but perceives the setting as “complicated” to activate.
\begin{quote}
    “[Filtering DMs is a setting that is] somewhat of a hard thing to turn on. I feel like you have to actively have an experience before that or have parental settings to not have unfiltered DMs. And in my case, I didn’t have parental settings and I did not know what [settings] Discord had. I didn’t heavily use Discord… When you open the app, they should probably list, ‘Do you want to accept all DM requests?’... and list a lot more information about teenage safety and make it easier to open...” P3
\end{quote}

P3 was not very experienced with Discord, so she was also not aware of various protective settings that she could use. She felt that it was “not beneficial” when she had to experience contact risks before seeking out DM filters on Discord. Thus, P3 desired more transparency and accessibility for teenagers’ safety tools. 

Additionally, some participants were not aware that they could turn off DMs or filter them into Message Requests while others like P2 was very experienced with Discord. His prior knowledge about DM filters and using Message Requests enabled him to protect himself from being contacted by risky users. Despite using Message Requests, his attitude towards DM filters is that they are not effective, remarking, “I would like to meet more people. Just in case, maybe it’s a nice person or they’re asking for something.” This reflected how teenagers, like P2, sometimes choose to allow contact from unfamiliar users, accepting risky messages while remaining optimistic about meeting “nice” people and helping others. 
\begin{figure*}[h]
    \centering
    \includegraphics[width=\textwidth]{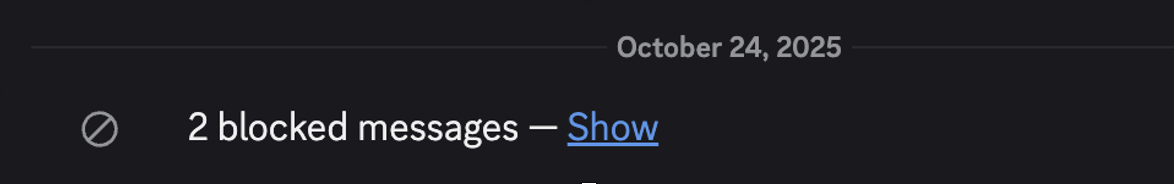}
    \caption{Screenshot of Discord interface when a user views a message in a text channel from a blocked user within a shared server.}
    \label{fig:6}
\end{figure*}

When in contact with risky users, many participants described blocking as a preferred method to restrict further interactions. P16 described blocking a risky user on Discord, but remained in the same server as them. In Figure 6, messages from blocked users are automatically hidden in shared chats, but users may select “Show” to reveal messages.
\begin{quote}
    “With the block feature, they don’t show the person’s messages anymore on servers you’re in with them. You have to manually click on their messages to actually see them. So that’s nice. I like that… Because you still have an option to not see what they’re saying… it helps with the context of the conversation.” P16
\end{quote}

This design granted participants like P16 to have control over how they engage with conversations without disengaging from the server. She believed she was able to safely navigate public conversations without being exposed to unwanted contact by restricting engagement through blocking and choosing to reveal potentially risky content when context is necessary. P16’s positive experience with blocking on Discord balanced her exposure to risky users by allowing her to selectively engage with their messages. 

However, a few participants felt that making the decision to block another user may be challenging. P4 described her difficulties with restricting contact from others and being “horrible” at blocking and “hesitant to block” a risky user because she felt she did not “necessarily have to.” This suggested that P4 perceived that others would feel bad if she blocked them.
\begin{quote}
    “Whenever I used to almost block someone, I imagined… going to text me and then seeing it’s blocked and then having an emotional breakdown. But then I realized these people, they can’t necessarily actually do something to me… And then I would just be able to block them easier. Now, as soon as anything like that would ever happen, immediately I block them.” P4
\end{quote}

She described that she developed a new mindset that prioritized her personal safety over empathy for risky others. This suggests that P4’s awareness of the limited consequences of blocking enabled her to reframe the act as a protective strategy, empowering her to disengage from risky interactions rather than remain vulnerable to them. In all, participants balanced risk exposure on Discord by selectively interacting with restrictive features throughout their risky social interactions so they could explore interpersonal connections safely.

\subsubsection{Engaging in Risky Settings During the Pursuit of Personal Interests}
Participants demonstrated agency in risky situations where they felt control over the interaction and could navigate the risk without being harmed. A few participants identified that talking to strangers in the context of shared interests or activities, such as gaming, could be a risky setting. However, P9 joined many Discord servers sought to interact with strangers because his friends at school disapproved of his interests in ‘obscure’ games. 
\begin{quote}
“They would laugh at me for [playing an obscure game]. I have like a main account and then an alt account. The alt account is for obscure games, rhythm games or stuff they would disapprove of a little bit. I usually use that account to go join those servers.” P9
\end{quote}

Through joining Discord servers on an alternate account, P9 explored his gaming interests freely while maintaining separation from his main social circle. His experience of using an alternate account is related to alternative personas and “dividing the self” by making himself appear more appropriate for social interactions \cite{bullinghamPresentationSelfOnline2013, goffmanPresentationSelfEveryday1959}. This suggests that Discord could provide a space for managing social impressions through alternative accounts, which may benefit teenaged users’ safer exploration of interests. Additionally, P9’s experience represented a controlled form of risk-taking that he preferred to the potential harmful judgment from his friends. Some participants perceived that interacting with strangers on Discord could be an exciting risk, even initiating political arguments with others.
\begin{quote}
    “…trying to have arguments with other people. I would join some right-wing servers and talk about something controversial like abortion. Say, ‘Abortion is good.’ It was kind of ‘rage baiting.’ Normally, they probably would just remove me. I was just a kid… Like Internet was new to me and talking to people was interesting and seeing their reactions.” P13
\end{quote}

“Rage baiting” is a form of manipulative content that capitalizes on inciting negative emotional reactions from the audience \cite{gruetWhatRagebaitingWhy, caspergrathwohlOxfordWordYear}. P13 entered political spaces and intentionally posted controversial statements to provoke outrage. Although moderators removed him for these actions, he found the resulting interactions engaging and entertaining. His experience illustrates how participation in online communities can involve calculated risk-taking, as he positioned himself as a potential target by expressing deliberately controversial ideas. Altogether, participants navigated potential risk exposure with confidence by being aware of risky settings while focusing on their personal interests. 

\subsection{Community-Level Risk Reduction in Discord Servers}
Discord servers are numerous and diverse, encompassing a wide range of community cultures. Participants navigated risks by engaging in vigilant governance to understand community behaviors and decide their participation in a community, whether by avoiding or addressing the risk. Selective participation was a strategy for teenagers to engage with Discord servers to become aware of potential community cultures that tolerated inappropriate jokes or content. Other participants protected themselves when participating in Discord servers by establishing boundaries using server features. In certain cases, teenagers contributed to community safety by engaging in moderation processes, reducing risk within Discord servers.

\subsubsection{Assessing Community Culture through Selective Participation in Servers}

Participants described that their motivations for selective participation allowed them to become familiar with a server’s culture and make decisions on their future interactions in the server’s community and conversations. Since Discord servers are designed to provide members with convenient access to text-based interactions, P9 often observes what people converse about in text channels, such as unacceptable jokes. 
\begin{quote}
    “Usually, you can pick up by how people talk or use text on the Internet [whether it is a safe place to be]. If someone jokes about pedophilia, that doesn’t seem like a great [safe] environment... I wouldn’t even consider staying in those servers when I did find that.” P9
\end{quote}

Upon learning how a server culture supports pedophilic jokes, P9 made an informed decision to leave. This shows how Discord servers afford their members the autonomy to leave a server at any time. Additionally, this structure enables participants, such as P9, to learn about a server’s culture by observing members’ attitudes and behaviors through messages without needing to engage directly or making connections with others. Many participants adopted this observational approach because it allowed them to evaluate potentially risky servers more confidently and to disengage from them as a form of self-protection. This could enhance members’ decision to leave a server easily and without any social pressures. 

However, assessment of servers simply through observation is limited because some participants described how it is necessary to interact with others for gaming activities, such as trading. P6 often joins Discord servers to enhance her Roblox game experiences by trading items with others. 
\begin{quote}
   “You don’t really know what kind of people are in that server. You never know if you’re with a scammer or hacker or a middle-aged man… and then you don’t know what’s gonna happen there ever. [Just actively recognize] when something weird is happening so you could leave or just have your guard up.” P6 
\end{quote}

P6 described not knowing what kind of members she would encounter when seeking connections to trade with in-game. Thus, she adopted selective participation to identify suspicious community culture by paying attention to trading activities and strategically disengage from potentially unsafe situations. Overall, many participants described how selectively participating allowed them to observe existing and ongoing conversations in text channels to decide whether to engage further. 

\subsubsection{Setting Boundaries During Discord Server Participation to Protect Privacy}
Participants were adept at controlling their risk exposure during participation in Discord’s communities through setting interpersonal boundaries. P14 described their experience using Discord’s activity status, which appears under a user’s profile so that other users to see what games or applications they are actively using \cite{ActivitySharingDiscord2025}. Activity status can be calibrated to be viewed in private, semi-private, or public servers. 
\begin{quote}
    “I keep my activity status off… so it doesn’t show what apps I’m on cause I don’t want people to be like, ‘Hey. I see that you’re on [this game]. What you doing there?’ I’m just playing Roblox or something. It’s like stalkers and I don’t really like it.” P14
\end{quote}

P14 chose to keep their activity status private because it felt invasive and intrusive for others to view what they are doing through Discord. Being able to calibrate personal activity status on Discord to reflect current activities and set restrictions on who can view it allowed participants like P14 to maintain their privacy while gaming. 

Channels in Discord servers are highly customizable with unique privacy and accessibility rules. For example, moderators can intentionally calibrate channels to enforce age restrictions on teen accounts (Figure 7). Teen accounts on Discord include parental settings that limit sensitive content visibility and friend requests \cite{buffyFamilyCenterTeens2025}. 
\begin{figure*}[h]
    \centering
    \includegraphics[width=\textwidth]{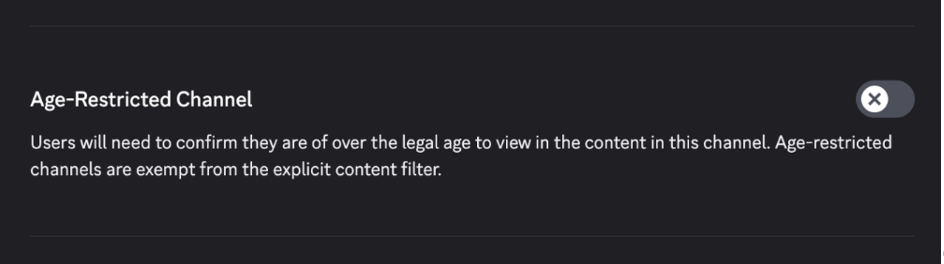}
    \caption{Screenshot of the “Age-Restricted” text channel setting that moderators could use to gatekeep explicit content from younger members.}
    \label{fig:7}
\end{figure*}

P3 uses a teen account and she self-moderates by electing for an “under 18” role in servers if available, where applying this role “basically just stops my account… So, I can’t go into that specific role’s page.” This setting allows teenaged members to openly browse most content within the server without restrictions, while gatekeeping age-inappropriate channels. Many participants were grateful for moderators’ proactive attention in implementing restrictive rules in channels with NSFW content, enabling teenagers using teen accounts to navigate content safely. 

However, some participants were not interested in using teen accounts due to channel restrictions. Rather than using teen account settings to protect themselves from risky content and contact, P9 actively set boundaries while engaging in a Discord call.
\begin{quote}
    “When somebody was asking, ‘Oh, where do you live?’ We just screamed, while [a friend], almost revealed where they lived or the city. We would yell or get a seagull sound effect and that would cue the person [to realize they] almost just spilled something important. Should probably be a bit more careful. Pretend that soundboard you did it on accident, ‘Oh, my bad!’ Some people find it hilarious and then you’d realize, ‘Oh, they’re fine.’ But the people who get mad, you can tell ‘Oh, that was a good idea!’” P9
\end{quote}

P9 described developing a system with friends to join voice channels together, enabling them to alert one another to risky situations in real time. They used Discord soundboards, typically designed for entertaining reactions \cite{DiscordSoundboardGuide2025}, as covert signals to self-moderate and recognize moments when personal information might be disclosed. This illustrates how creative engagement with server features can mitigate risks, as P9 fostered a sense of safety by observing strangers’ reactions to unexpected sounds and using this information to guide subsequent interactions. For many participants, collaborating with friends to monitor voice chats while playing with strangers and promptly addressing potential risks was their preferred safety strategy.

\subsubsection{Creating Community Safety by Participating in Server Moderation Processes}
Server moderation processes on Discord, such as moderating and deleting messages, managing membership, and banning users who violate server rules, are responsibilities overseen by moderators \cite{discordRoleAdministratorsModerators}. Participants recognized the importance of sharing information with moderators so that they can take action to maintain community safety. 

Some servers include “ticket bots,” which are implemented by moderators to allow members to get in contact with them through a ticketing system. Interacting with the bot allows a concerned member to present their issues to the moderation team privately. P16 talked about the usefulness of ticket bots when she was exposed to suspicious content through another member of the server. 
\begin{quote}
    “[Ticket bots are] nice, quick, and easy. I friended someone on this server and I took screenshots of what they said and their profile. I sent it into the ticket to Mods. And they’re like, ‘Yeah, we take this really seriously. It looks like they may have gotten hacked. We’ll handle this.’ And I think they banned the person from that server.” P16
\end{quote}

Creating and sending a ticket through the bot reassured P16 that the server was actively protecting her, as a human moderator responded to the report and addressed the emergent risk. Due to P16 reporting the safety risk, the issue was quickly resolved by moderators who removed the bad actor from the server and prevented others from getting the same suspicious messages. However, moderators may not always be available or active, making risky situations tough to handle. 
\begin{quote}
    “We had a troll in one of our Discord servers who sent explicit images the moment they joined… sending the same explicit gif, over and over until the server owner removed it.” P7
\end{quote}

P7 was disturbed by the presence of a troll entering a server she was in, but she had no control over content removal because she was not a moderator and did not have permissions to remove bad content. This experience was shared by other participants, who also found it unsettling that they could not remove harmful images themselves. As a result, some participants, like P7, were reliant on moderators to act after being informed of safety risks to feel safe. 

Other participants demonstrated how being moderators of servers themselves allowed them to have power and take actions when a member was threatening a server’s safety. P15 often plays Rust and developed strong connections with others through playing the game and forming a team. He considers Rust to be a game with a toxic and risky community, where it is acceptable for players to use harsh language towards each other. However, since P15 enjoys the game, he felt that it was important for him to create a safer server without toxicity.
\begin{quote}
    “The other day I kicked somebody from the team that was [making] really uncomfortable jokes... I said, ‘Hey, you can be more mature. You’re older than half the people here. Just don’t make those jokes and don’t say those words.’ And all of the moderators agreed with that. It’s a toxic community. But if it’s bad enough, people will stand up for it.” P15
\end{quote}

Despite the risk of interacting with a much older stranger, P15 challenged the toxic player who was causing severe issues for their team environment. When acting as a moderator, he felt comfortable speaking up and worked together with the other moderators to remove the bad actor and influence the servers’ attitude towards toxicity. Overall, participants took an active role in maintaining community safety using the tools and knowledge they had to moderate servers. 

\section{Discussion}
We reported teenagers’ varied experiences with navigating risky social interactions on Discord. Building upon the findings, our study \textbf{introduces teenagers’ vigilant strategies for online safety on Discord} to capture their safe participation and exploration of unfamiliar online spaces. We further demonstrate how teenagers’ \textbf{online safety strategies can be enhanced at the community level with vigilant governance}. Together, this shows \textbf{Discord-specific platform affordances enables and constrains safety management} for individuals and communities. Finally, our findings have implications for \textbf{safety by design, where we suggest platform designs} that aid teenagers’ participation in online communities beyond Discord.

\subsection{Teenagers’ Vigilance for Privacy and Security on Discord}
We foreground teenagers’ vigilance as necessary for navigating social interactions and interests on Discord. Despite frequent exposure to risky encounters, we highlight how participants’ intentions to continue using Discord remained unchanged, largely because their vigilance supported safer social interactions. Participants’ desire for connection and exploration required them to initiate or reciprocate contact with users who may appear suspicious. Existing work has focused primarily on adults’ Discord usage \cite{heslepMappingDiscordsDarkside2024, jiangModerationChallengesVoicebased2019, robinsonGovernanceDiscordPlatform2023}, leaving a substantial gap in understanding the everyday experiences of teenaged users. We contribute new insights into how teenagers practice vigilance and keep themselves safe in spaces where risks are heightened by drawing on both Discord’s tools and individual strategies for balancing risk with entertainment. 

\subsubsection{Using Discord Tools to Support Vigilant Strategies}
Our study shows how teenagers were vigilant by using Discord’s unique tools to feel safer when interacting with strangers. Beyond the common use of blocking and reporting \cite{aghaStrikeRootCodesigning2023, akterCalculatingConnectionVs2025, alsoubaiProfilingOfflineOnline2024, sanchez-zasOntologybasedApproachRealtime2023}, participants advocated using Discord tools to determine what additional interactions would be safe. For example, checking Discord profiles to understand basic user information helped participants like P10 avoid common phishing scams. Additionally, our study highlights teenagers’ vigilance in managing their own user profiles to align with their own boundaries, going beyond risk avoidance \cite{chouHowTeensNegotiate2023, rocheleauPrivacySafetySocial2022}. For instance, P14 re-calibrated the visibility of their activities when needed, preventing strangers from monitoring their gameplay while allowing others to chat with them. This offers a new perspective on how Discord tools supported teenagers’ vigilance, establishing boundaries that shaped safer interactions in real time.

Discord’s pending message feature allowed vigilant participants to navigate risky situations because it helped them recognize potential harm before engaging. Although pending messages are not a unique feature on Discord, as Instagram also uses pending messages for creators to interact with their audience \cite{arriagadaYouNeedLeast2020}, we contribute a new perspective to understand the purpose of pending messages for teenagers. Many participants like P3 strategically used pending messages and message restrictions as tools for risk vigilance. These features, however, have limitations. Some participants, as illustrated by P2, still wished to socialize with strangers on Discord and invited strangers from pending messages to talk with him despite potential risks. Nevertheless, for participants seeking to avoid unwanted contact, these controls provided an effective means of managing who could reach them. Together, these practices indicate that teenagers’ vigilance can meaningfully influence their safety when using Discord tools. We encourage parents to learn more about Discord tools and their child's online behaviors to support their vigilant navigation.

\subsubsection{Using Vigilant Attitudes to Navigate Emergent Risks}
Participants remained safe when interacting with unfamiliar or potentially risky strangers on Discord by adopting a mindset oriented toward learning about others without exposing themselves to harm. Although sharing personal information is often framed as a safety risk \cite{akterCalculatingConnectionVs2025, azzopardiAssessingRisksOnline2025, bryceRoleDisclosurePersonal2014, fogelInternetSocialNetwork2009, vanschaikSecurityPrivacyOnline2018}, P16 exemplified how selective self-disclosure can function as a strategy to determine who to form new connections with. She intentionally revealed limited personal details to observe how others responded, using these reactions to infer their intentions and assess their suitability as friends. This vigilant learning process—remaining aware of potential risks while still engaging—allowed her to assess unfamiliar individuals and maintain control over her safety. Additionally, some participants reported learning about Discord-related risks through YouTube and TikTok videos, as P10 described. This suggests that parents can support teenagers’ learning by sharing educational content to raise awareness of potential online risks. Such prior knowledge helped participants recognize and navigate emerging risks before encountering them directly, highlighting that safe Discord use requires teenagers to exercise vigilance and learn about potential harms both through experience and educational resources.

Our study highlights how vigilant attitudes enable teenagers to intentionally engage with risky social interactions for entertainment. Prior work has shown that teenagers’ perceptions of risk severity shape their subsequent engagement with online risks, as encountering low severity risks can help them develop social skills \cite{wisniewskiDearDiaryTeens2016}, but enjoyment from unsafe interactions such as sexting led to regret \cite{alsoubaiProfilingOfflineOnline2024}. However, participants in our study highlight that enjoying risky interactions allowed them to navigate Discord without being harmed. While research emphasizes that teenagers’ developmental stage may lead to unjustified risk-taking \cite{sunsteinAdolescentRisktakingSocial2008} and do not intentionally seek out risk \cite{wisniewskiDearDiaryTeens2016}, we found that teenagers justified risk-taking with a desire for entertainment. Participants like P13 intentionally provoked risky social situations by rage baiting another user. This proactively protected him against perceived risks through a lens of entertainment. Although rage bait is “deliberately designed to elicit anger or outrage” \cite{caspergrathwohlOxfordWordYear} and could negatively impact wellbeing \cite{tcherdakoffIveBecomeMore2025}, participants’ incitement of rage bait transformed a risky interaction with a stranger into an entertaining experience. Our findings indicate that teenagers’ desire for entertainment led them to knowingly become bad actors. Thus, we establish how teenagers can actively manipulate risky interactions for amusement on Discord.

\subsection{Teenagers’ Engagement with Discord Governance for Safer Online Communities}
Our research captures the adaptive and situational nature of Discord governance that allows teenagers to safely navigate online communities. We show how governance on Discord servers support teenagers’ strategies for navigating risks in real time when they directly affect moderation processes and be moderators themselves. However, we also show how teenagers’ vigilance is needed to overcome limitations in moderation for collective safety. Thus, we contribute an improved understanding of how teens navigate socialization in parts of Discord that operate without moderator oversight, an area that has received little attention despite its significance for online safety. Together, we establish how Discord servers are a site for understanding teenagers’ need for real-time governance and how it impacts their safety in risky online communities.

\subsubsection{Doing Real-Time Risk Navigation through Moderation}
We contribute a new understanding of how teenagers support Discord governance in real time. First, teenagers actively keep communities safe by engaging with bots designed for server moderation. Besides moderators using bots to facilitate justice and moderation \cite{doanDesignSpaceOnline2025}, we show that server members like our participants can use bots to alert community moderators to take swift action. For instance, when P16 encountered a suspicious user, she could have blocked or ignored the message, which are common methods used to deal with risks online \cite{parkTeensPrivacyAlgorithms2025, rocheleauPrivacySafetySocial2022, akterCalculatingConnectionVs2025}. Instead, she chose to report the interaction through a ticket bot. This action strengthened community-level safety by helping moderators organize, triage, and resolve risks more efficiently. Participants noted that such moderation was only possible because they already understood the bot’s features and how to use it in risky situations. This highlights the importance of accessible, interactive ticket bot systems that connect teenagers to the appropriate moderators and demonstrates how teens’ engagement with these tools directly contributes to the broader safety of their online communities.

Second, teenagers’ involvement as server moderators themselves illustrates how their vigilance scales from individual safety practices to community-level governance. Yoon et al.’s work examined a small sample of teenagers who became Discord moderators through formal training and rigorous certification through examination \cite{yoonItsGreatBecause2025, graggleAnnouncingDiscordModerator}. In contrast, our study shows that teenagers take on moderator responsibilities in their communities without any formal training or certification. For example, P15 moderated a risky community purely out of personal investment, actively confronting harmful actors despite having no professional preparation. By taking on these responsibilities, teenagers are not only navigating risk in real time, but are also shaping the safety and norms of their communities through lived experience and intrinsic motivation. Our findings surface an understudied form of teen-led governance in informal Discord communities, where vigilance is not merely self-protective but contributes directly to collective safety and moderation practices.

\subsubsection{Overcoming Governance Limitations through Vigilance}
Further, our study shows that teenage users navigate community limitations with vigilance by understanding how safety settings interact in unfamiliar situations and by creatively implementing alternative solutions. Human moderators are limited because maintaining online safety demands sustained attention \cite{wohnVolunteerModeratorsTwitch2019} and bots are constrained by rigid, context-insensitive filters \cite{jhaverPersonalizingContentModeration2023}. Notably, P7’s experience showed that when moderators are absent on servers, risky social interactions may escalate and prolong teenagers’ exposure to harm. In these moments, teenagers’ vigilant attitudes and situational understanding become a form of collective safety strategy, enabling them to recognize emerging risks from the community and alert others when necessary. Teenagers’ vigilant behaviors could compensate for governance limitations and help sustain safer community dynamics—even in volatile real-time spaces such as active voice calls.

We found that teen accounts on Discord are opt-in and designed to interact with channel permissions, helping prevent teenagers from accidentally accessing explicit content. P3’s experience with channels structured to protect minors illustrates how teenagers can safely explore new online communities while pursuing their interests. These permission structures show how server moderators proactively shape safer exploration environments—ensuring that teens can join new communities, pursue their interests, and engage with unfamiliar spaces without facing unnecessary risk. This reflects a broader pattern where governance can collectively support teen safety while enabling their participation.

Our study presents how servers afford effective mitigation of risky interactions in real time through teenagers’ proactive and inventive strategies. Actively playing games with strangers can be risky \cite{zhang-kennedyNosyLittleBrothers2016}, but our participants intentionally entered these spaces together, using each other’s presence as a collective safety strategy. Participants like P1 and P9 illustrate how governance may be socially created when risks are present, keeping conversations focused on gameplay or using soundboards to disengage risky topics. Additionally, this suggests how Discord safety moderation can be enacted in real time without formal moderation powers. Therefore, even without moderators in voice channels or tools to block inappropriate questions, teenagers may collaborate with their friends to creatively improvise strategies for collective safety. Overall, our findings suggest strengthening governance on Discord through vigilant moderation systems of people and tools that ensure reliable, transparent, and responsive protective oversight for teenagers’ safety needs in online communities. 

\subsection{Design Implications}
Our research suggests that designing tools that support teenagers’ strategies can enhance both individual and community-level safety on Discord. Additionally, our findings could impact teen-informed designs that can empower their communities to collectively manage risks. These approaches not only support teenagers’ agency in navigating online spaces but also foster safer, adaptive communities that evolve alongside their members and discourse. While our research is intentionally interpretive and Discord-specific, our insights support design implications for teen-centered safety features on other social platforms. Our findings can guide designers and researchers to implement solutions that support vigilant strategies—enhancing the ability of teenagers to monitor interactions, detect threats, and respond proactively.

Despite teenagers’ vigilant attitudes on Discord, navigating risky social interactions could be challenging due to inaccessible safety tools and unclear privacy features. Some teenagers, such as P3, initially struggled with safety because protective settings were difficult to understand and locate. Thus, onboarding workflows for new users and teen accounts that show where to find these settings and explain what they do—using real-life examples from other teenagers of how and when to use them—could help teenagers better navigate the platform. We synthesized our findings into \textbf{“Discord Safety Guidelines for Teens by Teens”} (Appendix A) to bring forth participants’ experiences and what other teenagers could learn from them. We encourage parents, educators, and other youth-safety stakeholders to share these guidelines with teens. Additionally, default settings could minimize disclosure and allow users to opt into greater visibility if they feel comfortable. An accessible privacy management tool would empower teenagers to maintain control over their presence on social platforms like Discord without making themselves vulnerable to risks. 

We found that Discord does not notify users about account takeovers. Even though such takeovers could be detected using machine-learning methods \cite{kawaseInternetFraudCase2019}, our study shows that the navigation of complex security risks relies on teenagers’ ability to monitor, assess, and respond to potential threats in real time by noticing suspicious cues on the platform. For example, P12 investigated a “pretty weird” verification process and identified it as suspicious by comparing it to known legitimate ones. Despite research suggesting that it is rare to experience account takeover by a familiar user \cite{thomasSoKHateHarassment2021}, our findings show how takeover can easily occur after interacting with a friend. After trying to play a game that a deceptive “close friend” recommended, P5 lost his Steam account because there was no signal to indicate the security risk. For such complex risks, automated tools, particularly explainable machine learning \cite{zhangExplainableEmpiricalRisk2024, reisExplainableMachineLearning2019}, could enhance moderation by detecting suspicious content and notifying teens about deceptive risks. 

Improving risk detection tools could transform vigilance into proactive safety, empowering teens to identify and address risks before harm occurs. Our study shows that risky content on Discord is not adequately filtered and typical teenage use of the platform complicates risk avoidance. Unlike content-based platforms such as TikTok and Instagram \cite{jhaverPersonalizingContentModeration2023, cresciPersonalizedInterventionsOnline2022}, where algorithmic filters can effectively moderate posts, teenagers on Discord encounter risky links or account requests through everyday gaming interactions. Our findings show that these navigating risks currently relies heavily on teenagers’ vigilance, like P10 and P12 recognizing suspicious details. Additionally, automated tools could provide context about a server’s culture to directly complement teenagers’ existing practice of monitoring conversations, like P6 upon joining a new server. 

Building on this, teenagers may benefit from nuanced designs that balance risky social interactions with safety to enhance their selective participation. Our findings about Discord servers enable teenagers to monitor community cultures and decide whether to participate, while its blocking feature allows teenagers to manage social interactions directly. For example, P16 described how blocking does not entirely remove the visibility of a user in shared servers, meaning teenagers must still navigate residual traces of unwanted contact in the form of hidden messages. This created an environment where teenagers like P16 must navigate ongoing shared experiences with vigilance. Furthermore, our findings suggest that this design could be improved by notifying users how blocking may impact their security, addressing concerns such as P4’s fear of unintended harm. Discord could be designed to provide clear, informative messages about how blocking choices affect teenagers’ visibility and security. Then, platforms could confirm that blocked users cannot further harass the teenager. 

Greater transparency about a stranger’s intentions could help teenagers on Discord reduce contact risks while pursuing social connections. Offering an interface that clarifies why a Message Request was initiated or how a profile was found could help teens interact safer. Providing this information as a nudge \cite{aghaTrickyVsTransparent2024, noggleManipulationSalienceNudges2018, obajemuEnforcingGoodDigital2024} could help teenagers infer others’ intentions while preserving their agency. Furthermore, platforms may consider highlighting a profile’s creation date to further signify a suspicious user, especially in the context of sharing game account information. Overall, improving teenagers’ understanding of unfamiliar users before contact can empower them to make safer choices in online social interactions.

Discord’s design features inadvertently reveal server membership, compromising privacy by signaling interests in potentially unwelcome contexts. Scholars emphasize the importance of designing privacy notices that enable users to customize their settings \cite{kitkowskaEnhancingPrivacyVisual2020} and controls specifically for teenagers \cite{aliTeensCoResearchersAdvocating2025}. In effort to preserve privacy, users may opt for alt accounts to explore their special interests without judgment \cite{goffmanPresentationSelfEveryday1959}. For participants like P9 who actively used an alt account due to negative social perception from peers, social platforms could provide a comprehensive, easily accessible privacy dashboard that allows teenagers to protect their interests. For example, allowing control over features that expose personal information and offer toggles for controlling the visibility of in-platform activities.

As teenagers continue to socialize with friends and strangers on Discord, their risk navigation strategies could be enhanced through safety by design. Safety by design embeds safer experiences into a platform by focusing on proactive and preventative measures \cite{SafetyDesignESafety}. Thus, designing for teenagers’ online safety on Discord may consider a multi-stakeholder perspective \cite{maLabelingDarkExploring2024}, where moderators and the platform should aid their navigation of risky online social interactions. Educators might also explore the safety by design approach for addressing problematic behavior, including educational interventions and mechanisms that encourage accountability and opportunities for behavioral change \cite{huangOpportunitiesTensionsChallenges2024}. Ensuring teenagers’ safety on Discord requires coordinated action from responsible adults, researchers, and platform designers, collaborating to create teen-centered strategies that strengthen their vigilance.
\section{Limitations and Future Work}
Recruitment for this study was constrained by IRB restrictions on directly engaging teenagers, making it challenging to obtain a diverse sample. Reaching participants via Discord proved particularly difficult, as the authors do not participate in teen-centric spaces. Although recruitment messages were posted on various Discord servers with moderator permission, server rules limiting self-promotion posed additional barriers. In some cases, Discord moderators responded with hostility, harassing and banning the first author from the server or immediately removing the author’s request. Others expressed skepticism about the study’s relevance or appropriateness for their communities, even after receiving official study information. Future research could broaden the sample by including participants from countries outside the United States or by targeting specific racial and gender demographics. 

We presented design implications with consideration of the Discord-specific context and the interpretive nature of the analysis, which may limit generalizability to other social platforms \cite{dourishImplicationsDesign2006}. Future research may consider quantitative analysis and investigating teenagers’ experiences with algorithmic moderation features on Discord to design tools that aid safer navigation of social interactions online. We further suggest researchers to investigate how teenagers’ vigilance is impacted beyond Discord.

\section{Conclusion}
Our study reveals how teenagers mitigate risks with vigilance when navigating social interactions with others on Discord. Through interviews and Reflexive Thematic Analysis, we indicate that teenagers interpret risks when developing interpersonal connections by identifying deceptive methods, noticing linguistic cues, applying restrictive features, and playfully engaging with risks in controlled ways. Additionally, they navigate servers strategically to mitigate community-level risks by selectively participating, setting privacy boundaries, and actively engaging in moderation processes. These strategies illustrate how teenagers enact vigilance both at the individual and community levels, proactively managing potential harms while maintaining social engagement and agency on Discord. We propose teen-centered design implications for social platforms to support their vigilant risk management, enhance privacy and safety, and enable meaningful social connection.

\begin{acks}
We extend our gratitude to each teenager involved in the study for being open to share their experiences with risky social interactions on Discord. We also appreciate their parents for taking time to understand and consent to our study on behalf of their teenagers. We thank our anonymous reviewers' suggestions for improving the paper. 

This material is based upon work supported by the U.S. National Science Foundation under award No. 2334934. Any opinions, findings and conclusions or recommendations expressed in this material are those of the author(s) and do not necessarily reflect the views of the U.S. National Science Foundation. 

This manuscript was minimally edited with AI. ChatGPT was used sparingly, only to assist with basic grammar checks and sometimes improve clarity. Text was not directly copied and pasted from another tool.
\end{acks}

\bibliographystyle{ACM-Reference-Format}
\bibliography{td}

@misc{ActivitySharingDiscord2025,
  title = {Activity {{Sharing}} on {{Discord FAQ}}},
  author = {, Buffy},
  year = 2025,
  month = aug,
  journal = {Discord},
  urldate = {2025-08-22},
  howpublished = {https://support.discord.com/hc/en-us/articles/7931156448919-Activity-Sharing-on-Discord-FAQ},
  langid = {american}
}

@article{adorjanNewPrivacyParadox2019,
  title = {A {{New Privacy Paradox}}? {{Youth Agentic Practices}} of {{Privacy Management Despite}} ``{{Nothing}} to {{Hide}}'' {{Online}}},
  shorttitle = {A {{New Privacy Paradox}}?},
  author = {Adorjan, Michael and Ricciardelli, Rosemary},
  year = 2019,
  journal = {Canadian Review of Sociology/Revue canadienne de sociologie},
  volume = {56},
  number = {1},
  pages = {8--29},
  issn = {1755-618X},
  doi = {10.1111/cars.12227},
  urldate = {2025-09-10},
  copyright = {\copyright{} 2019 Canadian Sociological Association/La Soci\'et\'e canadienne de sociologie},
  langid = {english}
}

@inproceedings{aghaCaseStudyUser2022,
  title = {A {{Case Study}} on {{User Experience Bootcamps}} with {{Teens}} to {{Co-Design Real-Time Online Safety Interventions}}},
  booktitle = {Extended {{Abstracts}} of the 2022 {{CHI Conference}} on {{Human Factors}} in {{Computing Systems}}},
  author = {Agha, Zainab and Zhang, Zinan and Obajemu, Oluwatomisin and Shirley, Luke and J. Wisniewski, Pamela},
  year = 2022,
  month = apr,
  series = {{{CHI EA}} '22},
  pages = {1--8},
  publisher = {Association for Computing Machinery},
  address = {New York, NY, USA},
  doi = {10.1145/3491101.3503563},
  urldate = {2025-07-15},
  isbn = {978-1-4503-9156-6}
}

@inproceedings{aghaConductingResponsibleResearch2020,
  title = {Towards {{Conducting Responsible Research}} with {{Teens}} and {{Parents}} Regarding {{Online Risks}}},
  booktitle = {Extended {{Abstracts}} of the 2020 {{CHI Conference}} on {{Human Factors}} in {{Computing Systems}}},
  author = {Agha, Zainab and Chatlani, Neeraj and Razi, Afsaneh and Wisniewski, Pamela},
  year = 2020,
  month = apr,
  series = {{{CHI EA}} '20},
  pages = {1--8},
  publisher = {Association for Computing Machinery},
  address = {New York, NY, USA},
  doi = {10.1145/3334480.3383073},
  urldate = {2025-09-09},
  isbn = {978-1-4503-6819-3}
}

@article{aghaStrikeRootCodesigning2023,
  title = {"{{Strike}} at the {{Root}}": {{Co-designing Real-Time Social Media Interventions}} for {{Adolescent Online Risk Prevention}}},
  shorttitle = {"{{Strike}} at the {{Root}}"},
  author = {Agha, Zainab and {Badillo-Urquiola}, Karla and Wisniewski, Pamela J.},
  year = 2023,
  month = apr,
  journal = {Proceedings of the ACM on Human-Computer Interaction},
  volume = {7},
  number = {CSCW1},
  pages = {1--32},
  issn = {2573-0142},
  doi = {10.1145/3579625},
  urldate = {2024-10-03},
  langid = {english}
}

@inproceedings{aghaTrickyVsTransparent2024,
  title = {Tricky vs. {{Transparent}}: {{Towards}} an {{Ecologically Valid}} and {{Safe Approach}} for {{Evaluating Online Safety Nudges}} for {{Teens}}},
  shorttitle = {Tricky vs. {{Transparent}}},
  booktitle = {Proceedings of the 2024 {{CHI Conference}} on {{Human Factors}} in {{Computing Systems}}},
  author = {Agha, Zainab and Park, Jinkyung and Wan, Ruyuan and Ali, Naima Samreen and Wang, Yiwei and Difranzo, Dominic and {Badillo-Urquiola}, Karla and Wisniewski, Pamela J.},
  year = 2024,
  month = may,
  series = {{{CHI}} '24},
  pages = {1--20},
  publisher = {Association for Computing Machinery},
  address = {New York, NY, USA},
  doi = {10.1145/3613904.3642313},
  urldate = {2025-06-06},
  isbn = {979-8-4007-0330-0}
}

@article{agostoDontBeDumb2017,
  title = {``{{Don}}'t Be Dumb---That's the Rule {{I}} Try to Live by'': {{A}} Closer Look at Older Teens' Online Privacy and Safety Attitudes},
  shorttitle = {``{{Don}}'t Be Dumb---That's the Rule {{I}} Try to Live by''},
  author = {Agosto, Denise E and Abbas, June},
  year = 2017,
  month = mar,
  journal = {New Media \& Society},
  volume = {19},
  number = {3},
  pages = {347--365},
  publisher = {SAGE Publications},
  issn = {1461-4448},
  doi = {10.1177/1461444815606121},
  urldate = {2025-06-06},
  langid = {english}
}

@article{akterCalculatingConnectionVs2025,
  title = {Calculating {{Connection}} vs. {{Risk}}: {{Understanding How Youth Negotiate Digital Privacy}} and {{Security}} with {{Peers Online}}},
  shorttitle = {Calculating {{Connection}} vs. {{Risk}}},
  author = {Akter, Mamtaj and Park, Jinkyung Katie and Headrick, Campbell Robinson and Page, Xinru and Wisniewski, Pamela J.},
  year = 2025,
  month = oct,
  journal = {Proc. ACM Hum.-Comput. Interact.},
  volume = {9},
  number = {7},
  pages = {CSCW383:1--CSCW383:26},
  doi = {10.1145/3757564},
  urldate = {2025-12-04}
}

@article{aliGettingMetaMultimodal2023,
  title = {Getting {{Meta}}: {{A Multimodal Approach}} for {{Detecting Unsafe Conversations}} within {{Instagram Direct Messages}} of {{Youth}}},
  shorttitle = {Getting {{Meta}}},
  author = {Ali, Shiza and Razi, Afsaneh and Kim, Seunghyun and Alsoubai, Ashwaq and Ling, Chen and De Choudhury, Munmun and Wisniewski, Pamela J. and Stringhini, Gianluca},
  year = 2023,
  month = apr,
  journal = {Proc. ACM Hum.-Comput. Interact.},
  volume = {7},
  number = {CSCW1},
  pages = {132:1--132:30},
  doi = {10.1145/3579608},
  urldate = {2025-06-29}
}

@article{aliTeensCoResearchersAdvocating2025,
  title = {Teens as {{Co-Researchers}}: {{Advocating}} for {{Disruptive Change}} to {{Engage Youth Meaningfully}} in {{Online Safety Research}} and the {{Design}} of {{Social Media}}},
  shorttitle = {Teens as {{Co-Researchers}}},
  author = {Ali, Naima Samreen and Ma, Renkai and Agha, Zainab and Park, Jinkyung Katie and Wisniewski, Pamela J.},
  year = 2025,
  month = oct,
  journal = {Proc. ACM Hum.-Comput. Interact.},
  volume = {9},
  number = {7},
  pages = {CSCW484:1--CSCW484:31},
  doi = {10.1145/3757665},
  urldate = {2025-12-02}
}

@inproceedings{aliUnderstandingDigitalLives2022,
  title = {Understanding the {{Digital Lives}} of {{Youth}}: {{Analyzing Media Shared}} within {{Safe Versus Unsafe Private Conversations}} on {{Instagram}}},
  shorttitle = {Understanding the {{Digital Lives}} of {{Youth}}},
  booktitle = {Proceedings of the 2022 {{CHI Conference}} on {{Human Factors}} in {{Computing Systems}}},
  author = {Ali, Shiza and Razi, Afsaneh and Kim, Seunghyun and Alsoubai, Ashwaq and Gracie, Joshua and De Choudhury, Munmun and Wisniewski, Pamela J. and Stringhini, Gianluca},
  year = 2022,
  month = apr,
  series = {{{CHI}} '22},
  pages = {1--14},
  publisher = {Association for Computing Machinery},
  address = {New York, NY, USA},
  doi = {10.1145/3491102.3501969},
  urldate = {2025-04-13},
  isbn = {978-1-4503-9157-3}
}

@article{alluhidanTeenTalkGood2024,
  title = {Teen {{Talk}}: {{The Good}}, the {{Bad}}, and the {{Neutral}} of {{Adolescent Social Media Use}}},
  shorttitle = {Teen {{Talk}}},
  author = {Alluhidan, Abdulmalik and Akter, Mamtaj and Alsoubai, Ashwaq and Park, Jinkyung Katie and Wisniewski, Pamela},
  year = 2024,
  month = nov,
  journal = {Proc. ACM Hum.-Comput. Interact.},
  volume = {8},
  number = {CSCW2},
  pages = {422:1--422:35},
  doi = {10.1145/3686961},
  urldate = {2025-06-12}
}

@article{alsoubaiFriendsBenefitsSextortion2022,
  title = {From '{{Friends}} with {{Benefits}}' to '{{Sextortion}}:' {{A Nuanced Investigation}} of {{Adolescents}}' {{Online Sexual Risk Experiences}}},
  shorttitle = {From '{{Friends}} with {{Benefits}}' to '{{Sextortion}}},
  author = {Alsoubai, Ashwaq and Song, Jihye and Razi, Afsaneh and Naher, Nurun and De Choudhury, Munmun and Wisniewski, Pamela J.},
  year = 2022,
  month = nov,
  journal = {Proc. ACM Hum.-Comput. Interact.},
  volume = {6},
  number = {CSCW2},
  pages = {411:1--411:32},
  doi = {10.1145/3555136},
  urldate = {2025-06-07}
}

@inproceedings{alsoubaiHumanCenteredApproachImproving2023,
  title = {A {{Human-Centered Approach}} to {{Improving Adolescent Real-Time Online Risk Detection Algorithms}}},
  booktitle = {Extended {{Abstracts}} of the 2023 {{CHI Conference}} on {{Human Factors}} in {{Computing Systems}}},
  author = {Alsoubai, Ashwaq},
  year = 2023,
  month = apr,
  series = {{{CHI EA}} '23},
  pages = {1--5},
  publisher = {Association for Computing Machinery},
  address = {New York, NY, USA},
  doi = {10.1145/3544549.3577045},
  urldate = {2025-05-16},
  isbn = {978-1-4503-9422-2}
}

@article{alsoubaiProfilingOfflineOnline2024,
  title = {Profiling the {{Offline}} and {{Online Risk Experiences}} of {{Youth}} to {{Develop Targeted Interventions}} for {{Online Safety}}},
  author = {Alsoubai, Ashwaq and Razi, Afsaneh and Agha, Zainab and Ali, Shiza and Stringhini, Gianluca and De Choudhury, Munmun and Wisniewski, Pamela J.},
  year = 2024,
  month = apr,
  journal = {Proceedings of the ACM on Human-Computer Interaction},
  volume = {8},
  number = {CSCW1},
  pages = {1--37},
  issn = {2573-0142},
  doi = {10.1145/3637391},
  urldate = {2025-03-19},
  langid = {english}
}

@article{arriagadaYouNeedLeast2020,
  title = {``{{You Need At Least One Picture Daily}}, If {{Not}}, {{You}}'re {{Dead}}'': {{Content Creators}} and {{Platform Evolution}} in the {{Social Media Ecology}}},
  shorttitle = {``{{You Need At Least One Picture Daily}}, If {{Not}}, {{You}}'re {{Dead}}''},
  author = {Arriagada, Arturo and Ib{\'a}{\~n}ez, Francisco},
  year = 2020,
  month = jul,
  journal = {Social Media + Society},
  volume = {6},
  number = {3},
  pages = {2056305120944624},
  publisher = {SAGE Publications Ltd},
  issn = {2056-3051},
  doi = {10.1177/2056305120944624},
  urldate = {2025-11-25},
  langid = {english}
}

@inproceedings{ashktorabDesigningCyberbullyingMitigation2016,
  title = {Designing {{Cyberbullying Mitigation}} and {{Prevention Solutions}} through {{Participatory Design With Teenagers}}},
  booktitle = {Proceedings of the 2016 {{CHI Conference}} on {{Human Factors}} in {{Computing Systems}}},
  author = {Ashktorab, Zahra and Vitak, Jessica},
  year = 2016,
  month = may,
  series = {{{CHI}} '16},
  pages = {3895--3905},
  publisher = {Association for Computing Machinery},
  address = {New York, NY, USA},
  doi = {10.1145/2858036.2858548},
  urldate = {2025-06-06},
  isbn = {978-1-4503-3362-7}
}

@inproceedings{azzopardiAssessingRisksOnline2025,
  title = {Assessing {{Risks}} in {{Online Information Sharing}}},
  booktitle = {Proceedings of the 2025 {{ACM SIGIR Conference}} on {{Human Information Interaction}} and {{Retrieval}}},
  author = {Azzopardi, Leif and Nicol, Emma and Briggs, Jo and Moncur, Wendy and Schafer, Burkhard and Nash, Callum and Duheric, Melissa},
  year = 2025,
  month = apr,
  series = {{{CHIIR}} '25},
  pages = {71--80},
  publisher = {Association for Computing Machinery},
  address = {New York, NY, USA},
  doi = {10.1145/3698204.3716447},
  urldate = {2025-12-03},
  isbn = {979-8-4007-1290-6}
}

@misc{backlinkoteamDiscordUserFunding2023,
  title = {Discord {{User}} and {{Funding Statistics}}: {{How Many People Use Discord}} in 2024?},
  shorttitle = {Discord {{User}} and {{Funding Statistics}}},
  author = {{Backlinko Team}},
  year = 2023,
  month = mar,
  journal = {Backlinko},
  urldate = {2025-05-05},
  howpublished = {https://backlinko.com/discord-users},
  langid = {american}
}

@article{balleysSearchingOneselfYouTube2020,
  title = {Searching for {{Oneself}} on {{YouTube}}: {{Teenage Peer Socialization}} and {{Social Recognition Processes}}},
  shorttitle = {Searching for {{Oneself}} on {{YouTube}}},
  author = {Balleys, Claire and Millerand, Florence and Tho{\"e}r, Christine and Duque, Nina},
  year = 2020,
  month = apr,
  journal = {Social Media + Society},
  volume = {6},
  number = {2},
  pages = {2056305120909474},
  publisher = {SAGE Publications Ltd},
  issn = {2056-3051},
  doi = {10.1177/2056305120909474},
  urldate = {2025-11-16}
}

@article{bassettToughTeensMethodological2008,
  title = {Tough {{Teens}}: {{The Methodological Challenges}} of {{Interviewing Teenagers}} as {{Research Participants}}},
  shorttitle = {Tough {{Teens}}},
  author = {Bassett, Raewyn and Beagan, Brenda L. and {Ristovski-Slijepcevic}, Svetlana and Chapman, Gwen E.},
  year = 2008,
  month = mar,
  journal = {Journal of Adolescent Research},
  volume = {23},
  number = {2},
  pages = {119--131},
  issn = {0743-5584, 1552-6895},
  doi = {10.1177/0743558407310733},
  urldate = {2025-06-15},
  copyright = {https://journals.sagepub.com/page/policies/text-and-data-mining-license},
  langid = {english}
}

@article{benvenutiTeensOnlineHow2024,
  title = {Teens Online: How Perceived Social Support Influences the Use of the {{Internet}} during Adolescence},
  shorttitle = {Teens Online},
  author = {Benvenuti, Martina and Panesi, Sabrina and Giovagnoli, Sara and Selleri, Patrizia and Mazzoni, Elvis},
  year = 2024,
  month = jun,
  journal = {European Journal of Psychology of Education},
  volume = {39},
  number = {2},
  pages = {629--650},
  issn = {1878-5174},
  doi = {10.1007/s10212-023-00705-5},
  urldate = {2025-08-29},
  langid = {english}
}

@inproceedings{beresDontYouKnow2021,
  title = {Don't {{You Know That You}}'re {{Toxic}}: {{Normalization}} of {{Toxicity}} in {{Online Gaming}}},
  shorttitle = {Don't {{You Know That You}}'re {{Toxic}}},
  booktitle = {Proceedings of the 2021 {{CHI Conference}} on {{Human Factors}} in {{Computing Systems}}},
  author = {Beres, Nicole A and Frommel, Julian and Reid, Elizabeth and Mandryk, Regan L and Klarkowski, Madison},
  year = 2021,
  month = may,
  pages = {1--15},
  publisher = {ACM},
  address = {Yokohama Japan},
  doi = {10.1145/3411764.3445157},
  urldate = {2024-10-09},
  isbn = {978-1-4503-8096-6},
  langid = {english}
}

@article{bhagatRoleIndividualsNeed2020,
  title = {The {{Role}} of {{Individuals}}' {{Need}} for {{Online Social Interactions}} and {{Interpersonal Incompetence}} in {{Digital Game Addiction}}},
  author = {Bhagat, Sarbottam and Jeong, Eui Jun and Kim, Dan J.},
  year = 2020,
  month = mar,
  journal = {International Journal of Human--Computer Interaction},
  volume = {36},
  number = {5},
  pages = {449--463},
  publisher = {Taylor \& Francis},
  issn = {1044-7318},
  doi = {10.1080/10447318.2019.1654696},
  urldate = {2025-09-11}
}

@article{boydConnectedConcernedVariation2013,
  title = {Connected and Concerned: {{Variation}} in Parents' Online Safety Concerns},
  shorttitle = {Connected and Concerned},
  author = {Boyd, Danah and Hargittai, Eszter},
  year = 2013,
  journal = {Policy \& Internet},
  volume = {5},
  number = {3},
  pages = {245--269},
  issn = {1944-2866},
  doi = {10.1002/1944-2866.POI332},
  urldate = {2025-09-08},
  langid = {english}
}

@article{boydSocialNetworkSites2007,
  title = {Social {{Network Sites}}: {{Definition}}, {{History}}, and {{Scholarship}}},
  shorttitle = {Social {{Network Sites}}},
  author = {Boyd, Danah M. and Ellison, Nicole B.},
  year = 2007,
  month = oct,
  journal = {Journal of Computer-Mediated Communication},
  volume = {13},
  number = {1},
  pages = {210--230},
  issn = {10836101},
  doi = {10.1111/j.1083-6101.2007.00393.x},
  urldate = {2025-11-16},
  copyright = {http://doi.wiley.com/10.1002/tdm\_license\_1.1},
  langid = {english}
}

@article{bozzolaUseSocialMedia2022,
  title = {The {{Use}} of {{Social Media}} in {{Children}} and {{Adolescents}}: {{Scoping Review}} on the {{Potential Risks}}},
  shorttitle = {The {{Use}} of {{Social Media}} in {{Children}} and {{Adolescents}}},
  author = {Bozzola, Elena and Spina, Giulia and Agostiniani, Rino and Barni, Sarah and Russo, Rocco and Scarpato, Elena and Di Mauro, Antonio and Di Stefano, Antonella Vita and Caruso, Cinthia and Corsello, Giovanni and Staiano, Annamaria},
  year = 2022,
  month = jan,
  journal = {International Journal of Environmental Research and Public Health},
  volume = {19},
  number = {16},
  pages = {9960},
  publisher = {Multidisciplinary Digital Publishing Institute},
  issn = {1660-4601},
  doi = {10.3390/ijerph19169960},
  urldate = {2025-09-07},
  copyright = {http://creativecommons.org/licenses/by/3.0/},
  langid = {english}
}

@incollection{braunDoingReflexiveThematic2022,
  title = {Doing {{Reflexive Thematic Analysis}}},
  booktitle = {Supporting {{Research}} in {{Counselling}} and {{Psychotherapy}}},
  author = {Braun, Virginia and Clarke, Victoria and Hayfield, Nikki and Davey, Louise and Jenkinson, Elizabeth},
  editor = {{Bager-Charleson}, Sofie and McBeath, Alistair},
  year = 2022,
  pages = {19--38},
  publisher = {Springer International Publishing},
  address = {Cham},
  doi = {10.1007/978-3-031-13942-0_2},
  urldate = {2025-01-23},
  isbn = {978-3-031-13941-3 978-3-031-13942-0},
  langid = {english}
}

@book{braunThematicAnalysisPractical2021,
  title = {Thematic {{Analysis}}: {{A Practical Guide}}},
  author = {Braun, Virginia and Clarke, Victoria},
  year = 2021,
  month = nov,
  urldate = {2025-02-19},
  langid = {english}
}

@article{braunUsingThematicAnalysis2006,
  title = {Using Thematic Analysis in Psychology},
  author = {Braun, Virginia and Clarke, Victoria},
  year = 2006,
  month = jan,
  journal = {Qualitative Research in Psychology},
  volume = {3},
  number = {2},
  pages = {77--101},
  issn = {1478-0887, 1478-0895},
  doi = {10.1191/1478088706qp063oa},
  urldate = {2024-10-07},
  langid = {english}
}

@article{bredaCriticalReviewResilience2018,
  title = {A Critical Review of Resilience Theory and Its Relevance for Social Work},
  author = {Breda, Van and D, Adrian},
  year = 2018,
  journal = {Social Work/Maatskaplike Werk},
  volume = {54},
  number = {1},
  pages = {1--18},
  publisher = {Authors of the articles},
  issn = {0037-8054},
  doi = {10.15270/54-1-611},
  urldate = {2025-09-03},
  langid = {english}
}

@article{brooksChildrenRiskFostering1994,
  title = {Children at Risk: {{Fostering}} Resilience and Hope},
  shorttitle = {Children at Risk},
  author = {Brooks, Robert B.},
  year = 1994,
  journal = {American Journal of Orthopsychiatry},
  volume = {64},
  number = {4},
  pages = {545--553},
  publisher = {American Orthopsychiatric Association, Inc.},
  address = {US},
  issn = {1939-0025},
  doi = {10.1037/h0079565}
}

@incollection{brownHatewareOutsourcingResponsibility2019,
  title = {Hateware and the {{Outsourcing}} of {{Responsibility}}},
  booktitle = {Digital {{Ethics}}: {{Rhetoric}} and {{Responsibility}} in {{Online Aggression}}},
  editor = {Brown, James and Hennis, Gregory and Reyman, Jessica and Sparby, Erika M.},
  year = 2019,
  month = jul,
  edition = {1},
  pages = {16},
  publisher = {Routledge},
  doi = {10.4324/9780429266140},
  urldate = {2025-06-22},
  isbn = {978-0-429-26614-0},
  langid = {english}
}

@article{bryceRoleDisclosurePersonal2014,
  title = {The Role of Disclosure of Personal Information in the Evaluation of Risk and Trust in Young Peoples' Online Interactions},
  author = {Bryce, Jo and Fraser, James},
  year = 2014,
  month = jan,
  journal = {Computers in Human Behavior},
  volume = {30},
  pages = {299--306},
  issn = {0747-5632},
  doi = {10.1016/j.chb.2013.09.012},
  urldate = {2025-06-07}
}

@misc{buffyFamilyCenterTeens2025,
  title = {Family {{Center}} for {{Teens}}},
  author = {{Buffy}},
  year = 2025,
  month = nov,
  journal = {Discord},
  urldate = {2025-12-03},
  howpublished = {https://support.discord.com/hc/en-us/articles/14155060633623-Family-Center-for-Teens},
  langid = {american}
}

@article{bullinghamPresentationSelfOnline2013,
  title = {`{{The}} Presentation of Self in the Online World': {{Goffman}} and the Study of Online Identities},
  shorttitle = {`{{The}} Presentation of Self in the Online World'},
  author = {Bullingham, Liam and Vasconcelos, Ana C.},
  year = 2013,
  month = feb,
  journal = {Journal of Information Science},
  volume = {39},
  number = {1},
  pages = {101--112},
  publisher = {SAGE Publications Ltd},
  issn = {0165-5515},
  doi = {10.1177/0165551512470051},
  urldate = {2025-11-25}
}

@misc{caspergrathwohlOxfordWordYear,
  title = {Oxford {{Word}} of the {{Year}} 2025},
  author = {{Casper Grathwohl}},
  journal = {Oxford University Press},
  urldate = {2025-12-01},
  langid = {british}
}

@article{cernikovaYouthInteractionOnline2018,
  title = {Youth Interaction with Online Strangers: Experiences and Reactions to Unknown People on the {{Internet}}},
  shorttitle = {Youth Interaction with Online Strangers},
  author = {Cernikova, Martina and Dedkova, Lenka and Smahel, David},
  year = 2018,
  month = jan,
  journal = {Information, Communication \& Society},
  publisher = {Routledge},
  issn = {1369-118X},
  urldate = {2025-07-28},
  copyright = {\copyright{} 2016 Informa UK Limited, trading as Taylor \& Francis Group},
  langid = {english}
}

@article{chandrasekharanCrossmodCrossCommunityLearningbased2019,
  title = {Crossmod: {{A Cross-Community Learning-based System}} to {{Assist Reddit Moderators}}},
  shorttitle = {Crossmod},
  author = {Chandrasekharan, Eshwar and Gandhi, Chaitrali and Mustelier, Matthew Wortley and Gilbert, Eric},
  year = 2019,
  month = nov,
  journal = {Proceedings of the ACM on Human-Computer Interaction},
  volume = {3},
  number = {CSCW},
  pages = {1--30},
  issn = {2573-0142},
  doi = {10.1145/3359276},
  urldate = {2025-02-26},
  langid = {english}
}

@article{charterisSnapchatYouthSubjectivities2018,
  title = {`{{Snapchat}}', Youth Subjectivities and Sexuality: Disappearing Media and the Discourse of Youth Innocence},
  shorttitle = {`{{Snapchat}}', Youth Subjectivities and Sexuality},
  author = {Charteris, Jennifer and Gregory, Sue and Masters, Yvonne},
  year = 2018,
  month = feb,
  journal = {Gender and Education},
  volume = {30},
  number = {2},
  pages = {205--221},
  issn = {0954-0253, 1360-0516},
  doi = {10.1080/09540253.2016.1188198},
  urldate = {2025-09-12},
  langid = {english}
}

@inproceedings{choiSocializationTacticsWikipedia2010,
  title = {Socialization Tactics in Wikipedia and Their Effects},
  booktitle = {Proceedings of the 2010 {{ACM}} Conference on {{Computer}} Supported Cooperative Work},
  author = {Choi, Boreum and Alexander, Kira and Kraut, Robert E. and Levine, John M.},
  year = 2010,
  month = feb,
  series = {{{CSCW}} '10},
  pages = {107--116},
  publisher = {Association for Computing Machinery},
  address = {New York, NY, USA},
  doi = {10.1145/1718918.1718940},
  urldate = {2025-08-28},
  isbn = {978-1-60558-795-0}
}

@article{chouHowTeensNegotiate2023,
  title = {How Teens Negotiate Privacy on Social Media Proactively and Reactively},
  author = {Chou, Hui-Lien and Chou, Chien},
  year = 2023,
  month = jun,
  journal = {New Media \& Society},
  volume = {25},
  number = {6},
  pages = {1290--1312},
  publisher = {SAGE Publications},
  issn = {1461-4448},
  doi = {10.1177/14614448211018797},
  urldate = {2025-09-04},
  langid = {english}
}

@article{coppOnlineSexualHarassment2021,
  title = {Online Sexual Harassment and Cyberbullying in a Nationally Representative Sample of Teens: {{Prevalence}}, Predictors, and Consequences},
  shorttitle = {Online Sexual Harassment and Cyberbullying in a Nationally Representative Sample of Teens},
  author = {Copp, Jennifer E. and Mumford, Elizabeth A. and Taylor, Bruce G.},
  year = 2021,
  month = dec,
  journal = {Journal of Adolescence},
  volume = {93},
  pages = {202--211},
  issn = {1095-9254},
  doi = {10.1016/j.adolescence.2021.10.003},
  langid = {english},
  pmid = {34801812}
}

@article{costelloVulnerableSelfdisclosureCodevelops2024,
  title = {Vulnerable Self-Disclosure Co-Develops in Adolescent Friendships: {{Developmental}} Foundations of Emotional Intimacy},
  shorttitle = {Vulnerable Self-Disclosure Co-Develops in Adolescent Friendships},
  author = {Costello, Meghan A. and Bailey, Natasha A. and Stern, Jessica A. and Allen, Joseph P.},
  year = 2024,
  month = sep,
  journal = {Journal of social and personal relationships},
  volume = {41},
  number = {9},
  pages = {2432--2454},
  issn = {0265-4075},
  doi = {10.1177/02654075241244821},
  urldate = {2025-09-11},
  pmcid = {PMC11781371},
  pmid = {39885900}
}

@inproceedings{cresciPersonalizedInterventionsOnline2022,
  title = {Personalized {{Interventions}} for {{Online Moderation}}},
  booktitle = {Proceedings of the 33rd {{ACM Conference}} on {{Hypertext}} and {{Social Media}}},
  author = {Cresci, Stefano and Trujillo, Amaury and Fagni, Tiziano},
  year = 2022,
  month = jun,
  pages = {248--251},
  publisher = {ACM},
  address = {Barcelona Spain},
  doi = {10.1145/3511095.3536369},
  urldate = {2025-02-27},
  isbn = {978-1-4503-9233-4},
  langid = {english}
}

@article{dasPlatformGovernancePresent2022,
  title = {Platform {{Governance}}: {{Past}}, {{Present}} and {{Future}}},
  shorttitle = {Platform {{Governance}}},
  author = {Das, Mithun and Dash, Abhisek and Jaiswal, Siddharth and Mathew, Binny and Saha, Punyajoy and Mukerjee, Animesh},
  year = 2022,
  month = may,
  journal = {GetMobile: Mobile Computing and Communications},
  volume = {26},
  number = {1},
  pages = {14--20},
  issn = {2375-0529, 2375-0537},
  doi = {10.1145/3539668.3539674},
  urldate = {2025-09-09},
  langid = {english}
}

@article{dearnleyReflectionUseSemistructured2005,
  title = {A Reflection on the Use of Semi-Structured Interviews},
  shorttitle = {A Reflection on the Use of Semi-Structured Interviews},
  author = {Dearnley, Christine},
  year = 2005,
  month = jul,
  journal = {Nurse Researcher},
  volume = {13},
  number = {1},
  pages = {19--28},
  issn = {1351-5578, 2047-8992},
  doi = {10.7748/nr2005.07.13.1.19.c5997},
  urldate = {2025-01-24},
  langid = {english}
}

@article{dedkovaParentalMediationOnline2023,
  title = {Parental Mediation of Online Interactions and Its Relation to Adolescents' Contacts with New People Online: The Role of Risk Perception},
  shorttitle = {Parental Mediation of Online Interactions and Its Relation to Adolescents' Contacts with New People Online},
  author = {Dedkova, Lenka and M{\'y}lek, Vojt{\v e}ch},
  year = 2023,
  month = dec,
  journal = {Information, Communication \& Society},
  volume = {26},
  number = {16},
  pages = {3179--3196},
  issn = {1369-118X, 1468-4462},
  doi = {10.1080/1369118X.2022.2146985},
  urldate = {2025-08-29},
  langid = {english}
}

@misc{discordDiscordOpensIts,
  title = {Discord {{Opens}} Its {{Social SDK Communication Features}}, {{Powering New Era}} of {{Social Play}}},
  author = {{Discord}},
  urldate = {2025-11-17},
  howpublished = {https://discord.com/press-releases/social-sdk-ingame-communications},
  langid = {english}
}

@misc{DiscordOurMission,
  title = {About {{Discord}} \textbar{} {{Our Mission}} and {{Story}}},
  urldate = {2025-09-06},
  howpublished = {https://discord.com/company},
  langid = {english}
}

@misc{discordRoleAdministratorsModerators,
  title = {Role of {{Administrators}} and {{Moderators}} \textbar{} {{Discord Safety}}},
  author = {Discord},
  urldate = {2025-08-26},
  howpublished = {https://discord.com/safety/360044103531-role-of-administrators-and-moderators-on-discord}
}

@misc{discordsafetyDiscordTransparencyReports,
  title = {Discord {{Transparency Reports}}},
  author = {{Discord Safety}},
  urldate = {2025-06-07},
  howpublished = {https://discord.com/safety-transparency-reports/2024-h1},
  langid = {english}
}

@misc{DiscordSoundboardGuide2025,
  title = {Discord {{Soundboard Guide}}: {{Using}}, {{Adding}}, and {{Managing Sounds}}},
  shorttitle = {Discord {{Soundboard Guide}}},
  author = {, Buffy},
  year = 2025,
  month = may,
  journal = {Discord},
  urldate = {2025-08-22},
  howpublished = {https://support.discord.com/hc/en-us/articles/12612888127767-Discord-Soundboard-Guide-Using-Adding-and-Managing-Sounds},
  langid = {american}
}

@inproceedings{doanDesignSpaceOnline2025,
  title = {The {{Design Space}} for {{Online Restorative Justice Tools}}: {{A Case Study}} with {{ApoloBot}}},
  shorttitle = {The {{Design Space}} for {{Online Restorative Justice Tools}}},
  booktitle = {Proceedings of the 2025 {{CHI Conference}} on {{Human Factors}} in {{Computing Systems}}},
  author = {Doan, Bich Ngoc (Rubi) and Seering, Joseph},
  year = 2025,
  month = apr,
  pages = {1--19},
  publisher = {ACM},
  address = {Yokohama Japan},
  doi = {10.1145/3706598.3713598},
  urldate = {2025-06-06},
  isbn = {979-8-4007-1394-1},
  langid = {english}
}

@article{domahidiDwellGamersInvestigating2014,
  title = {To Dwell among Gamers: {{Investigating}} the Relationship between Social Online Game Use and Gaming-Related Friendships},
  shorttitle = {To Dwell among Gamers},
  author = {Domahidi, Emese and Festl, Ruth and Quandt, Thorsten},
  year = 2014,
  journal = {Computers in Human Behavior},
  volume = {35},
  pages = {107--115},
  publisher = {Elsevier Science},
  address = {Netherlands},
  issn = {1873-7692},
  doi = {10.1016/j.chb.2014.02.023}
}

@article{doringConsensualSextingAdolescents2014,
  title = {Consensual Sexting among Adolescents: {{Risk}} Prevention through Abstinence Education or Safer Sexting?},
  shorttitle = {Consensual Sexting among Adolescents},
  author = {D{\"o}ring, Nicola},
  year = 2014,
  month = mar,
  journal = {Cyberpsychology: Journal of Psychosocial Research on Cyberspace},
  volume = {8},
  number = {1},
  issn = {1802-7962},
  doi = {10.5817/CP2014-1-9},
  urldate = {2025-09-08},
  copyright = {Copyright \copyright{} 2014 Nicola D\"oring},
  langid = {english}
}

@inproceedings{dourishImplicationsDesign2006,
  title = {Implications for Design},
  booktitle = {Proceedings of the {{SIGCHI Conference}} on {{Human Factors}} in {{Computing Systems}}},
  author = {Dourish, Paul},
  year = 2006,
  month = apr,
  series = {{{CHI}} '06},
  pages = {541--550},
  publisher = {Association for Computing Machinery},
  address = {New York, NY, USA},
  doi = {10.1145/1124772.1124855},
  urldate = {2025-12-01},
  isbn = {978-1-59593-372-0}
}

@article{elsaesserParentingDigitalAge2017,
  title = {Parenting in a Digital Age: {{A}} Review of Parents' Role in Preventing Adolescent Cyberbullying},
  shorttitle = {Parenting in a Digital Age},
  author = {Elsaesser, Caitlin and Russell, Beth and Ohannessian, Christine McCauley and Patton, Desmond},
  year = 2017,
  month = jul,
  journal = {Aggression and Violent Behavior},
  volume = {35},
  pages = {62--72},
  issn = {1359-1789},
  doi = {10.1016/j.avb.2017.06.004},
  urldate = {2025-09-08}
}

@article{farrukhYouthInternetSafety2014,
  title = {Youth {{Internet Safety}}: {{Risks}}, {{Responses}}, and {{Research Recommendations}}},
  author = {Farrukh, Adina and Sadwick, Rebecca and Villasenor, John},
  year = 2014,
  month = oct,
  journal = {Center for Technology Innovation at Brookings},
  langid = {english}
}

@inproceedings{farzanSocializingVolunteersOnline2012,
  title = {Socializing Volunteers in an Online Community: A Field Experiment},
  shorttitle = {Socializing Volunteers in an Online Community},
  booktitle = {Proceedings of the {{ACM}} 2012 Conference on {{Computer Supported Cooperative Work}}},
  author = {Farzan, Rosta and Kraut, Robert and Pal, Aditya and Konstan, Joseph},
  year = 2012,
  month = feb,
  series = {{{CSCW}} '12},
  pages = {325--334},
  publisher = {Association for Computing Machinery},
  address = {New York, NY, USA},
  doi = {10.1145/2145204.2145256},
  urldate = {2025-08-28},
  isbn = {978-1-4503-1086-4}
}

@misc{faverioTeensSocialMedia2025,
  title = {Teens and {{Social Media Fact Sheet}}},
  author = {Faverio, Olivia Sidoti {and} Michelle, Eugenie Park},
  year = 2025,
  month = jul,
  journal = {Pew Research Center},
  urldate = {2025-11-16},
  langid = {american}
}

@article{fergusonVideoGameViolence2014,
  title = {Video {{Game Violence Use Among}} ``{{Vulnerable}}'' {{Populations}}: {{The Impact}} of {{Violent Games}} on {{Delinquency}} and {{Bullying Among Children}} with {{Clinically Elevated Depression}} or {{Attention Deficit Symptoms}}},
  shorttitle = {Video {{Game Violence Use Among}} ``{{Vulnerable}}'' {{Populations}}},
  author = {Ferguson, Christopher J. and Olson, Cheryl K.},
  year = 2014,
  month = jan,
  journal = {Journal of Youth and Adolescence},
  volume = {43},
  number = {1},
  pages = {127--136},
  issn = {1573-6601},
  doi = {10.1007/s10964-013-9986-5},
  urldate = {2024-10-18},
  langid = {english}
}

@article{fireOnlineSocialNetworks2014,
  title = {Online {{Social Networks}}: {{Threats}} and {{Solutions}}},
  shorttitle = {Online {{Social Networks}}},
  author = {Fire, Michael and Goldschmidt, Roy and Elovici, Yuval},
  year = 2014,
  journal = {IEEE Communications Surveys \& Tutorials},
  volume = {16},
  number = {4},
  pages = {2019--2036},
  issn = {1553-877X},
  doi = {10.1109/COMST.2014.2321628},
  urldate = {2025-09-07}
}

@article{fogelInternetSocialNetwork2009,
  title = {Internet Social Network Communities: {{Risk}} Taking, Trust, and Privacy Concerns},
  shorttitle = {Internet Social Network Communities},
  author = {Fogel, Joshua and Nehmad, Elham},
  year = 2009,
  month = jan,
  journal = {Computers in Human Behavior},
  volume = {25},
  number = {1},
  pages = {153--160},
  issn = {0747-5632},
  doi = {10.1016/j.chb.2008.08.006},
  urldate = {2025-12-04}
}

@inproceedings{freedUnderstandingDigitalSafetyExperiences2023,
  title = {Understanding {{Digital-Safety Experiences}} of {{Youth}} in the {{U}}.{{S}}.},
  booktitle = {Proceedings of the 2023 {{CHI Conference}} on {{Human Factors}} in {{Computing Systems}}},
  author = {Freed, Diana and Bazarova, Natalie N. and Consolvo, Sunny and Han, Eunice J and Kelley, Patrick Gage and Thomas, Kurt and Cosley, Dan},
  year = 2023,
  month = apr,
  series = {{{CHI}} '23},
  pages = {1--15},
  publisher = {Association for Computing Machinery},
  address = {New York, NY, USA},
  doi = {10.1145/3544548.3581128},
  urldate = {2025-06-06},
  isbn = {978-1-4503-9421-5}
}

@book{goffmanPresentationSelfEveryday1959,
  title = {The Presentation of Self in Everyday Life},
  author = {Goffman, Erving},
  year = 1959,
  series = {The Presentation of Self in Everyday Life},
  publisher = {Doubleday},
  address = {Oxford, England}
}

@misc{gogginDiscordServersUsed2023,
  title = {Discord Servers Used in Child Abductions, Crime Rings, Sextortion},
  author = {Goggin, Ben},
  year = 2023,
  month = jun,
  journal = {NBC News},
  urldate = {2025-08-29},
  howpublished = {https://www.nbcnews.com/tech/social-media/discord-child-safety-social-platform-challenges-rcna89769},
  langid = {english}
}

@misc{graggleAnnouncingDiscordModerator,
  title = {Announcing the {{Discord Moderator Academy Exam}}},
  author = {{Graggle}},
  urldate = {2025-12-03},
  howpublished = {https://discord.com/blog/announcing-the-discord-moderator-academy-exam}
}

@incollection{greenfieldTeensInternetInterpersonal2006,
  title = {Teens on the {{Internet}}: {{Interpersonal Connection}}, {{Identity}}, and {{Information}}},
  shorttitle = {Teens on the {{Internet}}},
  booktitle = {Computers, {{Phones}}, and the {{Internet}}: {{Domesticating Information Technology}}},
  author = {Greenfield, Patricia M. and Gross, Elisheva F. and Subrahmanyam, Kaveri and Suzuki, Lalita K. and Tynes, Brendesha},
  editor = {Kraut, Robert and Brynin, Malcolm and Kiesler, Sara},
  year = 2006,
  month = jul,
  pages = {41},
  publisher = {Oxford University Press},
  doi = {10.1093/acprof:oso/9780195312805.003.0013},
  urldate = {2025-06-12},
  isbn = {978-0-19-531280-5}
}

@misc{gruetWhatRagebaitingWhy,
  title = {What Is Rage-Baiting and Why Is It Profitable?},
  author = {Gruet, Sam and Lawton, Megan},
  urldate = {2025-08-24},
  howpublished = {https://www.bbc.com/news/articles/c4gp555xy5ro}
}

@article{gruzdPrivacyConcernsSelfDisclosure2018,
  title = {Privacy {{Concerns}} and {{Self-Disclosure}} in {{Private}} and {{Public Uses}} of {{Social Media}}},
  author = {Gruzd, Anatoliy and {Hern{\'a}ndez-Garc{\'i}a}, {\'A}ngel},
  year = 2018,
  month = jul,
  journal = {Cyberpsychology, Behavior and Social Networking},
  volume = {21},
  number = {7},
  pages = {418--428},
  issn = {2152-2715},
  doi = {10.1089/cyber.2017.0709},
  urldate = {2025-08-07},
  pmcid = {PMC6719399},
  pmid = {29995525}
}

@article{haimeExploringMentalHealth2025,
  title = {Exploring {{Mental Health Content Moderation}} and {{Well-Being Tools}} on {{Social Media Platforms}}: {{Walkthrough Analysis}}},
  shorttitle = {Exploring {{Mental Health Content Moderation}} and {{Well-Being Tools}} on {{Social Media Platforms}}},
  author = {Haime, Zo{\"e} and Biddle, Lucy},
  year = 2025,
  month = may,
  journal = {JMIR Human Factors},
  volume = {12},
  number = {1},
  pages = {e69817},
  publisher = {JMIR Publications Inc., Toronto, Canada},
  doi = {10.2196/69817},
  urldate = {2025-09-08},
  copyright = {This is an open-access article distributed under the terms of the Creative Commons Attribution License (https://creativecommons.org/licenses/by/4.0/), which permits unrestricted use, distribution, and reproduction in any medium, provided the original work, first published JMIR Human Factors, is properly cited. The complete bibliographic information, a link to the original publication on https://humanfactors.jmir.org/, as well as this copyright and license information must be included.},
  langid = {english}
}

@inproceedings{hashishInvolvingChildrenContent2014,
  title = {Involving Children in Content Control: A Collaborative and Education-Oriented Content Filtering Approach},
  shorttitle = {Involving Children in Content Control},
  booktitle = {Proceedings of the {{SIGCHI Conference}} on {{Human Factors}} in {{Computing Systems}}},
  author = {Hashish, Yasmeen and Bunt, Andrea and Young, James E.},
  year = 2014,
  month = apr,
  series = {{{CHI}} '14},
  pages = {1797--1806},
  publisher = {Association for Computing Machinery},
  address = {New York, NY, USA},
  doi = {10.1145/2556288.2557128},
  urldate = {2025-09-10},
  isbn = {978-1-4503-2473-1}
}

@article{hePlatformGovernanceAlgorithmBased2025,
  title = {Platform {{Governance}} with {{Algorithm-Based Content Moderation}}: {{An Empirical Study}} on {{Reddit}}},
  shorttitle = {Platform {{Governance}} with {{Algorithm-Based Content Moderation}}},
  author = {He, Qinglai and Hong, Yili and Raghu, T. S.},
  year = 2025,
  month = jun,
  journal = {Information Systems Research},
  volume = {36},
  number = {2},
  pages = {1078--1095},
  publisher = {INFORMS},
  issn = {1047-7047},
  doi = {10.1287/isre.2021.0036},
  urldate = {2025-09-02}
}

@article{hernandez-serranoAnalysisDigitalSelfPresentation2022,
  title = {Analysis of {{Digital Self-Presentation Practices}} and {{Profiles}} of {{Spanish Adolescents}} on {{Instagram}} and {{TikTok}}},
  author = {{Hern{\'a}ndez-Serrano}, Mar{\'i}a Jos{\'e} and Jones, Barbara and {Ren{\'e}s-Arellano}, Paula and Ortu{\~n}o, Rosalynn A. Campos},
  year = 2022,
  month = jan,
  journal = {Journal of New Approaches in Educational Research},
  volume = {11},
  number = {1},
  pages = {49--63},
  issn = {2254-7339},
  doi = {10.7821/naer.2022.1.797},
  urldate = {2025-09-12},
  langid = {english}
}

@article{heslepMappingDiscordsDarkside2024,
  title = {Mapping {{Discord}}'s Darkside: {{Distributed}} Hate Networks on {{Disboard}}},
  shorttitle = {Mapping {{Discord}}'s Darkside},
  author = {Heslep, Daniel G and Berge, {\relax PS}},
  year = 2024,
  month = jan,
  journal = {New Media \& Society},
  volume = {26},
  number = {1},
  pages = {534--555},
  publisher = {SAGE Publications},
  issn = {1461-4448},
  doi = {10.1177/14614448211062548},
  urldate = {2025-03-19},
  langid = {english}
}

@inproceedings{huangOpportunitiesTensionsChallenges2024,
  title = {Opportunities, Tensions, and Challenges in Computational Approaches to Addressing Online Harassment},
  booktitle = {Proceedings of the 2024 {{ACM Designing Interactive Systems Conference}}},
  author = {Huang, Evey Jiaxin and Sarma, Abhraneel and Hwang, Sohyeon and Chandrasekharan, Eshwar and Chancellor, Stevie},
  year = 2024,
  month = jul,
  series = {{{DIS}} '24},
  pages = {1483--1498},
  publisher = {Association for Computing Machinery},
  address = {New York, NY, USA},
  doi = {10.1145/3643834.3661623},
  urldate = {2025-02-26},
  isbn = {979-8-4007-0583-0}
}

@inproceedings{jangTeensEngageMore2016,
  title = {Teens {{Engage More}} with {{Fewer Photos}}: {{Temporal}} and {{Comparative Analysis}} on {{Behaviors}} in {{Instagram}}},
  shorttitle = {Teens {{Engage More}} with {{Fewer Photos}}},
  booktitle = {Proceedings of the 27th {{ACM Conference}} on {{Hypertext}} and {{Social Media}}},
  author = {Jang, Jin Yea and Han, Kyungsik and Lee, Dongwon and Jia, Haiyan and Shih, Patrick C.},
  year = 2016,
  month = jul,
  pages = {71--81},
  publisher = {ACM},
  address = {Halifax Nova Scotia Canada},
  doi = {10.1145/2914586.2914602},
  urldate = {2025-08-31},
  isbn = {978-1-4503-4247-6},
  langid = {english}
}

@misc{jeffreygottfriedTeensVideoGames,
  title = {Teens and {{Video Games Today}} \textbar{} {{Pew Research Center}}},
  author = {{Jeffrey Gottfried} and {Olivia Sidoti}},
  urldate = {2025-11-14},
  howpublished = {https://www.pewresearch.org/internet/2024/05/09/teens-and-video-games-today/}
}

@article{jhaverDoesTransparencyModeration2019,
  title = {Does {{Transparency}} in {{Moderation Really Matter}}? {{User Behavior After Content Removal Explanations}} on {{Reddit}}},
  shorttitle = {Does {{Transparency}} in {{Moderation Really Matter}}?},
  author = {Jhaver, Shagun and Bruckman, Amy and Gilbert, Eric},
  year = 2019,
  month = nov,
  journal = {Proc. ACM Hum.-Comput. Interact.},
  volume = {3},
  number = {CSCW},
  pages = {150:1--150:27},
  doi = {10.1145/3359252},
  urldate = {2025-02-26}
}

@article{jhaverPersonalizingContentModeration2023,
  title = {Personalizing {{Content Moderation}} on {{Social Media}}: {{User Perspectives}} on {{Moderation Choices}}, {{Interface Design}}, and {{Labor}}},
  shorttitle = {Personalizing {{Content Moderation}} on {{Social Media}}},
  author = {Jhaver, Shagun and Zhang, Alice Qian and Chen, Quan Ze and Natarajan, Nikhila and Wang, Ruotong and Zhang, Amy X.},
  year = 2023,
  month = sep,
  journal = {Proceedings of the ACM on Human-Computer Interaction},
  volume = {7},
  number = {CSCW2},
  pages = {1--33},
  issn = {2573-0142},
  doi = {10.1145/3610080},
  urldate = {2024-05-11},
  langid = {english}
}

@article{jiangModerationChallengesVoicebased2019,
  title = {Moderation {{Challenges}} in {{Voice-based Online Communities}} on {{Discord}}},
  author = {Jiang, Jialun Aaron and Kiene, Charles and Middler, Skyler and Brubaker, Jed R. and Fiesler, Casey},
  year = 2019,
  month = nov,
  journal = {Proc. ACM Hum.-Comput. Interact.},
  volume = {3},
  number = {CSCW},
  pages = {55:1--55:23},
  doi = {10.1145/3359157},
  urldate = {2025-03-19}
}

@article{johnsonEmbracingDiscordRhetorical2022,
  title = {Embracing Discord? {{The}} Rhetorical Consequences of Gaming Platforms as Classrooms},
  shorttitle = {Embracing Discord?},
  author = {Johnson, Emily K. and Salter, Anastasia},
  year = 2022,
  month = sep,
  journal = {Computers and Composition},
  volume = {65},
  pages = {102729},
  issn = {8755-4615},
  doi = {10.1016/j.compcom.2022.102729},
  urldate = {2025-06-07}
}

@inproceedings{joshiContextualizingInternetMemes2024,
  title = {Contextualizing {{Internet Memes Across Social Media Platforms}}},
  booktitle = {Companion {{Proceedings}} of the {{ACM Web Conference}} 2024},
  author = {Joshi, Saurav and Ilievski, Filip and Luceri, Luca},
  year = 2024,
  month = may,
  series = {{{WWW}} '24},
  pages = {1831--1840},
  publisher = {Association for Computing Machinery},
  address = {New York, NY, USA},
  doi = {10.1145/3589335.3651970},
  urldate = {2025-11-16},
  isbn = {979-8-4007-0172-6}
}

@inproceedings{kawaseInternetFraudCase2019,
  title = {Internet {{Fraud}}: {{The Case}} of {{Account Takeover}} in {{Online Marketplace}}},
  shorttitle = {Internet {{Fraud}}},
  booktitle = {Proceedings of the 30th {{ACM Conference}} on {{Hypertext}} and {{Social Media}}},
  author = {Kawase, Ricardo and Diana, Francesca and Czeladka, Mateusz and Sch{\"u}ler, Markus and Faust, Manuela},
  year = 2019,
  month = sep,
  series = {{{HT}} '19},
  pages = {181--190},
  publisher = {Association for Computing Machinery},
  address = {New York, NY, USA},
  doi = {10.1145/3342220.3343651},
  urldate = {2025-09-09},
  isbn = {978-1-4503-6885-8}
}

@misc{kellyDarkSideDiscord,
  title = {The Dark Side of {{Discord}} for Teens \textbar{} {{CNN Business}}},
  author = {Kelly, Samantha Murphy},
  urldate = {2025-03-19},
  howpublished = {https://www.cnn.com/2022/03/22/tech/discord-teens/index.html}
}

@article{kennedyShiftOfflineOnline2016,
  title = {A Shift from Offline to Online: {{Adolescence}}, the Internet and Social Participation},
  shorttitle = {A Shift from Offline to Online},
  author = {Kennedy, Jessica and Lynch, Helen},
  year = 2016,
  month = apr,
  journal = {Journal of Occupational Science},
  volume = {23},
  number = {2},
  pages = {156--167},
  issn = {1442-7591, 2158-1576},
  doi = {10.1080/14427591.2015.1117523},
  urldate = {2025-09-12},
  langid = {english}
}

@article{kitkowskaEnhancingPrivacyVisual2020,
  title = {Enhancing {{Privacy}} through the {{Visual Design}} of {{Privacy Notices}}: {{Exploring}} the {{Interplay}} of {{Curiosity}}, {{Control}} and {{Affect}}},
  author = {Kitkowska, Agnieszka and University, Karlstad and Warner, Mark and University, Northumbria and Shulman, Yefim and University, Tel Aviv and W{\"a}stlund, Erik and University, Karlstad and Martucci, Leonardo A and University, Karlstad},
  year = 2020,
  journal = {USENIX Symposium on Usable Privacy and Security},
  langid = {english}
}

@inproceedings{kouHarmfulDesignMetaverse2023,
  title = {Harmful {{Design}} in the {{Metaverse}} and {{How}} to {{Mitigate}} It: {{A Case Study}} of {{User-Generated Virtual Worlds}} on {{Roblox}}},
  shorttitle = {Harmful {{Design}} in the {{Metaverse}} and {{How}} to {{Mitigate}} It},
  booktitle = {Proceedings of the 2023 {{ACM Designing Interactive Systems Conference}}},
  author = {Kou, Yubo and Gui, Xinning},
  year = 2023,
  month = jul,
  pages = {175--188},
  publisher = {ACM},
  address = {Pittsburgh PA USA},
  doi = {10.1145/3563657.3595960},
  urldate = {2025-09-09},
  isbn = {978-1-4503-9893-0},
  langid = {english}
}

@article{koungGenderedToxicityCompetitive2025,
  title = {Gendered {{Toxicity}} in {{Competitive Gaming}}: {{Women}}'s {{Perceptions}} and {{Responses}}},
  shorttitle = {Gendered {{Toxicity}} in {{Competitive Gaming}}},
  author = {Koung, Elena and Zhang, Zinan and Gui, Xinning and Kou, Yubo},
  year = 2025,
  month = oct,
  journal = {Proceedings of the ACM on Human-Computer Interaction},
  volume = {9},
  number = {6},
  pages = {415--446},
  issn = {2573-0142},
  doi = {10.1145/3748610},
  urldate = {2025-10-30},
  langid = {english}
}

@inproceedings{kouSystemMadeInherently2025,
  title = {``{{The System}} Is {{Made}} to {{Inherently Push Child Gambling}} in My {{Opinion}}'': {{Child Safety}}, {{Monetization}}, and {{Moderation}} on {{Roblox}}},
  shorttitle = {``{{The System}} Is {{Made}} to {{Inherently Push Child Gambling}} in My {{Opinion}}''},
  booktitle = {Proceedings of the 2025 {{CHI Conference}} on {{Human Factors}} in {{Computing Systems}}},
  author = {Kou, Yubo and Hernandez, Rie Helene (Lindy) and Gui, Xinning},
  year = 2025,
  month = apr,
  series = {{{CHI}} '25},
  pages = {1--18},
  publisher = {Association for Computing Machinery},
  address = {New York, NY, USA},
  doi = {10.1145/3706598.3713170},
  urldate = {2025-11-11},
  isbn = {979-8-4007-1394-1}
}

@inproceedings{krautInformationDevelopingRelationship2008,
  title = {Beyond {{Information}}: {{Developing}} the {{Relationship}} between the {{Individual}} and the {{Group}} in {{Online Communities}}},
  shorttitle = {Beyond {{Information}}},
  author = {Kraut, R. and Wang, Xiaoqing and Butler, B. and Joyce, Elisabeth and Burke, Moira},
  year = 2008,
  urldate = {2025-08-28}
}

@misc{kumarDiscordUsersMarket2025,
  title = {Discord {{Users}} \& {{Market Share Statistics}} 2025 ({{Data By Country}})},
  author = {Kumar, Naveen},
  year = 2025,
  month = aug,
  journal = {DemandSage},
  urldate = {2025-09-12},
  langid = {american}
}

@article{kumarNoTellingPasscodes2017,
  title = {'{{No Telling Passcodes Out Because They}}'re {{Private}}': {{Understanding Children}}'s {{Mental Models}} of {{Privacy}} and {{Security Online}}},
  shorttitle = {'{{No Telling Passcodes Out Because They}}'re {{Private}}'},
  author = {Kumar, Priya and Naik, Shalmali Milind and Devkar, Utkarsha Ramesh and Chetty, Marshini and Clegg, Tamara L. and Vitak, Jessica},
  year = 2017,
  month = dec,
  journal = {Proceedings of the ACM on Human-Computer Interaction},
  volume = {1},
  number = {CSCW},
  pages = {1--21},
  issn = {2573-0142},
  doi = {10.1145/3134699},
  urldate = {2025-08-23},
  langid = {english}
}

@article{kwonEffectsEscapeSelf2011,
  title = {The {{Effects}} of {{Escape}} from {{Self}} and {{Interpersonal Relationship}} on the {{Pathological Use}} of {{Internet Games}}},
  author = {Kwon, Jung-Hye and Chung, Chung-Suk and Lee, Jung},
  year = 2011,
  month = feb,
  journal = {Community Mental Health Journal},
  volume = {47},
  number = {1},
  pages = {113--121},
  issn = {1573-2789},
  doi = {10.1007/s10597-009-9236-1},
  urldate = {2025-09-11},
  langid = {english}
}

@article{laiOnlineStrangersOffline2020,
  title = {From Online Strangers to Offline Friends: A Qualitative Study of Video Game Players in {{Hong Kong}}},
  shorttitle = {From Online Strangers to Offline Friends},
  author = {Lai, Gina and Fung, Ka Yi},
  year = 2020,
  month = may,
  journal = {Media, Culture \& Society},
  volume = {42},
  number = {4},
  pages = {483--501},
  publisher = {SAGE Publications Ltd},
  issn = {0163-4437},
  doi = {10.1177/0163443719853505},
  urldate = {2025-09-05},
  langid = {english}
}

@misc{larsonMissingFremontTeen,
  title = {Missing {{Fremont}} Teen {{Katie Hong}}: {{What}} We Know \textbar{} {{KRON4}}},
  author = {Larson, Amy and Tran, Will},
  urldate = {2025-08-29},
  howpublished = {https://www.kron4.com/news/bay-area/missing-fremont-teen-katie-hong-what-we-know/}
}

@article{leeMappingCommunityAppeals2025,
  title = {Mapping {{Community Appeals Systems}}: {{Lessons}} for {{Community-led Moderation}} in {{Multi-Level Governance}}},
  author = {Lee, Juhoon and Doan, Bich Ngoc and Jee, Jonghyun and Seering, Joseph},
  year = 2025,
  journal = {Proc. ACM Hum.-Comput. Interact.},
  langid = {english}
}

@inproceedings{liEthicalIssuesVideo2025,
  title = {Ethical {{Issues}} in {{Video Games}}: {{Insights From Reddit Discussions}}},
  shorttitle = {Ethical {{Issues}} in {{Video Games}}},
  booktitle = {2025 {{IEEE}}/{{ACM}} 47th {{International Conference}} on {{Software Engineering}}: {{Software Engineering}} in {{Society}} ({{ICSE-SEIS}})},
  author = {Li, Yeqian and Aslam, Kousar},
  year = 2025,
  month = apr,
  pages = {185--196},
  issn = {2832-7616},
  doi = {10.1109/ICSE-SEIS66351.2025.00024},
  urldate = {2025-09-10}
}

@article{livingstone4CsClassifyingOnline2021,
  title = {The {{4Cs}}: {{Classifying Online Risk}} to {{Children}}},
  shorttitle = {The {{4Cs}}},
  author = {Livingstone, Sonia and Stoilova, Mariya},
  editor = {{Leibniz-Institut F\"ur Medienforschung \textbar{} Hans-Bredow-Institut (HBI)}},
  year = 2021,
  journal = {CO:RE Short Report Series on Key Topics},
  pages = {14 S.},
  publisher = {SSOAR -   GESIS Leibniz Institute for the Social Sciences},
  doi = {10.21241/SSOAR.71817},
  urldate = {2025-08-31},
  copyright = {Creative Commons Attribution 4.0 International},
  langid = {english}
}

@article{livingstoneCanPlatformLiteracy2025,
  title = {Can Platform Literacy Protect Vulnerable Young People against the Risky Affordances of Social Media Platforms?},
  author = {Livingstone, Sonia and Jessen, Reidar Schei and Stoilova, Mariya and St{\"a}nicke, Line Indrevoll and Graham, Richard and Staksrud, Elisabeth and Jensen, Tine},
  year = 2025,
  month = jun,
  journal = {Information, Communication \& Society},
  pages = {1--18},
  publisher = {Informa UK Limited},
  issn = {1369-118X, 1468-4462},
  doi = {10.1080/1369118x.2025.2518254},
  urldate = {2025-07-09},
  copyright = {http://creativecommons.org/licenses/by/4.0/},
  langid = {english}
}

@article{livingstoneMaximizingOpportunitiesMinimizing2017,
  title = {Maximizing {{Opportunities}} and {{Minimizing Risks}} for {{Children Online}}: {{The Role}} of {{Digital Skills}} in {{Emerging Strategies}} of {{Parental Mediation}}: {{Maximizing Opportunities}} and {{Minimizing Risks}}},
  shorttitle = {Maximizing {{Opportunities}} and {{Minimizing Risks}} for {{Children Online}}},
  author = {Livingstone, Sonia and {\'O}lafsson, Kjartan and Helsper, Ellen J. and {Lupi{\'a}{\~n}ez-Villanueva}, Francisco and Veltri, Giuseppe A. and Folkvord, Frans},
  year = 2017,
  month = feb,
  journal = {Journal of Communication},
  volume = {67},
  number = {1},
  pages = {82--105},
  issn = {00219916},
  doi = {10.1111/jcom.12277},
  urldate = {2024-09-02},
  copyright = {http://doi.wiley.com/10.1002/tdm\_license\_1},
  langid = {english}
}

@article{livingstoneTakingRisksWhen2007,
  title = {Taking Risks When Communicating on the {{Internet}}: The Role of Offline Social-Psychological Factors in Young People's Vulnerability to Online Risks},
  shorttitle = {Taking Risks When Communicating on the {{Internet}}},
  author = {Livingstone, Sonia and Helsper, Ellen J.},
  year = 2007,
  month = oct,
  journal = {Information, Communication \& Society},
  volume = {10},
  number = {5},
  pages = {619--644},
  issn = {1369-118X, 1468-4462},
  doi = {10.1080/13691180701657998},
  urldate = {2025-09-01},
  langid = {english}
}

@article{livingstoneWhenAdolescentsReceive2014,
  title = {When Adolescents Receive Sexual Messages on the Internet: {{Explaining}} Experiences of Risk and Harm},
  shorttitle = {When Adolescents Receive Sexual Messages on the Internet},
  author = {Livingstone, Sonia and G{\"o}rzig, Anke},
  year = 2014,
  month = apr,
  journal = {Computers in Human Behavior},
  volume = {33},
  pages = {8--15},
  issn = {0747-5632},
  doi = {10.1016/j.chb.2013.12.021},
  urldate = {2025-09-08}
}

@techreport{luriaYoungUsersStrategies2023,
  title = {Young {{Users}}' {{Strategies}} for {{Handling Unwanted Online Messages}}},
  author = {Luria, Michael},
  year = 2023,
  month = nov,
  institution = {CDT Research}
}

@article{maHowAdvertiserfriendlyMy2021,
  title = {"{{How}} Advertiser-Friendly Is My Video?": {{YouTuber}}'s {{Socioeconomic Interactions}} with {{Algorithmic Content Moderation}}},
  shorttitle = {"{{How}} Advertiser-Friendly Is My Video?},
  author = {Ma, Renkai and Kou, Yubo},
  year = 2021,
  month = oct,
  journal = {Proc. ACM Hum.-Comput. Interact.},
  volume = {5},
  number = {CSCW2},
  pages = {429:1--429:25},
  doi = {10.1145/3479573},
  urldate = {2025-10-11}
}

@inproceedings{maLabelingDarkExploring2024,
  title = {Labeling in the {{Dark}}: {{Exploring Content Creators}}' and {{Consumers}}' {{Experiences}} with {{Content Classification}} for {{Child Safety}} on {{YouTube}}},
  shorttitle = {Labeling in the {{Dark}}},
  booktitle = {Designing {{Interactive Systems Conference}}},
  author = {Ma, Renkai and Zhang, Zinan and Gui, Xinning and Kou, Yubo},
  year = 2024,
  month = jul,
  pages = {1518--1532},
  publisher = {ACM},
  address = {Copenhagen Denmark},
  doi = {10.1145/3643834.3661565},
  urldate = {2025-10-25},
  isbn = {979-8-4007-0583-0},
  langid = {english}
}

@inproceedings{maqsoodTheyThinkIts2021,
  title = {``{{They}} Think It's Totally Fine to Talk to Somebody on the Internet They Don't Know'': {{Teachers}}' Perceptions and Mitigation Strategies of Tweens' Online Risks},
  shorttitle = {``{{They}} Think It's Totally Fine to Talk to Somebody on the Internet They Don't Know''},
  booktitle = {Proceedings of the 2021 {{CHI Conference}} on {{Human Factors}} in {{Computing Systems}}},
  author = {Maqsood, Sana and Chiasson, Sonia},
  year = 2021,
  month = may,
  pages = {1--17},
  publisher = {ACM},
  address = {Yokohama Japan},
  doi = {10.1145/3411764.3445224},
  urldate = {2024-10-15},
  isbn = {978-1-4503-8096-6},
  langid = {english}
}

@article{marwickNetworkedPrivacyHow2014,
  title = {Networked Privacy: {{How}} Teenagers Negotiate Context in Social Media},
  shorttitle = {Networked Privacy},
  author = {Marwick, Alice E and {boyd}, danah},
  year = 2014,
  month = nov,
  journal = {New Media \& Society},
  volume = {16},
  number = {7},
  pages = {1051--1067},
  publisher = {SAGE Publications},
  issn = {1461-4448},
  doi = {10.1177/1461444814543995},
  urldate = {2025-09-08}
}

@article{mcdonaldReliabilityInterraterReliability2019,
  title = {Reliability and {{Inter-rater Reliability}} in {{Qualitative Research}}: {{Norms}} and {{Guidelines}} for {{CSCW}} and {{HCI Practice}}},
  shorttitle = {Reliability and {{Inter-rater Reliability}} in {{Qualitative Research}}},
  author = {McDonald, Nora and Schoenebeck, Sarita and Forte, Andrea},
  year = 2019,
  month = nov,
  journal = {Proceedings of the ACM on Human-Computer Interaction},
  volume = {3},
  number = {CSCW},
  pages = {1--23},
  issn = {2573-0142},
  doi = {10.1145/3359174},
  urldate = {2025-01-23},
  langid = {english}
}

@article{mcfaddenDiscordAppExposes2025,
  title = {Discord {{App Exposes Children}} to {{Abuse}} and {{Graphic Content}}, {{Lawsuit Says}}},
  author = {McFadden, Alyce},
  year = 2025,
  month = apr,
  journal = {The New York Times},
  issn = {0362-4331},
  urldate = {2025-08-29},
  chapter = {New York},
  langid = {american}
}

@article{mchughMostTeensBounce2017,
  title = {Most {{Teens Bounce Back}}: {{Using Diary Methods}} to {{Examine How Quickly Teens Recover}} from {{Episodic Online Risk Exposure}}},
  shorttitle = {Most {{Teens Bounce Back}}},
  author = {McHugh, Bridget Christine and Wisniewski, Pamela J. and Rosson, Mary Beth and Xu, Heng and Carroll, John M.},
  year = 2017,
  month = dec,
  journal = {Proc. ACM Hum.-Comput. Interact.},
  volume = {1},
  number = {CSCW},
  pages = {76:1--76:19},
  doi = {10.1145/3134711},
  urldate = {2025-11-21}
}

@misc{MessageRequests2025,
  title = {Message {{Requests}}},
  author = {, Kynthia},
  year = 2025,
  month = jul,
  journal = {Discord},
  urldate = {2025-08-24},
  howpublished = {https://support.discord.com/hc/en-us/articles/7924992471191-Message-Requests},
  langid = {american}
}

@inproceedings{nairSafeguardingTomorrowFortifying2024,
  title = {Safeguarding {{Tomorrow}} - {{Fortifying Child Safety}} in {{Digital Landscape}}},
  booktitle = {2024 {{International Conference}} on {{Computing}}, {{Sciences}} and {{Communications}} ({{ICCSC}})},
  author = {Nair, Sreejith Sreekandan and Lakshmikanthan, Govindarajan and Kendyala, Srinivasulu Harshavardhan and Dhaduvai, Vivek Sheetal},
  year = 2024,
  month = oct,
  pages = {1--6},
  doi = {10.1109/ICCSC62048.2024.10830389},
  urldate = {2025-09-09}
}

@article{noggleManipulationSalienceNudges2018,
  title = {Manipulation, Salience, and Nudges},
  author = {Noggle, Robert},
  year = 2018,
  month = mar,
  journal = {Bioethics},
  volume = {32},
  number = {3},
  pages = {164--170},
  issn = {1467-8519},
  doi = {10.1111/bioe.12421},
  langid = {english},
  pmid = {29283190}
}

@article{obajemuEnforcingGoodDigital2024,
  title = {Towards {{Enforcing Good Digital Citizenship}}: {{Identifying Opportunities}} for {{Adolescent Online Safety Nudges}}},
  shorttitle = {Towards {{Enforcing Good Digital Citizenship}}},
  author = {Obajemu, Oluwatomisin and Agha, Zainab and Chowdhury, Farzana A. and Wisniewski, Pamela J.},
  year = 2024,
  month = apr,
  journal = {Proc. ACM Hum.-Comput. Interact.},
  volume = {8},
  number = {CSCW1},
  pages = {136:1--136:37},
  doi = {10.1145/3637413},
  urldate = {2025-08-23}
}

@article{obokataAssociationsOnlineCommunication2023,
  title = {Associations between Online Communication with Strangers and Mild Delinquency in Junior High School Students},
  author = {Obokata, Akiko and Pauen, Sabina},
  year = 2023,
  month = jul,
  journal = {Current Psychology},
  volume = {42},
  number = {19},
  pages = {16533--16547},
  issn = {1936-4733},
  doi = {10.1007/s12144-022-03317-2},
  urldate = {2025-07-28},
  langid = {english}
}

@article{okeeffeImpactSocialMedia2011,
  title = {The Impact of Social Media on Children, Adolescents, and Families},
  author = {O'Keeffe, Gwenn Schurgin and {Clarke-Pearson}, Kathleen and {Council on Communications and Media}},
  year = 2011,
  month = apr,
  journal = {Pediatrics},
  volume = {127},
  number = {4},
  pages = {800--804},
  issn = {1098-4275},
  doi = {10.1542/peds.2011-0054},
  langid = {english},
  pmid = {21444588}
}

@article{padilla-walkerProtectiveRoleParental2018,
  title = {The {{Protective Role}} of {{Parental Media Monitoring Style}} from {{Early}} to {{Late Adolescence}}},
  author = {{Padilla-Walker}, Laura M. and Coyne, Sarah M. and Kroff, Savannah L. and {Memmott-Elison}, Madison K.},
  year = 2018,
  month = feb,
  journal = {Journal of Youth and Adolescence},
  volume = {47},
  number = {2},
  pages = {445--459},
  issn = {1573-6601},
  doi = {10.1007/s10964-017-0722-4},
  langid = {english},
  pmid = {28791572}
}

@inproceedings{parkTeensPrivacyAlgorithms2025,
  title = {Teens, {{Privacy}}, and {{Algorithms}}: {{Navigating}} and {{Co-Designing Solutions}} for {{Interpersonal Boundary Management}} on {{Social Media}}},
  shorttitle = {Teens, {{Privacy}}, and {{Algorithms}}},
  booktitle = {Proceedings of the 24th {{Interaction Design}} and {{Children}}},
  author = {Park, Jinkyung Katie and Ma, Renkai and Ali, Naima Samreen and Baptiste, Naulsberry Jean and Agha, Zainab and Wisniewski, Pamela J.},
  year = 2025,
  month = jun,
  series = {{{IDC}} '25},
  pages = {589--607},
  publisher = {Association for Computing Machinery},
  address = {New York, NY, USA},
  doi = {10.1145/3713043.3728840},
  urldate = {2025-06-28},
  isbn = {979-8-4007-1473-3}
}

@book{pattonQualitativeResearchEvaluation2014,
  title = {Qualitative {{Research}} \& {{Evaluation Methods}}: {{Integrating Theory}} and {{Practice}}},
  shorttitle = {Qualitative {{Research}} \& {{Evaluation Methods}}},
  author = {Patton, Michael Quinn},
  year = 2014,
  month = oct,
  publisher = {SAGE Publications},
  isbn = {978-1-4833-1481-5},
  langid = {english}
}

@article{perez-torresSocialMediaDigital2024,
  title = {Social Media: A Digital Social Mirror for Identity Development during Adolescence},
  shorttitle = {Social Media},
  author = {{P{\'e}rez-Torres}, Vanesa},
  year = 2024,
  month = jul,
  journal = {Current Psychology},
  volume = {43},
  number = {26},
  pages = {22170--22180},
  issn = {1936-4733},
  doi = {10.1007/s12144-024-05980-z},
  urldate = {2025-09-12},
  langid = {english}
}

@article{peterCharacteristicsMotivesAdolescents2006,
  title = {Characteristics and {{Motives}} of {{Adolescents Talking}} with {{Strangers}} on the {{Internet}}},
  author = {Peter, Jochen and Valkenburg, Patti M. and Schouten, Alexander P.},
  year = 2006,
  month = oct,
  journal = {CyberPsychology \& Behavior},
  volume = {9},
  number = {5},
  pages = {526--530},
  publisher = {Mary Ann Liebert, Inc., publishers},
  issn = {1094-9313},
  doi = {10.1089/cpb.2006.9.526},
  urldate = {2025-07-28}
}

@inproceedings{pinterAdolescentOnlineSafety2017,
  title = {Adolescent {{Online Safety}}: {{Moving Beyond Formative Evaluations}} to {{Designing Solutions}} for the {{Future}}},
  shorttitle = {Adolescent {{Online Safety}}},
  booktitle = {Proceedings of the 2017 {{Conference}} on {{Interaction Design}} and {{Children}}},
  author = {Pinter, Anthony T. and Wisniewski, Pamela J. and Xu, Heng and Rosson, Mary Beth and Caroll, Jack M.},
  year = 2017,
  month = jun,
  series = {{{IDC}} '17},
  pages = {352--357},
  publisher = {Association for Computing Machinery},
  address = {New York, NY, USA},
  doi = {10.1145/3078072.3079722},
  urldate = {2025-06-06},
  isbn = {978-1-4503-4921-5}
}

@inproceedings{pirExplorationHowChildren2024,
  title = {An {{Exploration}} of {{How Children Can Be Proactive}} for {{Their Own Digital Privacy}} and {{Security}} in the {{Perspective}} of {{North-Eastern Bangladesh}}},
  booktitle = {Human-{{Centric Smart Computing}}},
  author = {Pir, Rumel M. S. Rahman and Rabbi, Md. Forhad and Islam, M. Jahirul},
  editor = {Bhattacharyya, Siddhartha and Banerjee, Jyoti Sekhar and K{\"o}ppen, Mario},
  year = 2024,
  pages = {153--165},
  publisher = {Springer Nature},
  address = {Singapore},
  doi = {10.1007/978-981-99-7711-6_13},
  isbn = {978-981-99-7711-6},
  langid = {english}
}

@article{prinsterCommunityArchetypesEmpirical2024,
  title = {Community {{Archetypes}}: {{An Empirical Framework}} for {{Guiding Research Methodologies}} to {{Reflect User Experiences}} of {{Sense}} of {{Virtual Community}} on {{Reddit}}},
  shorttitle = {Community {{Archetypes}}},
  author = {Prinster, Gale H. and Smith, C. Estelle and Tan, Chenhao and Keegan, Brian C.},
  year = 2024,
  month = apr,
  journal = {Proceedings of the ACM on Human-Computer Interaction},
  volume = {8},
  number = {CSCW1},
  pages = {1--33},
  issn = {2573-0142},
  doi = {10.1145/3637310},
  urldate = {2025-08-28},
  langid = {english}
}

@inproceedings{raziLetsTalkSext2020,
  title = {Let's {{Talk}} about {{Sext}}: {{How Adolescents Seek Support}} and {{Advice}} about {{Their Online Sexual Experiences}}},
  shorttitle = {Let's {{Talk}} about {{Sext}}},
  booktitle = {Proceedings of the 2020 {{CHI Conference}} on {{Human Factors}} in {{Computing Systems}}},
  author = {Razi, Afsaneh and {Badillo-Urquiola}, Karla and Wisniewski, Pamela J.},
  year = 2020,
  month = apr,
  series = {{{CHI}} '20},
  pages = {1--13},
  publisher = {Association for Computing Machinery},
  address = {New York, NY, USA},
  doi = {10.1145/3313831.3376400},
  urldate = {2025-05-16},
  isbn = {978-1-4503-6708-0}
}

@article{redmilesJustWantFeel2019,
  title = {``{{I Just Want}} to {{Feel Safe}}'': {{A Diary Study}} of {{Safety Perceptions}} on {{Social Media}}},
  shorttitle = {``{{I Just Want}} to {{Feel Safe}}''},
  author = {Redmiles, Elissa M. and Bodford, Jessica and Blackwell, Lindsay},
  year = 2019,
  month = jul,
  journal = {Proceedings of the International AAAI Conference on Web and Social Media},
  volume = {13},
  pages = {405--416},
  issn = {2334-0770},
  doi = {10.1609/icwsm.v13i01.3356},
  urldate = {2025-08-29},
  copyright = {Copyright (c) 2019 Association for the Advancement of Artificial Intelligence},
  langid = {english}
}

@inproceedings{reisExplainableMachineLearning2019,
  title = {Explainable {{Machine Learning}} for {{Fake News Detection}}},
  booktitle = {Proceedings of the 10th {{ACM Conference}} on {{Web Science}}},
  author = {Reis, Julio C. S. and Correia, Andr{\'e} and Murai, Fabr{\'i}cio and Veloso, Adriano and Benevenuto, Fabr{\'i}cio},
  year = 2019,
  month = jun,
  series = {{{WebSci}} '19},
  pages = {17--26},
  publisher = {Association for Computing Machinery},
  address = {New York, NY, USA},
  doi = {10.1145/3292522.3326027},
  urldate = {2025-09-11},
  isbn = {978-1-4503-6202-3}
}

@article{renesesHeFlatteredMe2024,
  title = {``{{He}} Flattered Me''. {{A}} Comprehensive Look into Online Grooming Risk Factors: {{Merging}} Voices of Victims, Offenders and Experts through in-Depth Interviews},
  shorttitle = {``{{He}} Flattered Me''. {{A}} Comprehensive Look into Online Grooming Risk Factors},
  author = {Reneses, Mar{\'i}a and {Riberas-Guti{\'e}rrez}, Mar{\'i}a and {Bueno-Guerra}, Nereida},
  year = 2024,
  month = sep,
  journal = {Cyberpsychology: Journal of Psychosocial Research on Cyberspace},
  volume = {18},
  number = {4},
  issn = {1802-7962},
  doi = {10.5817/CP2024-4-3},
  urldate = {2025-04-10},
  copyright = {Copyright \copyright{} 2024 Mar\'ia Reneses, Mar\'ia Riberas-Guti\'errez, Nereida Bueno-Guerra},
  langid = {english}
}

@article{robinsonGovernanceDiscordPlatform2023,
  title = {Governance on, with, behind, and beyond the {{Discord}} Platform: A Study of Platform Practices in an Informal Learning Context},
  shorttitle = {Governance on, with, behind, and beyond the {{Discord}} Platform},
  author = {Robinson, Bradley},
  year = 2023,
  month = jan,
  journal = {Learning, Media and Technology},
  volume = {48},
  number = {1},
  pages = {81--94},
  publisher = {Routledge},
  issn = {1743-9884},
  doi = {10.1080/17439884.2022.2052312},
  urldate = {2025-06-07}
}

@article{rocheleauPrivacySafetySocial2022,
  title = {Privacy and {{Safety}} on {{Social Networking Sites}}: {{Autistic}} and {{Non-Autistic Teenagers}}' {{Attitudes}} and {{Behaviors}}},
  shorttitle = {Privacy and {{Safety}} on {{Social Networking Sites}}},
  author = {Rocheleau, Jessica N. and Chiasson, Sonia},
  year = 2022,
  month = jan,
  journal = {ACM Trans. Comput.-Hum. Interact.},
  volume = {29},
  number = {1},
  pages = {1:1--1:39},
  issn = {1073-0516},
  doi = {10.1145/3469859},
  urldate = {2025-08-23}
}

@article{roehlImposterParticipantsOvercoming2022,
  title = {Imposter {{Participants}}: {{Overcoming Methodological Challenges Related}} to {{Balancing Participant Privacy}} with {{Data Quality When Using Online Recruitment}} and {{Data Collection}}},
  shorttitle = {Imposter {{Participants}}},
  author = {Roehl, Jacqueline and Harland, Darci},
  year = 2022,
  month = nov,
  journal = {The Qualitative Report},
  issn = {21603715},
  doi = {10.46743/2160-3715/2022.5475},
  urldate = {2025-09-03}
}

@misc{SafetyDesignESafety,
  title = {Safety by {{Design}} \textbar{} {{eSafety Commissioner}}},
  urldate = {2025-11-29},
  howpublished = {https://www.esafety.gov.au/industry/safety-by-design},
  langid = {english}
}

@article{sampasa-kanyingaSocialNetworkingSites2015,
  title = {Social Networking Sites and Mental Health Problems in Adolescents: {{The}} Mediating Role of Cyberbullying Victimization},
  shorttitle = {Social Networking Sites and Mental Health Problems in Adolescents},
  author = {{Sampasa-Kanyinga}, H. and Hamilton, H. A.},
  year = 2015,
  month = nov,
  journal = {European Psychiatry},
  volume = {30},
  number = {8},
  pages = {1021--1027},
  issn = {0924-9338, 1778-3585},
  doi = {10.1016/j.eurpsy.2015.09.011},
  urldate = {2025-11-29},
  langid = {english}
}

@article{sanchez-zasOntologybasedApproachRealtime2023,
  title = {Ontology-Based Approach to Real-Time Risk Management and Cyber-Situational Awareness},
  author = {{S{\'a}nchez-Zas}, Carmen and Villagr{\'a}, V{\'i}ctor A. and {Vega-Barbas}, Mario and {Larriva-Novo}, Xavier and Moreno, Jos{\'e} Ignacio and Berrocal, Julio},
  year = 2023,
  month = apr,
  journal = {Future Generation Computer Systems},
  volume = {141},
  pages = {462--472},
  issn = {0167-739X},
  doi = {10.1016/j.future.2022.12.006},
  urldate = {2025-09-05}
}

@article{sasInformingChildrenPrivacy2023,
  title = {Informing {{Children}} about {{Privacy}}: {{A Review}} and {{Assessment}} of {{Age-Appropriate Information Designs}} in {{Kids-Oriented F2P Video Games}}},
  shorttitle = {Informing {{Children}} about {{Privacy}}},
  author = {Sas, Martin and Denoo, Maarten and M{\"u}hlberg, Jan Tobias},
  year = 2023,
  month = oct,
  journal = {Proc. ACM Hum.-Comput. Interact.},
  volume = {7},
  number = {CHI PLAY},
  pages = {390:425--390:463},
  doi = {10.1145/3611036},
  urldate = {2025-08-23}
}

@article{savoiaAdolescentsExposureOnline2021,
  title = {Adolescents' {{Exposure}} to {{Online Risks}}: {{Gender Disparities}} and {{Vulnerabilities Related}} to {{Online Behaviors}}},
  shorttitle = {Adolescents' {{Exposure}} to {{Online Risks}}},
  author = {Savoia, Elena and Harriman, Nigel Walsh and Su, Max and Cote, Tyler and Shortland, Neil},
  year = 2021,
  month = jan,
  journal = {International Journal of Environmental Research and Public Health},
  volume = {18},
  number = {11},
  pages = {5786},
  publisher = {Multidisciplinary Digital Publishing Institute},
  issn = {1660-4601},
  doi = {10.3390/ijerph18115786},
  urldate = {2025-09-08},
  copyright = {http://creativecommons.org/licenses/by/3.0/},
  langid = {english}
}

@article{schlosserSelfdisclosureSelfpresentationSocial2020,
  title = {Self-Disclosure versus Self-Presentation on Social Media},
  author = {Schlosser, Ann E},
  year = 2020,
  month = feb,
  journal = {Current Opinion in Psychology},
  series = {Privacy and {{Disclosure}}, {{Online}} and in {{Social Interactions}}},
  volume = {31},
  pages = {1--6},
  issn = {2352-250X},
  doi = {10.1016/j.copsyc.2019.06.025},
  urldate = {2025-11-16}
}

@inproceedings{seeringDyadicInteractionsConsidering2019,
  title = {Beyond {{Dyadic Interactions}}: {{Considering Chatbots}} as {{Community Members}}},
  shorttitle = {Beyond {{Dyadic Interactions}}},
  booktitle = {Proceedings of the 2019 {{CHI Conference}} on {{Human Factors}} in {{Computing Systems}}},
  author = {Seering, Joseph and Luria, Michal and Kaufman, Geoff and Hammer, Jessica},
  year = 2019,
  month = may,
  series = {{{CHI}} '19},
  pages = {1--13},
  publisher = {Association for Computing Machinery},
  address = {New York, NY, USA},
  doi = {10.1145/3290605.3300680},
  urldate = {2025-03-18},
  isbn = {978-1-4503-5970-2}
}

@misc{shewaleDiscordUsersStatistics2025,
  title = {Discord {{Users Statistics}} 2025: {{Key Stats}}, {{Growth And More}}},
  shorttitle = {Discord {{Users Statistics}} 2025},
  author = {Shewale, Rohit},
  year = 2025,
  month = apr,
  urldate = {2025-09-04},
  chapter = {Social},
  langid = {american}
}

@inproceedings{shutskoUserGeneratedShortVideo2020,
  title = {User-{{Generated Short Video Content}} in {{Social Media}}. {{A Case Study}} of {{TikTok}}},
  booktitle = {Social {{Computing}} and {{Social Media}}. {{Participation}}, {{User Experience}}, {{Consumer Experience}}, and {{Applications}} of {{Social Computing}}},
  author = {Shutsko, Aliaksandra},
  editor = {Meiselwitz, Gabriele},
  year = 2020,
  pages = {108--125},
  publisher = {Springer International Publishing},
  address = {Cham},
  doi = {10.1007/978-3-030-49576-3_8},
  isbn = {978-3-030-49576-3},
  langid = {english}
}

@article{simona-nicoletavoicuPREVENTIVEONLINESAFETY2023,
  title = {{{PREVENTIVE ONLINE SAFETY EDUCATION FOR TEENAGERS}}},
  author = {{Simona-Nicoleta VOICU} and {Ioan CR\u ACIUN}},
  year = 2023,
  journal = {International Journal of Legal and Social Order},
  volume = {3},
  number = {1},
  pages = {571--579},
  publisher = {Centrul de Cercetare \&\#238;n Drept SARA},
  issn = {2810-4188, 2821-4161},
  urldate = {2025-09-08},
  langid = {english}
}

@inproceedings{singhTheyBasicallyDestroyed2017,
  title = {"{{They}} Basically like Destroyed the School One Day": {{On Newer App Features}} and {{Cyberbullying}} in {{Schools}}},
  shorttitle = {"{{They}} Basically like Destroyed the School One Day"},
  booktitle = {Proceedings of the 2017 {{ACM Conference}} on {{Computer Supported Cooperative Work}} and {{Social Computing}}},
  author = {Singh, Vivek K. and Radford, Marie L. and Huang, Qianjia and Furrer, Susan},
  year = 2017,
  month = feb,
  series = {{{CSCW}} '17},
  pages = {1210--1216},
  publisher = {Association for Computing Machinery},
  address = {New York, NY, USA},
  doi = {10.1145/2998181.2998279},
  urldate = {2025-06-11},
  isbn = {978-1-4503-4335-0}
}

@techreport{sonialivingstoneRisksSafetyInternet2011,
  title = {Risks and Safety on the Internet: The Perspective of {{European}} Children: Full Findings and Policy Implications from the {{EU Kids Online}} Survey of 9-16 Year Olds and Their Parents in 25 Countries},
  author = {{Sonia Livingstone} and {Leslie Haddon} and {Anke G\"orzig} and {Kjartan \'Olafsson}},
  year = 2011,
  address = {London, UK},
  institution = {EU Kids Online Network}
}

@article{sparbyDigitalSocialMedia2017,
  title = {Digital {{Social Media}} and {{Aggression}}: {{Memetic Rhetoric}} in 4chan's {{Collective Identity}}},
  shorttitle = {Digital {{Social Media}} and {{Aggression}}},
  author = {Sparby, Derek M.},
  year = 2017,
  month = sep,
  journal = {Computers and Composition},
  volume = {45},
  pages = {85--97},
  issn = {8755-4615},
  doi = {10.1016/j.compcom.2017.06.006},
  urldate = {2025-08-28}
}

@article{staksrudCHILDRENONLINERISK2009,
  title = {{{CHILDREN AND ONLINE RISK}}: {{Powerless}} Victims or Resourceful Participants?},
  shorttitle = {{{CHILDREN AND ONLINE RISK}}},
  author = {Staksrud, Elisabeth and Livingstone, Sonia},
  year = 2009,
  month = apr,
  journal = {Information, Communication \& Society},
  volume = {12},
  number = {3},
  pages = {364--387},
  issn = {1369-118X, 1468-4462},
  doi = {10.1080/13691180802635455},
  urldate = {2024-10-04},
  langid = {english}
}

@article{stanickeChoseBadYouths2023,
  title = {`{{I}} Chose the Bad': {{Youth}}'s Meaning Making of Being Involved in Self-harm Content Online during Adolescence},
  shorttitle = {`{{I}} Chose the Bad'},
  author = {St{\"a}nicke, Line Indrevoll},
  year = 2023,
  month = feb,
  journal = {Child \& Family Social Work},
  volume = {28},
  number = {1},
  pages = {160--170},
  publisher = {Wiley},
  issn = {1356-7500, 1365-2206},
  doi = {10.1111/cfs.12950},
  urldate = {2025-07-11},
  copyright = {http://creativecommons.org/licenses/by-nc-nd/4.0/},
  langid = {english}
}

@article{sumnerTemporalGeographicPatterns2019,
  title = {Temporal and {{Geographic Patterns}} of {{Social Media Posts About}} an {{Emerging Suicide Game}}},
  author = {Sumner, Steven A. and Galik, Stacey and Mathieu, Jennifer and Ward, Megan and Kiley, Thomas and Bartholow, Brad and Dingwall, Alison and Mork, Peter},
  year = 2019,
  month = jul,
  journal = {The Journal of Adolescent Health: Official Publication of the Society for Adolescent Medicine},
  volume = {65},
  number = {1},
  pages = {94--100},
  issn = {1879-1972},
  doi = {10.1016/j.jadohealth.2018.12.025},
  langid = {english},
  pmcid = {PMC7164676},
  pmid = {30819581}
}

@article{sunsteinAdolescentRisktakingSocial2008,
  title = {Adolescent Risk-Taking and Social Meaning: {{A}} Commentary},
  shorttitle = {Adolescent Risk-Taking and Social Meaning},
  author = {Sunstein, Cass R.},
  year = 2008,
  month = mar,
  journal = {Developmental Review},
  series = {Current {{Directions}} in {{Risk}} and {{Decision Making}}},
  volume = {28},
  number = {1},
  pages = {145--152},
  issn = {0273-2297},
  doi = {10.1016/j.dr.2007.11.003},
  urldate = {2025-09-03}
}

@article{taddeiPrivacyTrustControl2013,
  title = {Privacy, Trust and Control: {{Which}} Relationships with Online Self-Disclosure?},
  shorttitle = {Privacy, Trust and Control},
  author = {Taddei, Stefano and Contena, Bastianina},
  year = 2013,
  month = may,
  journal = {Computers in Human Behavior},
  volume = {29},
  number = {3},
  pages = {821--826},
  issn = {0747-5632},
  doi = {10.1016/j.chb.2012.11.022},
  urldate = {2025-09-04}
}

@article{tcherdakoffIveBecomeMore2025,
  title = {``{{I}}'ve {{Become More Myself}}'': {{Challenges}} and {{Benefits}} of {{Engaging}} with {{ADHD Short-Form Video Content}} and {{Communities}}},
  shorttitle = {``{{I}}'ve {{Become More Myself}}''},
  author = {Tcherdakoff, Nathalie Alexandra ``Alex'' and Cox, Anna L. and Bird, Jon and Marshall, Paul},
  year = 2025,
  month = nov,
  journal = {ACM Trans. Comput.-Hum. Interact.},
  issn = {1073-0516},
  doi = {10.1145/3778352},
  urldate = {2025-12-04}
}

@article{thomasInstagramToolStudy2020,
  title = {Instagram as a Tool for Study Engagement and Community Building among Adolescents: {{A}} Social Media Pilot Study},
  shorttitle = {Instagram as a Tool for Study Engagement and Community Building among Adolescents},
  author = {Thomas, Veronica L and Chavez, Marisol and Browne, Erica N and Minnis, Alexandra M},
  year = 2020,
  month = jan,
  journal = {DIGITAL HEALTH},
  volume = {6},
  pages = {2055207620904548},
  issn = {2055-2076, 2055-2076},
  doi = {10.1177/2055207620904548},
  urldate = {2025-09-09},
  langid = {english}
}

@inproceedings{thomasSoKHateHarassment2021,
  title = {{{SoK}}: {{Hate}}, {{Harassment}}, and the {{Changing Landscape}} of {{Online Abuse}}},
  shorttitle = {{{SoK}}},
  booktitle = {2021 {{IEEE Symposium}} on {{Security}} and {{Privacy}} ({{SP}})},
  author = {Thomas, Kurt and Akhawe, Devdatta and Bailey, Michael and Boneh, Dan and Bursztein, Elie and Consolvo, Sunny and Dell, Nicola and Durumeric, Zakir and Kelley, Patrick Gage and Kumar, Deepak and McCoy, Damon and Meiklejohn, Sarah and Ristenpart, Thomas and Stringhini, Gianluca},
  year = 2021,
  month = may,
  pages = {247--267},
  issn = {2375-1207},
  doi = {10.1109/SP40001.2021.00028},
  urldate = {2025-03-05}
}

@article{valkenburgOnlineCommunicationAdolescents2011,
  title = {Online {{Communication Among Adolescents}}: {{An Integrated Model}} of {{Its Attraction}}, {{Opportunities}}, and {{Risks}}},
  shorttitle = {Online {{Communication Among Adolescents}}},
  author = {Valkenburg, Patti M. and Peter, Jochen},
  year = 2011,
  month = feb,
  journal = {Journal of Adolescent Health},
  volume = {48},
  number = {2},
  pages = {121--127},
  issn = {1054-139X},
  doi = {10.1016/j.jadohealth.2010.08.020},
  urldate = {2025-08-05}
}

@article{vanderhofChildsRightProtection2020,
  title = {The {{Child}}'s {{Right}} to {{Protection}} against {{Economic Exploitation}} in the {{Digital World}}},
  author = {Van Der Hof, Simone and Lievens, E. and Milkaite, I. and Verdoodt, V. and Hannema, T. and Liefaard, T.},
  year = 2020,
  month = dec,
  journal = {The International Journal of Children's Rights},
  volume = {28},
  number = {4},
  pages = {833--859},
  issn = {0927-5568, 1571-8182},
  doi = {10.1163/15718182-28040003},
  urldate = {2025-08-31},
  copyright = {https://creativecommons.org/licenses/by-nc/4.0/}
}

@article{vandoninckOnlineRisksCoping2013,
  title = {Online {{Risks}}: {{Coping}} Strategies of Less Resilient Children and Teenagers across {{Europe}}},
  shorttitle = {Online {{Risks}}},
  author = {Vandoninck, Sofie and {d'Haenens}, Leen and Roe, Keith},
  year = 2013,
  month = feb,
  journal = {Journal of Children and Media},
  volume = {7},
  number = {1},
  pages = {60--78},
  issn = {1748-2798, 1748-2801},
  doi = {10.1080/17482798.2012.739780},
  urldate = {2025-09-08},
  langid = {english}
}

@article{vanouytselAdolescentsSexySelfPresentation2020,
  title = {Adolescents' {{Sexy Self-Presentation}} on {{Instagram}}: {{An Investigation}} of {{Their Posting Behavior Using}} a {{Prototype Willingness Model Perspective}}},
  shorttitle = {Adolescents' {{Sexy Self-Presentation}} on {{Instagram}}},
  author = {Van Ouytsel, Joris and Walrave, Michel and Ojeda, M{\'o}nica and Del Rey, Rosario and Ponnet, Koen},
  year = 2020,
  month = jan,
  journal = {International Journal of Environmental Research and Public Health},
  volume = {17},
  number = {21},
  pages = {8106},
  publisher = {Multidisciplinary Digital Publishing Institute},
  issn = {1660-4601},
  doi = {10.3390/ijerph17218106},
  urldate = {2025-09-12},
  copyright = {http://creativecommons.org/licenses/by/3.0/},
  langid = {english}
}

@article{vanschaikSecurityPrivacyOnline2018,
  title = {Security and Privacy in Online Social Networking: {{Risk}} Perceptions and Precautionary Behaviour},
  shorttitle = {Security and Privacy in Online Social Networking},
  author = {{van Schaik}, Paul and Jansen, Jurjen and Onibokun, Joseph and Camp, Jean and Kusev, Petko},
  year = 2018,
  month = jan,
  journal = {Computers in Human Behavior},
  volume = {78},
  pages = {283--297},
  issn = {0747-5632},
  doi = {10.1016/j.chb.2017.10.007},
  urldate = {2025-12-04}
}

@article{wangProtectionPunishmentRelating2021,
  title = {Protection or {{Punishment}}? {{Relating}} the {{Design Space}} of {{Parental Control Apps}} and {{Perceptions}} about {{Them}} to {{Support Parenting}} for {{Online Safety}}},
  shorttitle = {Protection or {{Punishment}}?},
  author = {Wang, Ge and Zhao, Jun and Van Kleek, Max and Shadbolt, Nigel},
  year = 2021,
  month = oct,
  journal = {Proceedings of the ACM on Human-Computer Interaction},
  volume = {5},
  number = {CSCW2},
  pages = {1--26},
  issn = {2573-0142},
  doi = {10.1145/3476084},
  urldate = {2024-10-03},
  langid = {english}
}

@techreport{websterEuropeanOnlineGrooming2012,
  title = {European {{Online Grooming Project}} - {{Final Report}}},
  shorttitle = {European {{Online Grooming Project}} - {{Final Report}}},
  author = {Webster, Stephen and Davidson, Julia and Bifulco, Antonia and Gottschalk, Peter and Caretti, Vincenzo and Pham, Thierry and {Grove-Hills}, Julie and Turley, Caroline and Tompkins, Charlotte and Ciulla, Stefano and Milazzo, Vanessa and Schimmenti, Adriano and Craparo, Giuseppe},
  year = 2012,
  month = mar,
  pages = {152}
}

@book{weinsteinTheirScreensWhat2022,
  title = {Behind Their Screens: What Teens Are Facing (and Adults Are Missing)},
  shorttitle = {Behind Their Screens},
  author = {Weinstein, Emily and James, Carrie},
  year = 2022,
  publisher = {the MIT press},
  address = {Cambridge (Mass.)},
  isbn = {978-0-262-04735-7},
  langid = {english},
  lccn = {302.231 083}
}

@inproceedings{wisniewskiAdolescentOnlineSafety2014,
  title = {Adolescent Online Safety: The "Moral" of the Story},
  shorttitle = {Adolescent Online Safety},
  booktitle = {Proceedings of the 17th {{ACM}} Conference on {{Computer}} Supported Cooperative Work \& Social Computing},
  author = {Wisniewski, Pamela J. and Xu, Heng and Rosson, Mary Beth and Carroll, John M.},
  year = 2014,
  month = feb,
  series = {{{CSCW}} '14},
  pages = {1258--1271},
  publisher = {Association for Computing Machinery},
  address = {New York, NY, USA},
  doi = {10.1145/2531602.2531696},
  urldate = {2025-06-06},
  isbn = {978-1-4503-2540-0}
}

@inproceedings{wisniewskiDearDiaryTeens2016,
  title = {Dear {{Diary}}: {{Teens Reflect}} on {{Their Weekly Online Risk Experiences}}},
  shorttitle = {Dear {{Diary}}},
  booktitle = {Proceedings of the 2016 {{CHI Conference}} on {{Human Factors}} in {{Computing Systems}}},
  author = {Wisniewski, Pamela and Xu, Heng and Rosson, Mary Beth and Perkins, Daniel F. and Carroll, John M.},
  year = 2016,
  month = may,
  pages = {3919--3930},
  publisher = {ACM},
  address = {San Jose California USA},
  doi = {10.1145/2858036.2858317},
  urldate = {2024-10-04},
  isbn = {978-1-4503-3362-7},
  langid = {english}
}

@incollection{wisniewskiMovingFearRestriction2025,
  title = {Moving {{Beyond Fear}} and {{Restriction}} to {{Promoting Adolescent Resilience}} and {{Intentional Technology Use}}},
  booktitle = {Handbook of {{Children}} and {{Screens}}: {{Digital Media}}, {{Development}}, and {{Well-Being}} from {{Birth Through Adolescence}}},
  author = {Wisniewski, Pamela and Park, Jinkyung and {Badillo-Urquiola}, Karla and Gabrielli, Joy and Doty, Jennifer L. and Hartikainen, Heidi},
  editor = {Christakis, Dimitri A. and Hale, Lauren},
  year = 2025,
  pages = {403--410},
  publisher = {Springer Nature Switzerland},
  address = {Cham},
  doi = {10.1007/978-3-031-69362-5_55},
  urldate = {2025-06-29},
  isbn = {978-3-031-69362-5},
  langid = {english}
}

@inproceedings{wisniewskiParentalControlVs2017,
  title = {Parental {{Control}} vs. {{Teen Self-Regulation}}: {{Is}} There a Middle Ground for Mobile Online Safety?},
  shorttitle = {Parental {{Control}} vs. {{Teen Self-Regulation}}},
  booktitle = {Proceedings of the 2017 {{ACM Conference}} on {{Computer Supported Cooperative Work}} and {{Social Computing}}},
  author = {Wisniewski, Pamela and Ghosh, Arup Kumar and Xu, Heng and Rosson, Mary Beth and Carroll, John M.},
  year = 2017,
  month = feb,
  series = {{{CSCW}} '17},
  pages = {51--69},
  publisher = {Association for Computing Machinery},
  address = {New York, NY, USA},
  doi = {10.1145/2998181.2998352},
  urldate = {2024-10-15},
  isbn = {978-1-4503-4335-0}
}

@inproceedings{wisniewskiPreventativeVsReactive2015,
  title = {"{{Preventative}}" vs. "{{Reactive}}": {{How Parental Mediation Influences Teens}}' {{Social Media Privacy Behaviors}}},
  shorttitle = {"{{Preventative}}" vs. "{{Reactive}}"},
  booktitle = {Proceedings of the 18th {{ACM Conference}} on {{Computer Supported Cooperative Work}} \& {{Social Computing}}},
  author = {Wisniewski, Pamela and Jia, Haiyan and Xu, Heng and Rosson, Mary Beth and Carroll, John M.},
  year = 2015,
  month = feb,
  series = {{{CSCW}} '15},
  pages = {302--316},
  publisher = {Association for Computing Machinery},
  address = {New York, NY, USA},
  doi = {10.1145/2675133.2675293},
  urldate = {2025-09-08},
  isbn = {978-1-4503-2922-4}
}

@article{wisniewskiPrivacyParadoxAdolescent2018,
  title = {The {{Privacy Paradox}} of {{Adolescent Online Safety}}: {{A Matter}} of {{Risk Prevention}} or {{Risk Resilience}}?},
  shorttitle = {The {{Privacy Paradox}} of {{Adolescent Online Safety}}},
  author = {Wisniewski, Pamela},
  year = 2018,
  month = mar,
  journal = {IEEE Security \& Privacy},
  volume = {16},
  number = {2},
  pages = {86--90},
  issn = {1558-4046},
  doi = {10.1109/MSP.2018.1870874},
  urldate = {2025-06-07}
}

@inproceedings{wohnVolunteerModeratorsTwitch2019,
  title = {Volunteer {{Moderators}} in {{Twitch Micro Communities}}: {{How They Get Involved}}, the {{Roles They Play}}, and the {{Emotional Labor They Experience}}},
  shorttitle = {Volunteer {{Moderators}} in {{Twitch Micro Communities}}},
  booktitle = {Proceedings of the 2019 {{CHI Conference}} on {{Human Factors}} in {{Computing Systems}}},
  author = {Wohn, Donghee Yvette},
  year = 2019,
  month = may,
  series = {{{CHI}} '19},
  pages = {1--13},
  publisher = {Association for Computing Machinery},
  address = {New York, NY, USA},
  doi = {10.1145/3290605.3300390},
  urldate = {2025-12-03},
  isbn = {978-1-4503-5970-2}
}

@article{wolakOnlinePredatorsTheir2010,
  title = {Online ``Predators'' and Their Victims: {{Myths}}, Realities, and Implications for Prevention and Treatment},
  shorttitle = {Online ``Predators'' and Their Victims},
  author = {Wolak, Janis and Finkelhor, David and Mitchell, Kimberly J. and Ybarra, Michele L.},
  year = 2010,
  journal = {Psychology of Violence},
  volume = {1},
  number = {S},
  pages = {13--35},
  publisher = {Educational Publishing Foundation},
  address = {US},
  issn = {2152-081X},
  doi = {10.1037/2152-0828.1.S.13}
}

@article{wuHowYouQuantify2023,
  title = {"{{How Do You Quantify How Racist Something Is}}?": {{Color-Blind Moderation}} in {{Decentralized Governance}}},
  shorttitle = {"{{How Do You Quantify How Racist Something Is}}?},
  author = {Wu, Qunfang and Semaan, Bryan},
  year = 2023,
  month = oct,
  journal = {Proc. ACM Hum.-Comput. Interact.},
  volume = {7},
  number = {CSCW2},
  pages = {239:1--239:27},
  doi = {10.1145/3610030},
  urldate = {2025-09-12}
}

@inproceedings{yardiSocialTechnicalChallenges2011,
  title = {Social and Technical Challenges in Parenting Teens' Social Media Use},
  booktitle = {Proceedings of the {{SIGCHI Conference}} on {{Human Factors}} in {{Computing Systems}}},
  author = {Yardi, Sarita and Bruckman, Amy},
  year = 2011,
  month = may,
  series = {{{CHI}} '11},
  pages = {3237--3246},
  publisher = {Association for Computing Machinery},
  address = {New York, NY, USA},
  doi = {10.1145/1978942.1979422},
  urldate = {2024-10-07},
  isbn = {978-1-4503-0228-9}
}

@article{yauItsJustLot2019,
  title = {``{{It}}'s {{Just}} a {{Lot}} of {{Work}}'': {{Adolescents}}' {{Self-Presentation Norms}} and {{Practices}} on {{Facebook}} and {{Instagram}}},
  shorttitle = {``{{It}}'s {{Just}} a {{Lot}} of {{Work}}''},
  author = {Yau, Joanna C. and Reich, Stephanie M.},
  year = 2019,
  journal = {Journal of Research on Adolescence},
  volume = {29},
  number = {1},
  pages = {196--209},
  issn = {1532-7795},
  doi = {10.1111/jora.12376},
  urldate = {2025-09-12},
  copyright = {\copyright{} 2018 Society for Research on Adolescence},
  langid = {english}
}

@article{ybarraHowRiskyAre2008,
  title = {How {{Risky Are Social Networking Sites}}? {{A Comparison}} of {{Places Online Where Youth Sexual Solicitation}} and {{Harassment Occurs}}},
  shorttitle = {How {{Risky Are Social Networking Sites}}?},
  author = {Ybarra, Michele L. and Mitchell, Kimberly J.},
  year = 2008,
  month = feb,
  journal = {Pediatrics},
  volume = {121},
  number = {2},
  pages = {e350-e357},
  issn = {0031-4005},
  doi = {10.1542/peds.2007-0693},
  urldate = {2025-11-29}
}

@article{yoonItsGreatBecause2025,
  title = {"{{It}}'s {{Great Because It}}'s {{Ran By Us}}": {{Empowering Teen Volunteer Discord Moderators}} to {{Design Healthy}} and {{Engaging Youth-Led Online Communities}}},
  shorttitle = {"{{It}}'s {{Great Because It}}'s {{Ran By Us}}"},
  author = {Yoon, Jina and Zhang, Amy X. and Seering, Joseph},
  year = 2025,
  month = may,
  journal = {Proc. ACM Hum.-Comput. Interact.},
  volume = {9},
  number = {2},
  pages = {CSCW121:1--CSCW121:30},
  doi = {10.1145/3711114},
  urldate = {2025-06-06}
}

@inproceedings{zhang-kennedyNosyLittleBrothers2016,
  title = {From {{Nosy Little Brothers}} to {{Stranger-Danger}}: {{Children}} and {{Parents}}' {{Perception}} of {{Mobile Threats}}},
  shorttitle = {From {{Nosy Little Brothers}} to {{Stranger-Danger}}},
  booktitle = {Proceedings of the {{The}} 15th {{International Conference}} on {{Interaction Design}} and {{Children}}},
  author = {{Zhang-Kennedy}, Leah and Mekhail, Christine and Abdelaziz, Yomna and Chiasson, Sonia},
  year = 2016,
  month = jun,
  series = {{{IDC}} '16},
  pages = {388--399},
  publisher = {Association for Computing Machinery},
  address = {New York, NY, USA},
  doi = {10.1145/2930674.2930716},
  urldate = {2025-12-03},
  isbn = {978-1-4503-4313-8}
}

@article{zhangBurstCollaborativeCuration2025,
  title = {Burst: {{Collaborative Curation}} in {{Connected Social Media Communities}}},
  shorttitle = {Burst},
  author = {Zhang, Yutong and Kang, Taeuk and Yeh, Sydney and Baddepudi, Anavi and Popowski, Lindsay and Piccardi, Tiziano and Bernstein, Michael S.},
  year = 2025,
  month = oct,
  journal = {Proc. ACM Hum.-Comput. Interact.},
  volume = {9},
  number = {7},
  pages = {CSCW382:1--CSCW382:29},
  doi = {10.1145/3757563},
  urldate = {2025-11-16}
}

@inproceedings{zhangDangerousPlaygroundsChild2025,
  title = {Dangerous {{Playgrounds}}: {{Child Players}}' {{Encounters}} with {{Design-Mediated Risks}} on {{User Generated Game Platforms}} and {{Their Safety Practices}}},
  shorttitle = {Dangerous {{Playgrounds}}},
  booktitle = {Proceedings of the 24th {{Interaction Design}} and {{Children}}},
  author = {Zhang, Zinan and Gui, Xinning and Yu, Junnan and Bai, Sunhye and Kou, Yubo},
  year = 2025,
  month = jun,
  series = {{{IDC}} '25},
  pages = {296--313},
  publisher = {Association for Computing Machinery},
  address = {New York, NY, USA},
  doi = {10.1145/3713043.3728858},
  urldate = {2025-07-22},
  isbn = {979-8-4007-1473-3}
}

@article{zhangExplainableEmpiricalRisk2024,
  title = {Explainable Empirical Risk Minimization},
  author = {Zhang, Linli and Karakasidis, Georgios and Odnoblyudova, Arina and Dogruel, Leyla and Tian, Yu and Jung, Alex},
  year = 2024,
  month = mar,
  journal = {Neural Computing and Applications},
  volume = {36},
  number = {8},
  pages = {3983--3996},
  issn = {1433-3058},
  doi = {10.1007/s00521-023-09269-3},
  urldate = {2025-09-11},
  langid = {english}
}

@article{zhaoUnderstandingTeenagePerceptions2022,
  title = {Understanding {{Teenage Perceptions}} and {{Configurations}} of {{Privacy}} on {{Instagram}}},
  author = {Zhao, Dorothy and Inaba, Mikako and {Monroy-Hern{\'a}ndez}, Andr{\'e}s},
  year = 2022,
  month = nov,
  journal = {Proc. ACM Hum.-Comput. Interact.},
  volume = {6},
  number = {CSCW2},
  pages = {550:1--550:28},
  doi = {10.1145/3555608},
  urldate = {2025-09-12}
}

\appendix
\section{Discord Safety Guidelines for Teens by Teens}
\begin{enumerate}
    \item Upon joining a new server, get familiar with how the server is moderated. This helps you know who to contact when there is a present risk.
    \begin{enumerate}
        \item Moderators may be a person, a team of people, a Discord bot.
    \end{enumerate}
    \item Learn about a server’s culture before fully diving in by reading general chats.
    \begin{enumerate}
        \item When doing so, ask yourself:
        \begin{enumerate}
            \item Do these people seem like they could be good friends?
            \item Do they make strange jokes or try to make others feel bad?
        \end{enumerate}
    \end{enumerate}
    \item You can always report and block people for anything they do that is wrong towards you or others.
    \item Your safety is not only your responsibility. Especially if your experiences have been disturbing, make sure moderators, your parents, or trusted friends know what you are going through.
\end{enumerate}

\section{Interview Protocol}
Experiences using Discord:
\begin{enumerate}
    \item How long have you been using Discord and what made you start using it?
    \item What do you usually do when talking with new people on Discord?
    \item Can you share an experience on Discord that you found difficult or unexpected?
\end{enumerate}
Risky Social Interactions on Discord:
\begin{enumerate}
    \item Can you describe any experiences on Discord that you found concerning or uncomfortable?
    \begin{enumerate}
        \item How do you deal with this?
    \end{enumerate}
    \item Have there been any upsetting feelings over something that happened on Discord?
    \item Is there a time when you had an uncomfortable experience with another user?
    \item If a connection on Discord isn’t working out, how do you usually handle it?
    \item Can you describe a time you met someone on Discord who was difficult to deal with?
\end{enumerate}
Discord Safety Design Questions:
\begin{enumerate}
    \item Can you share what makes you feel safe in servers? Do you use any tools or settings?
    \item If you could change or add anything to how safety works on Discord, what would that be?
    \item Can you describe anything you learned from using Discord’s safety tools?
\end{enumerate}

\section{Participant Demographics}

\begin{table}[H]
\centering
\caption{Participant Demographics}
\label{tab:participant-demographics}
\begin{tabular}{llll}
\hline
\textbf{ID} & \textbf{Age} & \textbf{Gender} & \textbf{Racial or Ethnic Identity} \\
\hline
P1  & 17 & Female      & Asian/Korean \\
P2  & 14 & Male        & Asian \\
P3  & 15 & Female      & Asian \\
P4  & 15 & Female      & Asian \\
P5  & 14 & Male        & White \\
P6  & 15 & Female      & Asian \\
P7  & 16 & Female      & White \\
P8  & 13 & Female      & White \\
P9  & 15 & Male        & Asian \\
P10 & 14 & Male        & Asian \\
P11 & 14 & Female      & Asian \\
P12 & 16 & Male        & Asian \\
P13 & 17 & Male        & Asian \\
P14 & 14 & Non-binary  & White \\
P15 & 17 & Male        & White \\
P16 & 14 & Female      & Mexican/European \\
\hline
\end{tabular}

\raggedright
\footnotesize
\textsuperscript{a} Participants were encouraged to self-describe their background along racial or ethnic lines that they preferred.
\end{table}

\end{document}